\documentclass[sigconf]{acmart}

\usepackage[english]{babel}
\usepackage{subfigure}
\usepackage{blindtext}

\renewcommand\footnotetextcopyrightpermission[1]{} 
\setcopyright{none}

\acmDOI{}

\acmISBN{}

\acmConference[To be submitted for review to NetAI Workshop at SIGCOMM]{}
\acmYear{2020}
\copyrightyear{}

\acmPrice{}

\begin{document}
\title{NetML: A Challenge for Network Traffic Analytics}


\author{Onur Barut}
 \affiliation{
   \institution{University of Massachusetts Lowell}
   \streetaddress{220 Pawtucket Street}
   \city{Lowell} 
   \state{MA} 
   \postcode{01854}
 }
\author{Yan Luo}
 \affiliation{
   \institution{University of Massachusetts Lowell}
   \streetaddress{220 Pawtucket Street}
   \city{Lowell} 
   \state{MA} 
   \postcode{01854}
 }
\author{Tong Zhang}
 \affiliation{
   \institution{Intel Corporation}
   \streetaddress{2200 Mission College Boulevard}
   \city{Santa Clara} 
   \state{CA} 
   \postcode{95054}
 }
\author{Weigang Li}
 \affiliation{
   \institution{Intel Corporation}
   \streetaddress{No. 880 Zi Xing Road, Shanghai Zizhu Science Park, Min Hang District, Shanghai, China}
   \city{Santa Clara} 
   \state{CA} 
   \postcode{200241}
 }
 \author{Peilong Li}
 \affiliation{
   \institution{Elizabethtown College}
   \streetaddress{1 Alpha Dr}
   \city{Elizabethtown} 
   \state{PA} 
   \postcode{17022}
 }

\renewcommand{\shortauthors}{B.et al.}

\begin{abstract}

Classifying network traffic is the basis for important network applications.
Prior research in this area has faced challenges on the availability of representative datasets, and many of the results cannot be readily reproduced. Such a problem is exacerbated by emerging data-driven machine learning based approaches. To address this issue, we provide three open datasets containing almost 1.3M labeled flows in total, with flow features and anonymized raw packets, for the research community. We focus on broad aspects in network traffic analysis, including both malware detection and application classification. We release the datasets in the form of an open challenge called NetML\footnote{https://github.com/ACANETS/NetML-Competition2020} and implement several machine learning methods including random-forest, SVM and MLP. As we continue to grow NetML, we expect the datasets to serve as a common platform for AI driven, reproducible research on network flow analytics.

\end{abstract}

\maketitle

\section{Introduction}
Recent advances in technology have provided researchers a lot of data available for analysis. In particular, many studies have been published in Computer Vision (CV), Natural Language Processing (NLP), and lately Network Traffic Analysis (NTA) investigating big data with Artificial Intelligence (AI) approaches \cite{Taylor_2016_eta, Wang_2018_datanet, Yao_2019_attentionLSTM, Gao_2020_cicids2017_nsl_kdd}. Even though many benchmarks and challenges are available for CV \cite{imagenet_cvpr09} and NLP fields \cite{senseval}, there is a lack of common ground for network traffic analysis for researchers to compare their results with others.

In the twenty-first century and digital era, network traffic classification has become very important task with rapidly increasing amounts of internet devices connected together. Network traffic analysis, which is the classification of network traffic into appropriate classes, has become vital for many applications such as quality of service (QoS) control, resource allocation and malware detection. Particularly from cybersecurity perspective, a growing number of connected devices are subject to cyber attackers and therefore it is a must to take required precautions against these ever-increasing attempts of intrusions. However, it is very difficult to find comprehensive datasets for researchers to evaluate their proposed techniques. In addition to that, having an appropriate dataset is another major challenge too.

Network traffic analysis (NTA) techniques have evolved from port-based approaches up to machine learning based techniques over time. The first and easiest port-based approach has become obsolete since newer applications mostly use dynamic port allocation instead of using standard registered port numbers. Then, network researchers begin using payload or data packet inspection (DPI). As the volume of encrypted traffic increases, payload-based approaches fail which directed the researchers to employ machine learning methods with flow statistical features since they do not rely on port numbers or payload itself.

There have been a plethora of research attempting to classify and analyze network flows using a variety of datasets. However, unlike the open datasets such as ImageNet \cite{imagenet_cvpr09} and COCO \cite{lin2014microsoft_coco} in computer vision research, it is very difficult to find comprehensive datasets for researchers in networking domain to evaluate their proposed techniques, and for the community to reproduce prior art. Many agree that a comprehensive, up-to-date and open dataset for flow analytics is dispensable for the network research community.

In this paper, we address a common ground for NTA and introduce NetML Network Traffic Analytics datasets with three specific collection of labeled network flows and their features for researchers as a benchmarking platform to evaluate their approaches and contribute to NTA research. In order to be comprehensive, we compile datasets for both malware detection and application type categorization, two representative applications. For malware detection tasks, we introduce a new dataset NetML with raw traffic captures obtained from Stratosphere IPS \cite{stratosphere} website with almost $500k$ network flow samples belonging to $20$ different types of malware and benign classes. We also create another dataset for malware detection which is obtained from the raw traffic captures from well-known CICIDS2017 dataset \cite{CICIDS2017}. We generate around $550k$ flow samples for $7$ different malware types and a benign class. Lastly, we introduce non-vpn2016 dataset with around $163k$ flow samples whose raw traffic data are acquired from ISCX-VPN-nonVPN2016 dataset \cite{iscx_vpn2016} containing different levels of annotations with a number of classes ranging from $7$ to $31$. On top of those data, we provide preliminary data analysis on flow features and baseline results achieved using popular machine learning algorithms including Random Forest (RF), Support Vector Machine (SVM) and Multi-layer Perceptron (MLP). We host the dataset and release the implementations of the algorithms on github and encourage the community to collaborate and grow the collection.

\begin{sloppypar}
Our contribution in this paper is twofold: First, we provide a novel malware detection NetML dataset and curate two other datasets originated from open source captures (named as CICIDS2017 for malware detection and non-vpn2016 dataset for traffic classification). Second, we provide a platform for researchers to participate in our firstly announced network traffic analytics challenge and baseline results obtained by RF, SVM and MLP models. We choose only Metadata features (such as number of packets, number of bytes, time duration etc.) to expedite the efficiency of baseline classifiers; however, we strongly encourage researchers to investigate TLS, DNS and HTTP features as well. 
\end{sloppypar}

This paper is organized as follows: Section 2 gives an overview on literature search for malware detection and traffic classification tasks. Section 3 explains the dataset collection and preparation process along with a description of extracted flow features. Preliminary dataset analysis results with the flow features are explained in Section 4 while the results achieved by the proposed baseline models are given in Section 5. Finally, a brief information about the NetML Challenge and workshop is provided in Section 6 and this study is concluded in Section 7. In appendix, detailed graphs for flow feature analysis are available.

\section{Related Work}
Increasing attention on network traffic analytics brings many studies published recently. We will focus several of them within two categories: malware detection and application or traffic classification.

\subsection{Malware Detection}
Random Forest as ensemble classifier \cite{Zhang_2008_7ofLuNet} and Support Vector Machine (SVM) with Radial Basis Function (RBF) kernel (also called Gaussian kernel) as kernel machines \cite{Ahmad_2018_5ofLuNet} are two effective ways among many machine learning approaches for network traffic analysis. However, most of the researchers use their own data with their own features for NTA, which makes it difficult to compare one study with another. For instance, Anderson et al. \cite{AndersonBlake_2016} use flow metadata, sequence of packet length and times, byte distribution and unencrypted TLS header information to detect malware in TLS encrypted traffic. Similarly, features extracted by Moore et al. \cite{Moore_2013_DiscriminatorsFU} is a broad example of total 249 features used in several studies but only the different subsets of those features are used in other studies meaning that there is not a unique best set of features for NTA.

KDD-Cup 99 \cite{KDDcup99} and NSL-KDD datasets \cite{nsl_kdd} are common datasets widely used for malware detection. Several studies have been conducted on those datasets to investigate the ensemble model accuracies such as C4.5 algorithm, decision tree, random forest, SVM and AdaBoost \cite{Wang_2009_df, Farid_2010_naive, Hu_2008_adaboost, Ektefa_2010_datamining_ids}. After a two-level classification algorithm that uses naive Bayesian as the primary classifier in the first stage, and in the second stage they use nominal-binary filtering and cross-validation for testing is proposed by \cite{Panda_2010_twolevel}, further optimized two-level detection algorithm using balanced forest in the first stage and random forest for prediction in the second stage is implemented by \cite{Panda_2012_twolevel}.

However, those two datasets are out of date and prone to recently emerged attacks. CICIDS2017 dataset \cite{CICIDS2017} is relatively up-to-date and more suitable for detecting current attacks. Authors introduce CICIDS2017 dataset with their results obtained using CICFlowMeter \cite{flowmeter} features by comparing several techniques including k-nearest neighbor (kNN), random forest and ID3 and report best F1 score $0.98$ for ID3 algorithm and $0.97$ for random forest whose executing time is one-third of ID3. Gao et al. \cite{Gao_2020_cicids2017_nsl_kdd} compare several machine learning and deep learning techniques including Random Forest, SVM and Deep Neural Networks on NSL-KDD and CICIDS2017 datasets. They perform both binary malware detection and multi-class classification and find out that RF model achieves impressive malware detection accuracy and comparable classification accuracy to other deep learning models in multi-class classification task. 

\subsection{Traffic Classification}
There are several researches on application classification in network flow. For example, Conti et al. \cite{Conti_2016_useractions} use Random forest with their own dataset and their own features and achieve 95\% accuracy on user action classification. Taylor et al. \cite{Taylor_2016_eta} conclude that using statistical features of flow vectors of raw packet lengths such as minimum, maximum, mean, median, standard deviation on binary SVM and Random Forest results in ~95\% accuracy, but when using multi-class classifier, performance drops to ~42\% for SVM and ~87\% for RF. Lashkari et al. \cite{iscx_vpn2016} introduce UNB ISCX VPN-nonVPN2016 dataset and analyze it with kNN and C4.5 algorithms.

Many researchers adopt VPN-nonVPN2016 dataset and report their results. For example, Yamansavascilar et al. \cite{Baris_2017_iscx} utilize this dataset for application type classification and compare the results with their own dataset using the same classes. They report that kNN works best for VPN-nonVPN dataset and random forest produces most accurate results on their own dataset. They also investigate feature selection approaches and report 2\% boost in the accuracy of their own dataset. Wang et al. \cite{Wang_2018_datanet} use packet-level analysis and used deep learning approach for protocol classification. Later, Yao et al. \cite{Yao_2019_attentionLSTM} implement attention-based LSTM (Long Short-Term Memory) to classify protocol types in VPN-nonVPN dataset. On the other hand, Qin et al. \cite{netai_19} propose Hybrid Flow Clustering (HFC) method to explore flow features and link patterns in an unsupervised profiling view and achieve impressive classification performance on network traffic data obtained from three different open sources.

Previous studies show that the demand for analyzing network flow is increasing and there is not a single trivial solution for application classification or malware detection in network analysis. Therefore, there is an urgent need of a generic set of features that are known to be useful for malware detection and application classification with a more comprehensive and up-to-date dataset.

\section{NetML Data Collection}
We now describe our collection of data from several sources. We explain how we obtained raw data firstly for malware detection task and then application type classification task.

Stratosphere IPS \cite{stratosphere} and Canadian Institute of Cybersecurity (CIC) \cite{iscx_vpn2016, CICIDS2017} are two sources of raw traffic data (i.e pcap files) open to the public for research. CIC provides several different types of dataset for Intrusion Detection Systems (IDS) along with an application type classification dataset while Stratosphere IPS only focus on numerous normal and malware captures. Three datasets are compiled from these two sources of raw traffic data, referred as NetML, CICIDS2017 and non-vpn2016 for the rest of the paper. Raw capture data are preprocessed if needed and flow features are extracted using an Intel proprietary flow feature extraction tool. 

In this paper, we focus broadly on flow classification tasks for various purposes and at different granularity. We study both malware detection, which is closely related to cybersecurity, and more general flow type identification. We use "top-level", "mid-level" and "fine-grained" to denote the granularity levels. At "top-level", malware detection is a binary benign or malware classification problem, while for flow identification task, we annotate major application categories such as chat, email, video etc. The "mid-level" identifies specific applications such as facebook, skype, hangouts etc. Finally, the "fine-grained" level performs multi-class classification to identify specific types of malware (e.g. Adload, portScan etc.) and detailed types of application traffic such as facebook\_audio, skype\_audio etc.

\subsection{Dataset Preparation}
We use an accelerated feature extraction library provided by our sponsor Intel. It computes the features of a flow up to 200 packets in both directions. Given a raw traffic capture (pcap) file as input to the feature extraction tool, flow features are extracted in JSON format, and each flow sample is listed line by line in the output file. List of computed features by the feature extraction tool is given in Table \ref{flow_features}.

\begin{figure}[tp]
\centering
\includegraphics[width=8.5cm,height=8cm,keepaspectratio]{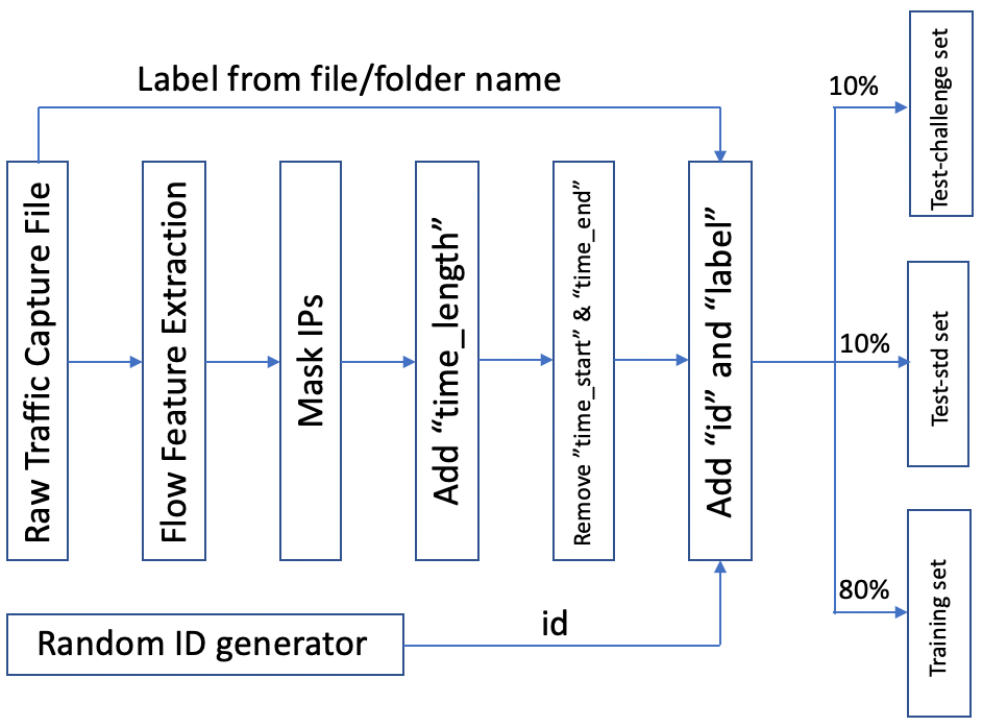}
\caption{Flow diagram of dataset preparation}
\label{data_preparation}
\end{figure}

We process the raw packet traces by following the steps depicted in Figure \ref{data_preparation}. We use an accelerated feature extraction library provided by Intel to compute the features of a flow up to $200$ packets in both directions. Given a raw traffic capture file as input to the feature extraction tool, flow features are extracted in JSON format, and each flow sample is listed line by line in the output file. Metadata features are extracted for each flow sample and TLS, DNS and HTTP features are extracted only if the flow sample contains packets for the given protocols. After feature extraction, we perform a little of preprocessing to prepare the dataset for a fair evaluation. Firstly, IPs of source \textit{sa} and destination \textit{da} are replaced with \textit{IP\_masked} string to mask the IPs so that the flows are not traceable by the competitors during the challenge. Secondly, \textit{time\_start} and \textit{time\_end} features are removed and the time interval between the two is added as a new feature called as \textit{time\_length}. Thirdly, a unique \textit{id} number to identify each flow sample and the label information of the flow according to the raw traffic packet capture file is added to output JSON files to obtain the NetML, CICIDS2017 and non-vpn2016 datasets with $484056$, $551372$ and $163711$ flows were obtained, respectively. Finally, the data are split into three sets, namely test-challenge, test-std, and training sets by allocating randomly selected $10\%$ of any class to test-challenge set, another randomly selected $10\%$ to test-std set, and the remaining $80\%$ to training set. The whole process for the data preparation is summarized in Figure \ref{data_preparation}.

\begin{table}[thpb]
\small
\caption{Description of flow features extracted by FFEL}
\label{flow_features}
\begin{center}
\begin{tabular}{|l|}
\hline
\textbf{Metadata Features} \\
\hline
\textbf{sa}: source address \\
\textbf{da}: destination address \\
\textbf{pr}: protocol (6 or 17) \\
\textbf{src\_port}: source port \\
\textbf{dst\_port}: destination port \\
\textbf{bytes\_out}: total bytes out \\
\textbf{num\_pkts\_out}: total packets out \\
\textbf{bytes\_in}: total bytes in \\
\textbf{num\_pkts\_in}: total packets in \\
\textbf{time\_start}: time stamp of first packet \\
\textbf{time\_end}: time stamp of last packet \\ 
\textbf{intervals\_ccnt[]}: compact histogram of pkt arriving intervals \\
\textbf{ack\_psh\_rst\_syn\_fin\_cnt[]}: histogram of tcp flag counting \\
\textbf{hdr\_distinct}: number of distinct values of header lengths \\
\textbf{hdr\_ccnt[]}: compact histogram of header lengths \\
\textbf{pld\_distinct}: number of distinct values of payload length \\
\textbf{pld\_ccnt[]}: compact histogram of payload lengths \\
\textbf{hdr\_mean}: mean value of header lengths \\
\textbf{hdr\_bin\_40}: \# of pkts with header lengths between 28 and 40 \\
\textbf{pld\_bin\_128}: \# of pkts whose payload lengths are below 128 \\
\textbf{pld\_bin\_inf}: \# of pkts whose payload lengths are above 1024 \\
\textbf{pld\_max}: max value of payload length \\
\textbf{pld\_mean}: mean value of payload length \\
\textbf{pld\_medium}: medium value of payload length \\
\textbf{pld\_var}: variance value of payload length \\ 
\textbf{rev\_...}: flow features of the reverse flow \\
\hline
\textbf{TLS Features} \\
\hline
\textbf{tls\_cnt}: number of tls packets \\
\textbf{tls\_len[]}: array of tls payload length \\
\textbf{tls\_cs\_cnt}: number of ciphersuits \\
\textbf{tls\_cs[]}: array of ciphersuits value \\
\textbf{tls\_ext\_cnt}: number of tls extensions \\
\textbf{tls\_ext\_types[]}: array of tls extensions \\
\textbf{tls\_key\_exchange\_len}: length of tls key exchange \\
\textbf{tls\_svr\_...}: tls features advertised by the server \\
\hline
\textbf{DNS Features} \\
\hline
\textbf{dns\_query\_cnt}: number of dns query records \\
\textbf{dns\_query\_name\_len[]}: array of dns query name lengths \\
\textbf{dns\_query\_name[]}: array of dns query names \\
\textbf{dns\_query\_type[]}: array of dns query types \\ 
\textbf{dns\_query\_class[]}: array of dns query classes \\
\textbf{dns\_answer\_cnt}: number of dns answer records \\
\textbf{dns\_answer\_ttl[]}: array of dns answer “time to live” values \\
\textbf{dns\_answer\_ip[]}: array of dns answer ips \\
\hline
\textbf{HTTP Features} \\
\hline
\textbf{http\_method}: http request method \\
\textbf{http\_uri}: http request uri \\
\textbf{http\_host}: http request hostname \\
\textbf{http\_code}: http code \\
\textbf{http\_content\_type}: http content\_type \\
\textbf{http\_content\_len}: http content length \\
\hline
\end{tabular}
\end{center}
\end{table}

\subsection{Malware Detection Datasets}
Our first task to focus on is detecting malware flows in network traffic through routers.  The details about how to create our own dataset from these two sources are described in the following two sections. 

\subsubsection{NetML}
\begin{figure}[tp]
\centering
\includegraphics[width=8.5cm,height=13cm,keepaspectratio]{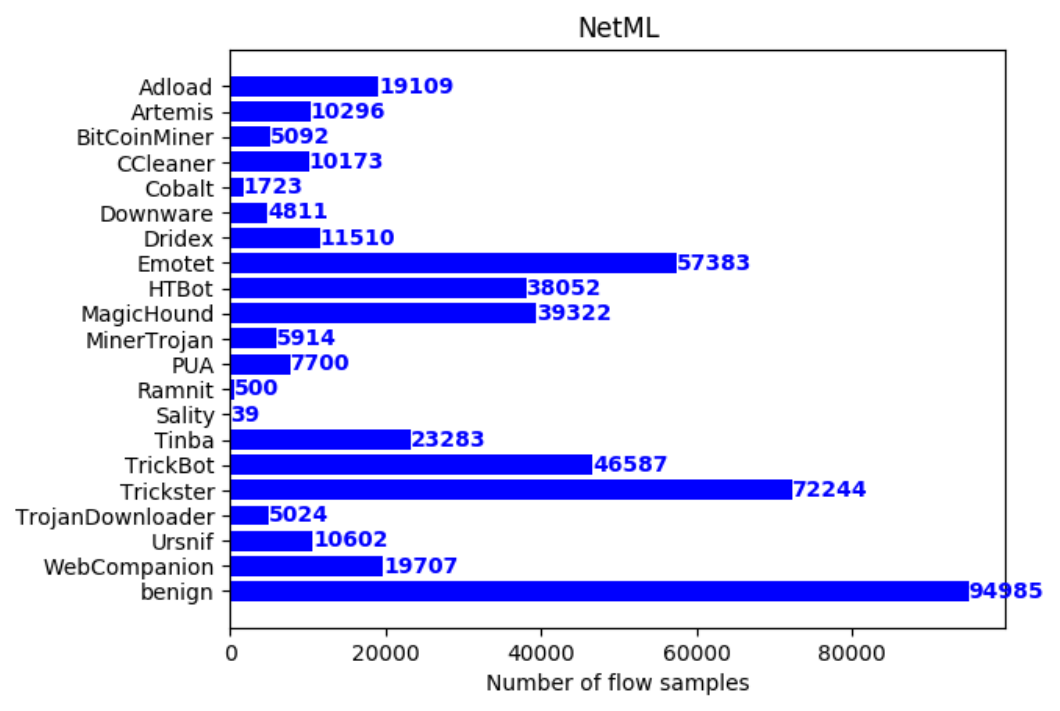}
\caption{Class distribution of NetML dataset}
\label{NetML_numbers}
\end{figure}

This dataset is created for malware detection task by obtaining $30$ out of more than $300$ raw traffic data from Stratosphere IPS. Selected capture files are listed in Table \ref{NetML_files}. Each capture file is assumed to contain flows belonging to a single class. For example, all of the flows extracted from \textit{capture\_win15.pcap} file are labeled as \textit{malware} in the top-level annotations and \textit{Artemis} in the fine-grained annotations. Fine-grained class distribution of NetML dataset is given in Figure \ref{NetML_numbers}. Top-level annotations contain benign or malware classes and fine-grained annotations contain $20$ different malware types and a benign class.

\begin{table}[thpb]
\footnotesize
\caption{Files used to create NetML dataset}
\label{NetML_files}
\begin{center}
\begin{tabular}{|l|l|l|}
\hline
\textbf{Top-level} & \textbf{Fine-grained} & \textbf{Filename} \\
\hline
benign & benign & 2013-12-17\_capture1.pcap \\
benign & benign & 2017-04-18\_win-normal.pcap \\
benign & benign & 2017-04-19\_win-normal.pcap \\
benign & benign & 2017-04-25\_win-normal.pcap \\
benign & benign & 2017-04-28\_normal.pcap \\
benign & benign & 2017\_04\_30-normal.pcap \\
benign & benign & 2017-04-30\_win-normal.pcap \\
benign & benign & 2017-05-01\_normal.pcap \\
benign & benign & 2017-05-02\_kali-normal.pcap \\
\hline
malware & Adload & 2018-05-03\_win12.pcap \\
malware & Artemis & capture\_win15.pcap \\
malware & BitCoinMiner & 2018-04-04\_win16.pcap \\
malware & CCleaner & 2017-12-18\_win2.pcap \\
malware & CCleaner & 2018-01-30\_win17.pcap \\
malware & Cobalt & 2018-04-03\_win11.pcap \\
malware & Downware & 2018-02-23\_win10.pcap \\
malware & Dridex & 2018-04-03\_win12.pcap \\
malware & Emotet & 2017-06-24\_win3.pcap \\
malware & HTBot & 2018-04-04\_win20.pcap \\
malware & MagicHound & 2017-11-22\_win4.pcap \\
malware & MinerTrojan & 2018-03-27\_win4.pcap \\
malware & PUA & 2018-02-16\_win8.pcap \\
malware & PUA & 2018-02-23\_win11.pcap \\
malware & Ramnit & 2018-04-03\_win6.pcap \\
malware & Sality & 2017-11-23\_win16.pcap \\
malware & Tinba & capture\_win1.pcap \\
malware & TrickBot & capture\_win11.pcap \\
malware & Trickster & 2018-01-29\_win7.pcap \\
malware & TrojanDownloader & 2018-03-27\_win23.pcap \\
malware & Ursnif & capture\_win12.pcap \\
malware & WebCompanion & 2018-03-01\_win9.pcap \\
\hline
\end{tabular}
\end{center}
\end{table}

\subsubsection{CICIDS2017}
\begin{figure}[tp]
\centering
\includegraphics[width=8.5cm,height=8cm,keepaspectratio]{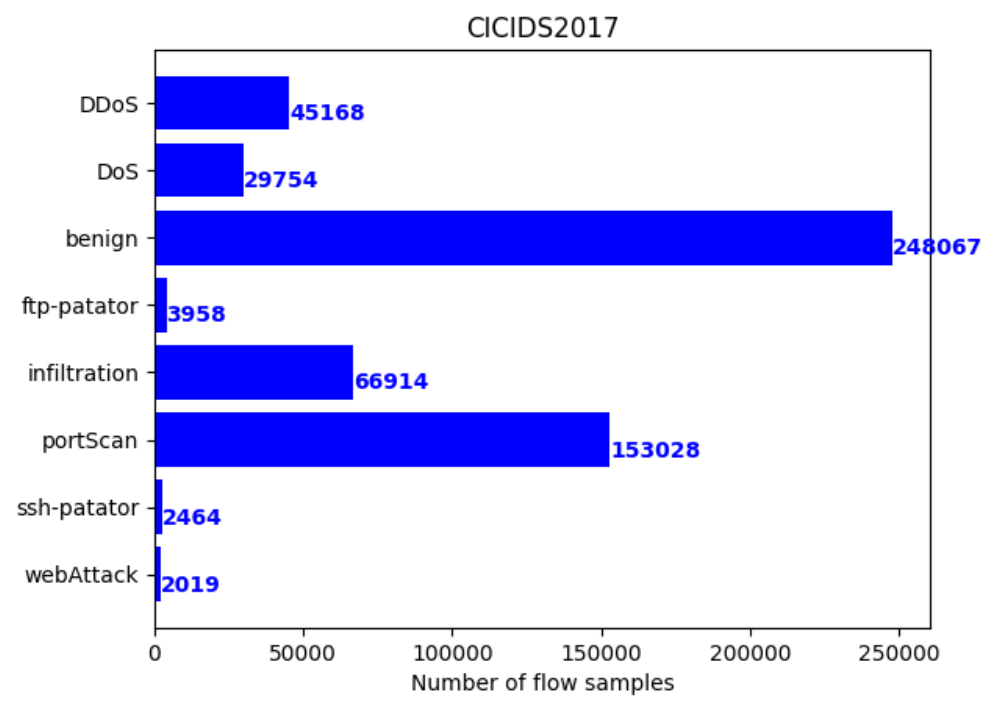}
\caption{Class distribution of CICIDS2017 dataset}
\label{CICIDS2017_numbers}
\end{figure}

CICIDS2017 \cite{CICIDS2017} is a set of raw traffic captures for different types of malware attacks and normal flows. 
Each trace file contains the whole network traffic throughout the day and $5$ traces for each day of a week is collected. However, only some specific time intervals with packet exchange between the predetermined IP addresses contain the flows we want to extract. Therefore, we download CICIDS2017 dataset from the original source and filter the flows-of-interest according to the descriptions in the dataset website, and extract flow features using the flow feature extraction tool. Our method was unable to extract flow features for “Botnet” and “Heartbleed” attacks therefore they are excluded in our final CICIDS2017 dataset. Extracted malware types and which file is used are given in Table \ref{CICIDS2017_files} and the fine-grained class distribution is shown in Figure \ref{CICIDS2017_numbers}. Similar to the other malware detection dataset, top-level annotations in CICIDS2017 dataset contain benign or malware classes and fine-grained annotations contain $7$ different malware types and a benign class.

\begin{table}[thpb]
\footnotesize
\caption{Files used to create CICIDS2017 dataset}
\label{CICIDS2017_files}
\begin{center}
\begin{tabular}{|l|l|l|}
\hline
\textbf{Top-level} & \textbf{Fine-grained} & \textbf{Filename} \\
\hline
benign & benign & Monday-WorkingHours.pcap \\
\hline
malware & DDoS & *Friday-WorkingHours.pcap \\
malware & DoS & *Wednesday-WorkingHours.pcap \\
malware & ftp-patator & *Tuesday-WorkingHours.pcap \\
malware & infiltration & *Thursday-WorkingHours.pcap \\
malware & portScan & *Friday-WorkingHours.pcap \\
malware & ssh-patator & *Tuesday-WorkingHours.pcap \\
malware & webAttack & *Thursday-WorkingHours.pcap \\
\hline
\multicolumn{3}{l}{(*): These files are firstly filtered to obtain the flow-of-interest} \\
\multicolumn{3}{l}{then those files are used to extract flow features and give labels.} \\
\end{tabular}
\end{center}
\end{table}

\subsection{Traffic Classification Dataset: non-vpn2016}
The third dataset focuses on application type classification and is called non-vpn2016. It is obtained by extracting flow features using only the non-vpn raw traffic capture files from the CIC website as our feature extraction tool does not support VLAN processing yet.

\begin{sloppypar}
Three levels of annotations are assigned to this dataset: top-level, mid-level and fine-grained. Top-level annotations are a general grouping of those traffic capture data and $7$ classes are selected including P2P, audio, chat, email, file\_transfer, tor, video. Mid-level annotations contain $18$ type of applications (facebook, skype etc.) while fine-grained annotations identify $31$ lower-level classes in an application (facebook\_audio, facebook\_chat, skype\_audio, skype\_chat etc.). Table \ref{non-vpn2016_files} gives the list of files used to create non-vpn2016 dataset.
\end{sloppypar}

\begin{table}[thpb]
\small
\caption{Files used to create non-vpn2016 dataset}
\label{non-vpn2016_files}
\begin{center}
\begin{tabular}{|l|l|}
\hline
\textbf{Top, Mid, Fine-grained} & \textbf{Filename} \\
\hline
chat, aim, aim\_chat & AIMchat*.pcapng \\
chat, aim, aim\_chat & aim\_chat\_3*.pcap \\
email, email, email & email*.pcap \\
audio, facebook, facebook\_audio & facebook\_audio*.pcap \\
audio, facebook, facebook\_audio & facebook\_audio*.pcapng \\
chat, facebook, facebook\_chat & facebookchat*.pcapng \\
chat, facebook, facebook\_chat & facebook\_chat\_4*.pcap \\
video, facebook, facebook\_video & facebook\_video*.pcap \\
video, facebook, facebook\_video & facebook\_video*.pcapng \\
file\_transfer, ftps, ftps\_down & ftps\_down\_1*.pcap \\
file\_transfer, ftps, ftps\_up & ftps\_up\_2*.pcap \\
chat, gmail, gmail\_chat & gmailchat*.pcapng \\
audio, hangouts, hangouts\_audio & hangouts\_audio*.pcap \\
audio, hangouts, hangouts\_audio & hangouts\_audio*.pcapng \\
chat, hangouts, hangouts\_chat & hangout\_chat\_4b.pcap \\
chat, hangouts, hangouts\_chat & hangouts\_chat\_4a.pcap \\
video, hangouts, hangouts\_video & hangouts\_video*.pcap \\
video, hangouts, hangouts\_video & hangouts\_video*.pcapng \\
chat, icq, icq\_chat & ICQchat*.pcapng \\
chat, icq, icq\_chat & icq\_chat\_3*.pcap \\
video, netflix, netflix & netflix*.pcap \\
file\_transfer, scp, scp\_down & scpDown*.pcap \\
file\_transfer, scp, scp\_up & scpUp*.pcap \\
file\_transfer, sftp, sftp\_down & scpDown*.pcap \\
file\_transfer, sftp, sftp\_down & scp\_down\_3*.pcap \\
file\_transfer, sftp, sftp\_up & scpUp1.pcap \\
file\_transfer, sftp, sftp\_up & scp\_up\_2*.pcap \\
audio, skype, skype\_audio & skype\_audio*.pcap \\
audio, skype, skype\_audio & skype\_audio*.pcapng \\
chat, skype, skype\_chat & skype\_chat1*.pcap \\
file\_transfer, skype, skype\_file & skype\_file*.pcap \\
file\_transfer, skype, skype\_file & skype\_file*.pcapng \\
video, skype, skype\_video & skype\_video*.pcap \\
video, skype, skype\_video & skype\_video*.pcapng \\
audio, spotify, spotify & spotify*.pcap \\
tor, facebook, tor\_facebook & torFacebook.pcap \\
tor, google, tor\_google & torGoogle.pcap \\
P2P, torrent, torrent & Torrent01.pcapng \\
tor, twitter, tor\_twitter & torTwitter.pcap \\
video, vimeo, tor\_vimeo & torVimeo*.pcap \\
video, youtube, tor\_youtube & torYoutube*.pcap \\
video, vimeo, vimeo & vimeo*.pcap \\
audio, voipbuster, voipbuster & voipbuster*.pcapng \\
audio, voipbuster, voipbuster & voipbuster\_4*.pcap \\
video, youtube, youtube & youtube*.pcap \\
\hline
\end{tabular}
\end{center}
\end{table}

\begin{figure}[tp]
\centering
\includegraphics[width=8.5cm,height=8cm,keepaspectratio]{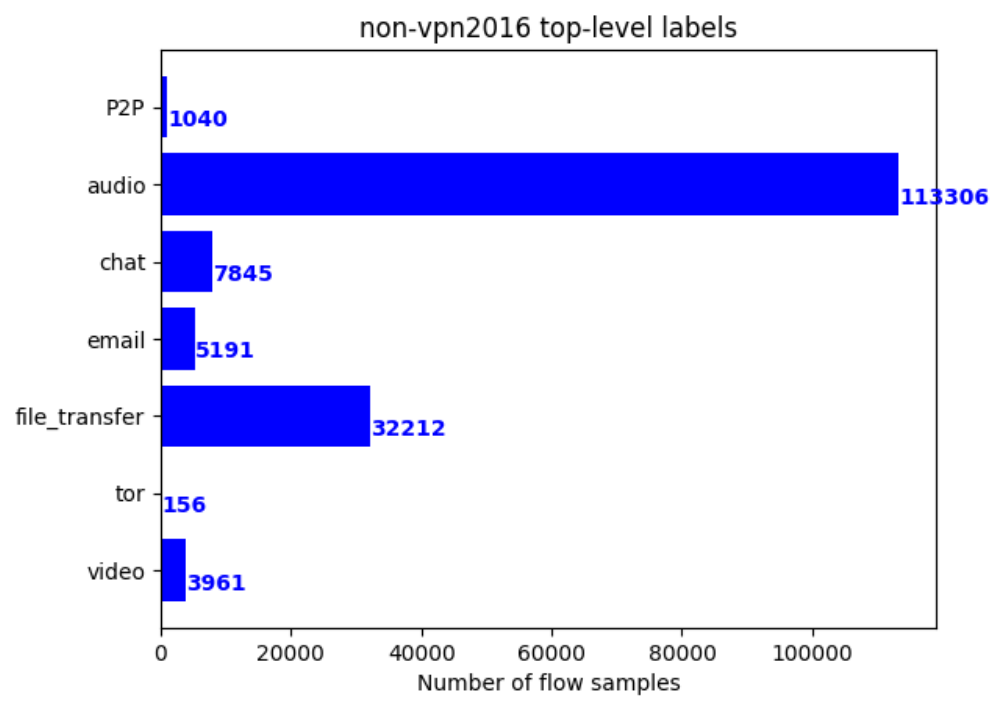}
\caption{Class distribution of non-vpn2016 dataset with top-level labeling}
\label{non-vpn_top_numbers}
\end{figure}

\begin{figure}[tp]
\centering
\includegraphics[width=8.5cm,height=16cm,keepaspectratio]{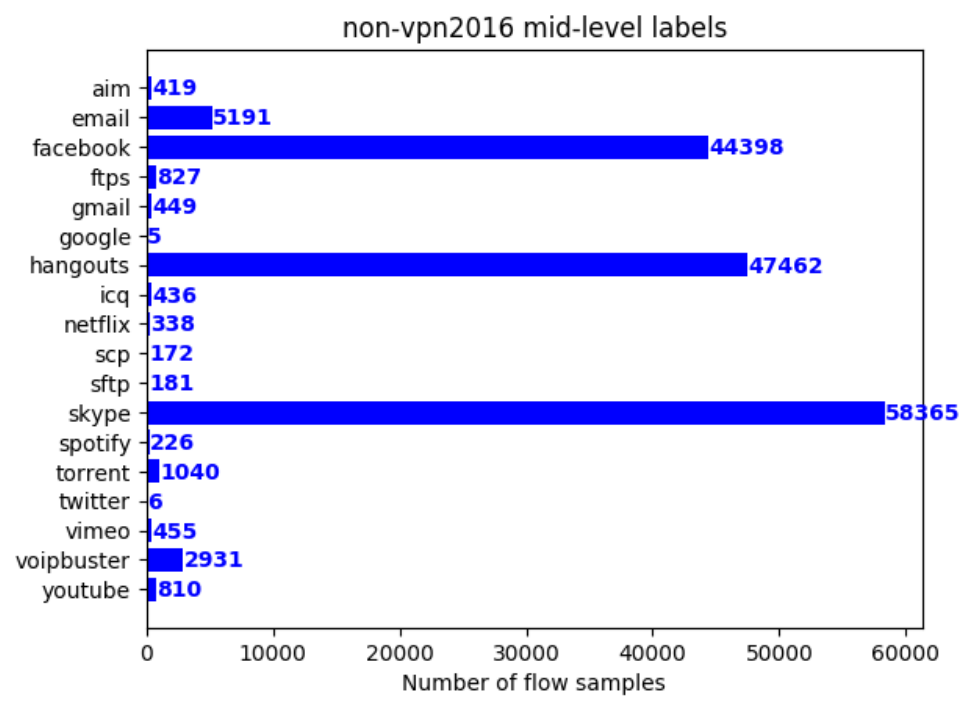}
\caption{Class distribution of non-vpn2016 dataset with mid-level labeling}
\label{non-vpn_mid_numbers}
\end{figure}

\begin{figure}[tp]
\centering
\includegraphics[width=8.5cm,height=22cm,keepaspectratio]{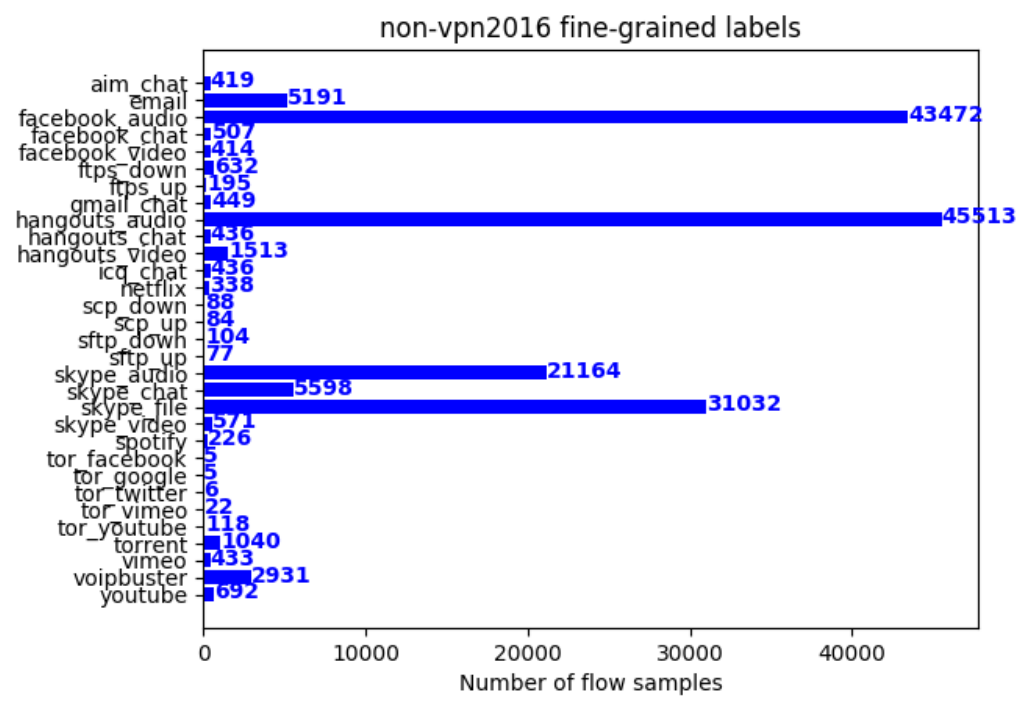}
\caption{Class distribution of non-vpn2016 dataset with fine-grained labeling}
\label{non-vpn_fine_numbers}
\end{figure}

The numbers of output flow samples after feature extraction for top-level, mid-level and fine-grained annotations are provided in Figure \ref{non-vpn_top_numbers}-\ref{non-vpn_fine_numbers}. 

The imbalance of number of samples in a dataset is an important but mostly inevitable. For example, we observe in Figure \ref{non-vpn_fine_numbers} that facebook\_audio, hangouts\_audio. skype\_audio and skype\_file classes are oversampled in the dataset. Such significantly imbalanced class distribution may introduce a bias to the predictions of the trained models.

\section{NetML Dataset Analysis}
\label{data_analysis}
\begin{sloppypar}
In this section we an analysis of the flow features for the training sets of all three datasets. We extract four set of features: (1) Metadata, (2) TLS, (3) DNS, (4) HTTP. Metadata features are protocol-independent features such as number of packets, bytes inbound and bytes outbound, time length of a flow etc. On the other side, there are protocol-specific features such as TLS, DNS and HTTP. Number of ciphersuites and extensions supported by the client or server are a subset of the TLS features. Similarly, DNS query name and DNS answer IP can be given as example for DNS features. Finally, HTTP code and HTTP method are two examples for HTTP features. While Metadata features can be extracted for any kind of flow, protocol-specific features can only be extracted if the given flow contains packets with one of these protocols. Therefore, Table \ref{data_percentage} summarizes number of flows in each dataset containing Metadata, TLS, DNS and HTTP features.
\end{sloppypar}

The training sets for NetML, CICIDS2017 and non-vpn2016 datasets include $387268$, $441116$ and $131065$ flow samples, respectively.

\begin{table}[th]
\small
\caption{Number of flow features for each training set}
\label{data_percentage}
\begin{center}
\begin{tabular}{|l|c|c|c|c|}
\hline
& \textbf{Metadata} & \textbf{TLS} & \textbf{DNS} & \textbf{HTTP} \\
\hline
\textbf{NetML} & 387,268 & 114,396 & 60,271 & 46,503 \\
\hline
\textbf{CICIDS2017} & 441,116 & 74,836 & 93,224 & 40,339 \\
\hline
\textbf{non-vpn2016} & 131,065 & 1,262 & 8,179 & 548 \\
\hline
\end{tabular}
\end{center}
\vspace{-4mm}
\end{table}

\subsection{Metadata Features}
Metadata features are mostly composed of statistical features or histograms computed in a flow. Return values are a constant size of array for the histogram-like features such as \textit{intervals\_ccnt[]} and \textit{hdr\_ccnt[]}. The averaged mean values for each of these features dimensions are calculated and given in Figure \ref{metadata_arrays_1_malware},\ref{metadata_arrays_1_nonvpn} and Figure \ref{metadata_arrays_2_malware},\ref{metadata_arrays_2_nonvpn}. Horizontal axis gives the index number in the returned array and vertical axis represents the mean value of that dimension in the given dataset. 


Unlike histogram arrays, the other Metadata features such as \textit{bytes\_in} or \textit{src\_prt} return a single value such as $160$ or $6006$. Class probability distribution for those Metadata features are provided in Figures \ref{metadata_single_1_malware}-\ref{metadata_single_4_nonvpn}. Horizontal axis represents the returned value of the feature while vertical axis gives the number of occurrences for that value. For all figures, 100 equally sized bins are used to plot the distributions.


Distribution of Metadata features shows us which features are more discriminating in the given context. For example, \textit{hdr\_bin\_40} feature tends to have smaller values in malware classes when compared to benign flows. Similarly, for the non-vpn2016 set, the same feature mostly contains smaller values especially for P2P, chat, email and tor classes. On the other hand, there is one outlier example in \textit{time\_length} feature of NetML training set which has the value around 7 million therefore causing the histogram to be awkwardly plotted. However, we are not going to clean the data for the scope of this study and leave it for the future studies.

Metadata features can be extracted for any flow samples. In this study, $31$ different Metadata features are extracted using the flow feature extraction library. For NetML training set, all of the flow samples, i.e. $387268$ produces all of the Metadata features. Similarly, CICIDS2017 training set contains $441116$ flows with Metadata features and non-vpn2016 training set contains $131065$ flows with all those Metadata features listed in Table \ref{flow_features}.

\subsection{TLS Features}
Several client-based and server-based TLS-specific features are used in our study. Some of these TLS features are the list of offered ciphersuites, the list of advertised extensions and the key exchange length advertised by client and server. In total, $14$ different TLS features are extracted. For NetML training set, $114396$ flow samples contain TLS features which corresponds to nearly $30\%$ of the whole training set. For CICIDS2017, $74836$ or in other words almost $15\%$ TLS-contained flow samples are extracted while for non-vpn2016 it is only $1262$ corresponding to less than $1\%$ of the whole training set.

Figure \ref{tls_client_single_malware},\ref{tls_client_single_nonvpn} and Figure \ref{tls_server_single_malware},\ref{tls_server_single_nonvpn} depict the single-valued feature class distribution for NetML, CICIDS2017 and non-vpn2016 training sets for client advertised and server advertised features, respectively. Similar to Metadata single-valued features, horizontal axis represents the value returned by the feature extraction tool and vertical axis gives the number of occurrences of that value in the corresponding data set.

First two rows of Figure \ref{tls_arrays_malware},\ref{tls_arrays_nonvpn} show size of the return array values on horizontal axis and number of occurrences on vertical axis. On the other hand, the remaining $4$ rows shows the offered ciphersuites and supported TLS extensions for three training sets. Horizontal axis values represent a unique index number of ciphersuite or extension type with the number of occurrences on vertical axis. Index value to ciphersuite of extension type mapping for those $4$ features are provided in the appendix.

We observe $122$ unique hex codes in the lists of offered ciphersuites and $38$ unique hex codes in the lists of supported TLS extensions in NetML training set. Similarly, $11$ unique hex codes in the lists if offered ciphersuites and supported TLS extensions are obtained from the CICIDS2017 training set and $16$ unique ciphersuites and $12$ TLS extensions are observed in the non-vpn2016 training set. Commonly used hex values are provided in Figures \ref{common_tls_malware} and \ref{common_tls_nonvpn} for all three datasets. It is interesting to note that P2P class of non-vpn2016 dataset does not contain any flows with TLS features. We do not compare these hex code inter-datasets for the scope of this paper. Finally, a single integer value is returned to represent the other TLS count features and key exchange lengths. 

\subsection{DNS Features}
\begin{sloppypar}
Our feature extraction tool returns several DNS query and answer features listed in Table \ref{flow_features}. Similar to TLS features, there are both single-valued and array-like features in DNS involved flows. For example, \textit{dns\_query\_cnt} and \textit{dns\_answer\_cnt} features have a single integer value returned while the other DNS query and answer features return array of integers for \textit{dns\_answer\_ttl} and strings for the rest. For NetML training set, $60271$ flow samples contain DNS features which corresponds to around $15\%$ of the whole training set. For CICIDS2017, $93224$ or in other words almost $21\%$ of the training set contains DNS features while for non-vpn2016 it is only $8179$ corresponding to less than $7\%$ of the whole training set.
\end{sloppypar}

Figure \ref{dns_single_malware},\ref{dns_single_nonvpn} show single-values DNS features. We can see that \textit{dns\_query\_cnt} feature returns always $1$ no matter what the input is, therefore can be discarded in the classification algorithm. On the other hand, Figure \ref{dns_arrays_malware},\ref{dns_arrays_nonvpn} show the class distribution for array-like DNS flow features. For \textit{dns\_query\_name} and \textit{dns\_answer\_ip} features, horizontal axis represents the index value for the query name of answer IP while vertical axis gives the number of occurrences of that particular feature while the other DNS feature plots gives the length of the returned array versus number of occurrences. Similar to \textit{dns\_query\_cnt}, \textit{dns\_query\_class} feature returns always $1$ therefore can be ignored for the further analysis.

\begin{sloppypar}
We analyze the commonly used \textit{dns\_answer\_ip} and \textit{dns\_query\_name} features. Figures \ref{common_dns_malware} and \ref{common_dns_nonvpn} shows top-5 most commonly used values for those features acquired from NetML, CICIDS2017 and non-vpn2016 training sets, respectively. Interestingly, most of the flows with DNS feature have an empty space for the \textit{dns\_answer\_ip} feature for all classes of three datasets except P2P of non-vpn2016.
\end{sloppypar}

\subsection{HTTP Features}
For each flow containing HTTP packets, we extract 6 different HTTP features listed in Table \ref{flow_features}. \textit{http\_method}, \textit{http\_code} and \textit{http\_content\_len} are the only features which are represented as a single integer value while the others contain string datatype which needs to be processed before classification. For \textit{http\_uri}, \textit{http\_host} and \textit{http\_content\_type} features, each value is stored with an index mapping to an integer number.

Figures \ref{common_http_malware} and \ref{common_http_nonvpn} shows which \textit{http\_content\_type}, \textit{http\_host} and \textit{http\_uri} values are commonly used for each top-level annotations of the three datasets. It is clear to observe that \textit{http\_content\_type} feature is very important to classify benign samples from malware flows such that for both malware detection datasets \textit{text/html} is the mostly observed in malware flows. In other words, if a flow contains some other value for this particular feature our model could make a strong guess by predicting the given sample does not belong to a malware flow. Similarly, for the application classification dataset, each class has different most commonly used values for this feature. It should be noted that for the same reasons observed in TLS features, Tor class in non-vpn dataset is observed not to contain any HTTP related features in the limited amount of collected samples.

\section{NetML Challenge Baselines}
In this section, we demonstrate how the above three proposed datasets can be used for flow analysis. We present the implementation and evaluation of several classical machine learning methods (called baselines). We use training dataset to train the machine learning models and use the test-std set to assess their effectiveness. Unless stated otherwise all the numbers presented in this section are obtained from the test-std sets.

\subsection{Baselines}
We select three basic classification models including Random Forest, Support Vector Machine (SVM) and Multi-layer Perceptron (MLP) classifiers as our baselines since Random Forest is easy to train, SVM is widely used and proved to be useful in many applications and MLP offers a great potential for deep learning that can achieve significant improvements in the accuracy. These baseline models are trained using the SK-learn library of python scripting language. As explained in Section \ref{data_analysis}, protocol specific (TLS, DNS and HTTP) features are not available for all of the flows in the datasets. Therefore, we use only Metadata features to obtain baseline results.

We select three basic classification models including Random Forest, Support Vector Machine (SVM) and Multi-layer Perceptron (MLP) classifiers as our baselines since Random Forest is easy to train, SVM is widely used and proved to be useful in many applications and MLP offers a great potential for deep learning that can achieve significant improvements in the accuracy for many other fields. These baselines are trained using the SK-learn library of pyhton scripting language.

\subsubsection{Random Forest Classifier}
The random forest classifier consists of many individual decision tree classifiers working as in an ensemble model. Each decision tree classifier in the random forest produces a class estimate and the top-rated class is chosen as the final decision. We pick this classifier model as baseline since it is very easy and fast to implement such a simple but surprisingly very accurate classifier. We select $100$ estimators and set maximum depth to $10$ to design the random forest classifier.

\subsubsection{Support Vector Machine Classifier}
Support Vector Machines (SVM) are one of the widely accepted supervised machine learning algorithms. The idea is to define a hyper-plane that best separates the two support vectors for two datasets where support vectors are the closest sample point of a class to decision boundary and therefore the most important element of this classifier. One of the advantages of SVM classifier is that they are proven to be accurate for many different problems. On the other hand, the biggest disadvantage is that the training time gets very high with the large datasets as those we provide in this study. However, this problem can be evaded by setting the number of samples in the training set to small enough and reserving the rest as validation set. Design parameters for SVM classifier are set to default values, i.e. regularization parameter \textit{C} is $1.0$ and chosen kernel is radial basis function (rbf).

\subsubsection{Multi-Layer Perceptron Classifier}
A perceptron is a single node with multiple inputs and a single output. The output is obtained by passing the linear combination of each input to an activation function, which is usually determined to be sigmoid function. Multi-layer Perceptron classifier is formed with many perceptrons in multiple linear layers. The most widely applied MLP classifiers contain input layer, hidden layer, and output layer.

MLP classifiers are also referred as Artificial Neural Networks (ANN) for Feed-Forward Neural Networks (FFNN). Training of MLP classifier is implemented with a back propagation technique which lead the model to learn robust representations in the training set. Deep neural networks are different implementations of MLPs which contain many hidden layers. 

Artificial Neural Network based approaches are widely proposed in many recent studies and they are proven to be more accurate than traditional machine learning methods such as Random Forest and SVM classifiers in many applications. Therefore, in this study, we select MLP as another baseline, but we only focus on the basic implementation of ANN with a single hidden layer. Hyper-parameters for the designed baseline are $121$ units in the hidden layer, L2 regularization factor \textit{alpha} as $0.0001$. Optimizer is chosen to be \textit{adam} to train the model.

\subsection{Preprocessing}
As explained in Section \ref{data_analysis}, TLS, DNS and HTTP flow features are not available for all of the flows in the datasets. Therefore, we use only Metadata features to obtain baseline results.

\begin{figure}[tp]
\centering
\includegraphics[width=8.5cm,height=22cm,keepaspectratio]{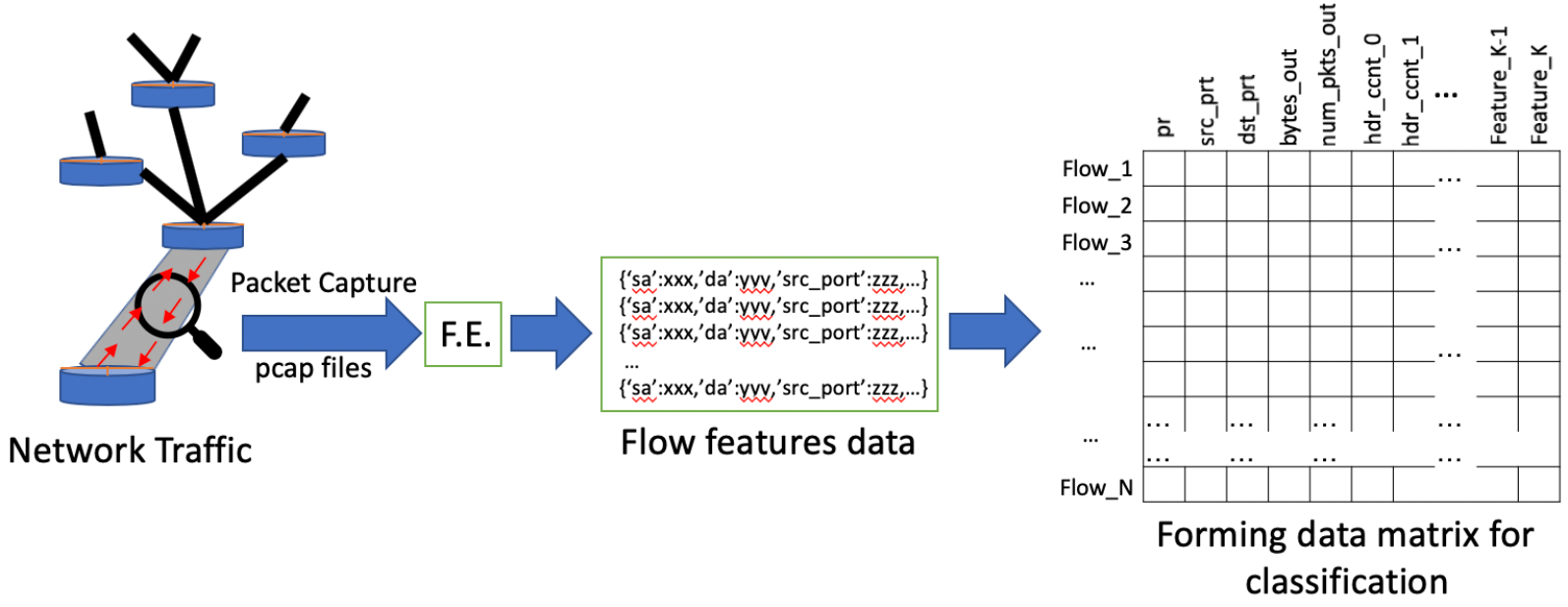}
\caption{Input data processing for Machine Learning model training/prediction. - F.E.: Feature Extraction}
\label{datamatrix}
\end{figure}

\begin{sloppypar}
The baseline classifiers require the input data as a two-dimensional array whose columns represent features and rows stand for flow samples. Therefore, we need to convert the given data into matrix format. Firstly, we discard source and destination IP address features as they are already masked. All other Metadata features in the datasets contain numeric values and hence it is straight-forward to import these datasets into a matrix form that will be used to train the classifiers. For array-like features such as hdr\_ccnt[], we just treat each of those dimensions in this array as a separate feature such as \textit{hdr\_ccnt\_0}, \textit{hdr\_ccnt\_1} etc. up to \textit{hdr\_ccnt\_k} for k+1 dimensional hdr\_ccnt[] feature and place each of those into seperate columns in the final data matrix. Figure \ref{datamatrix} represents how the network traffic is captured and processed to build sample-feature matrices.
\end{sloppypar}

Before training, the data is standardized in the preprocessing step so that data in each feature column follows standard normal distribution. For Random Forest and MLP classifiers, $80\%$ of the training set is kept for training the model and the remaining $20\%$ is used as validation set. However, only $10\%$ of the training set is used to train the SVM model while the remaining $90\%$ is used for validation because the time complexity of the SVM algorithm is too high with the given size of training set.

\subsection{Results}
We provide the classification results of 7 different scenarios in this section. All of the baseline models are trained data in the training set and results are obtained from the test-std set.

Malware detection problem is a binary classification task which can be best described by using detection rate and false alarm rate of malware. Our NetML and CICIDS2017 datasets used with top-level annotations are a binary detection problem and therefore we use detection rate and false alarm rate to analyze the model performance. Detection rate is indicated by True Positive Rate (TPR). TPR and False Alarm Rate (FAR) are calculated by the equation (1).

$$ TPR = \frac{TP}{TP+FN}, FAR = \frac{FP}{TN+FP} \eqno{(1)}$$

Fine-grained malware classification problems with NetML and CICIDS2017 datasets used with fine-grained annotations and traffic classification problems of non-vpn2016 dataset for all top-level, mid-level and fine-grained annotations are multi-class classification tasks and best described by F1 and mean average precision (mAP) scores which are calculated according to equation (2)

$$ F1 = \frac{2*precision*recall}{(precision+recall)}, mAP = \frac{1}{N}\sum_{i=1}^{N}AP_i \eqno{(2)}$$

where $N$ is the number of samples and $AP_i$ is the average precision $i^{th}$ class.

\begin{table}[th]
\small
\caption{Baseline results for malware detection}
\label{malware_results}
\begin{center}
\begin{tabular}{|l|c|c|c|}
\hline
& \textbf{RF} & \textbf{SVM} & \textbf{MLP} \\
& \textbf{TPR} | \textbf{FAR} & \textbf{TPR} | \textbf{FAR} & \textbf{TPR} | \textbf{FAR} \\
\hline
\textbf{NetML} & 0.9937 | 0.0092 & 0.9624 | 0.0137 & 0.9887 | 0.0171 \\
\hline
\textbf{CICIDS} & 0.9859 | 0.0044 & 0.9780 | 0.0028 & 0.9872 | 0.0069 \\
\hline
\end{tabular}
\end{center}
\end{table}

\subsubsection{Malware detection}

Table \ref{malware_results} shows the TPR and FAR scores for both malware detection datasets which are NetML and CICIDS2017 used with top-level annotations. Interestingly, Random Forest model which is the most trivial approach works best among other baseline models for the binary classification problem. For NetML dataset, highest TPR and lowest FAR are achieved by RF model with scores $0.9937$ and $0.0092$, respectively. SVM model has the lowest TPR with $0.9624$; however, it performs better in terms of FAR when compared to MLP model whose TPR is better but FAR is the worst. For CICIDS2017 dataset, we observe that MLP model has the highest TPR rate $0.9872$. However, unlike the case for NetML dataset, SVM model achieves the lowest FAR rate with $0.0028$ for CICIDS2017 dataset.

\subsubsection{Multi-class classification}

To gain further insights, we use fine-grained annotations for each malware type and construct a multi-class classification problem. Similarly, traffic classification dataset non-vpn2016 has multiple classes for each level of annotations. Table \ref{multiclass_results} gives the F1 and mAP scores for multi-class class classification problems.

\begin{table}[th]
\footnotesize
\setlength{\tabcolsep}{2pt}
\caption{Baseline results for multi-class classification scenarios}
\label{multiclass_results}
\begin{center}
\begin{tabular}{|l|c|c|c|}
\hline
& \textbf{RF} & \textbf{SVM} & \textbf{MLP} \\
& \textbf{F1} | \textbf{mAP} & \textbf{F1} | \textbf{mAP} & \textbf{F1} | \textbf{mAP} \\
\hline
\textbf{NetML-f} & 0.7442 | 0.4217 & 0.6959 | 0.3536 & 0.7314 | 0.4116 \\
\hline
\textbf{CICIDS-f} & 0.9872 | 0.8682 & 0.9850 | 0.8621 & 0.9889 | 0.8966 \\
\hline
\textbf{non-vpn-t} & 0.6273 | 0.3257 & 0.5868 | 0.1934 & 0.6066 | 0.2304 \\
\textbf{non-vpn-m} & 0.3693 | 0.3223 & 0.3441 | 0.1398 & 0.3609 | 0.2041 \\
\textbf{non-vpn-f} & 0.2486 | 0.2127 & 0.2036 | 0.0768 & 0.2359 | 0.1404 \\
\hline
\end{tabular}
\end{center}
\begin{center}
    \begin{tabular}{c}
    \footnotesize
t, m and f stand for top, mid and fine-grained annotations, respectively. \\
    \end{tabular}
\end{center}
\vspace{-4mm}
\end{table}

\textbf{NetML - fine-grained multi-class classification}

Among all baselines, similar to the malware detection task, even though MLP model produces comparable results, RF model performs as the most accurate model with F1 score $0.7442$ and mAP $0.4217$. 

\textbf{CICIDS2017 - fine-grained multi-class classification}

Random forest and SVM models perform comparable for multi-class malware classification problem. Similar to binary malware detection case, MLP model has the highest F1 score $0.9889$ and mAP score $0.8966$.

\begin{figure}[t]
\subfigure{\includegraphics[height=1.65in,width=1.65in]{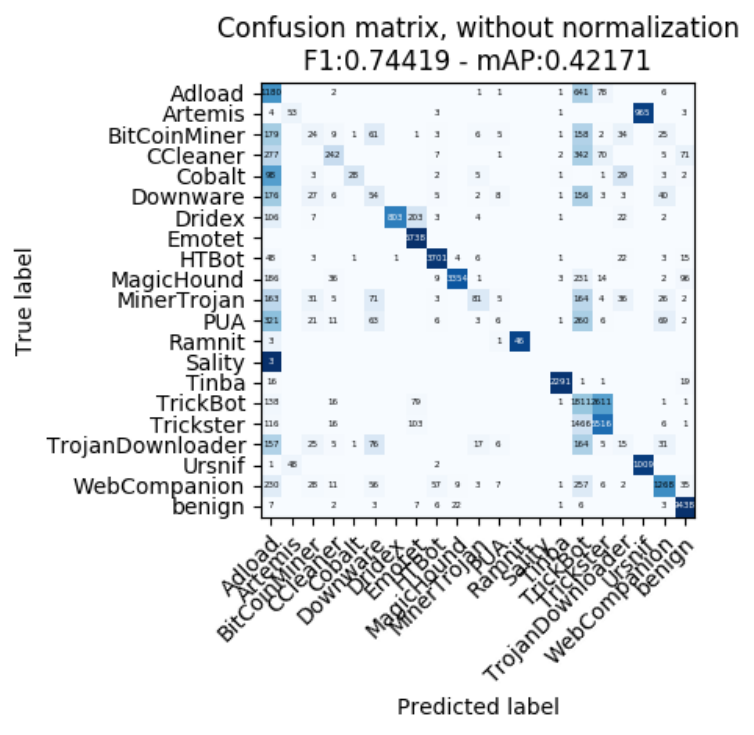}}
\subfigure{\includegraphics[height=1.65in,width=1.65in]{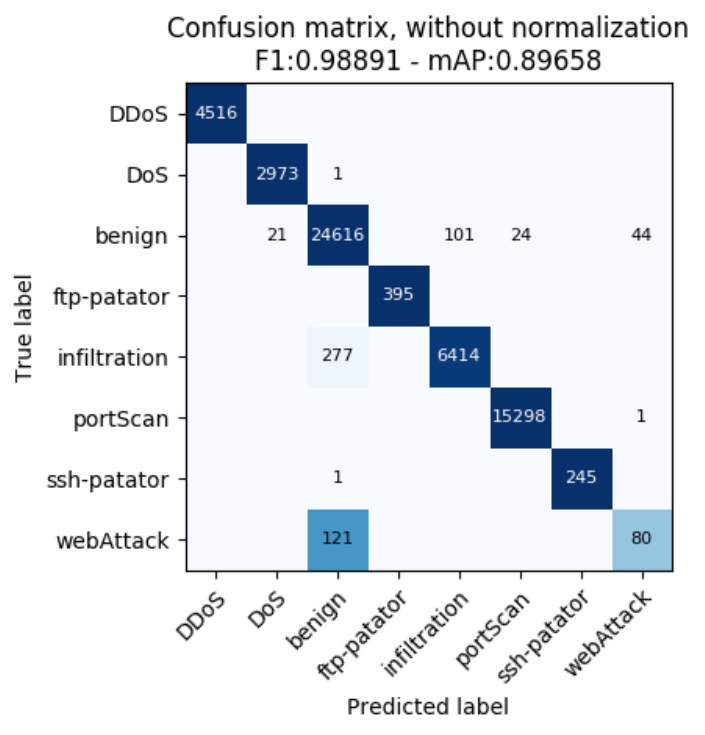}}
\caption{Confusion matrices of NetML test-std results obtained by Random Forest model (left) and CICIDS2017 test-std results obtained by MLP (right)}
\label{results_malware_fine}
\end{figure}

Figure \ref{results_malware_fine} presents confusion matrices of best performing models for both NetML and CICIDS2017 dataset results obtained from test-std set. We observe that for NetML dataset \textit{Adload} and \textit{TrickBot} classes are commonly confused by other class predictions. Similarly, \textit{TrickBot} and \textit{Trickster} classes are mostly confused by the model. Interestingly, \textit{Artemis} class is mostly predicted as \textit{Ursnif} and there is no other class predicted as \textit{Ursnif}. For CICIDS2017 dataset, we observe some misclassification between \textit{infiltration} and \textit{benign} classes. But most importantly, \textit{webAttack} class is barely detected and usually misclassified by the model as \textit{benign} class. For \textit{DDoS} class, we observe $100\%$ accuracy and for \textit{ssh-patator} there is only one flow sample missed by the model and all others are correctly detected with no false alarm.

It is worth noting that multi-class classification results achieve lower FAR rate in exchange with a slightly reduced detection rate when considered as binary malware detection. For example, multi-class classifier with Random Forest model for NetML dataset produces $0.0051$ FAR with $0.9922$ TPR. Similarly, for CICIDS2017 dataset, comparable TPR $0.9865$ is achieved with better FAR $0.0067$ using MLP model.

\textbf{non-vpn2016 - top-level classification}

We observe weak performance of all three baseline models for non-vpn2016 dataset. Although the F1 and mAP score are poor, the best performing model is again Random Forest model with F1 score $0.6273$ and mAP score $0.3257$.

\textbf{non-vpn2016 - mid-level classification}

We also use mid-level annotations to classify the network traffic according to the application type. Similar to protocol classification with top-level annotations, the scores are not impressive; however, Random Forest model is the most accurate baseline model among all with F1 score $0.3693$ and mAP score $0.3223$.

\textbf{non-vpn2016 - fine-grained classification}

Lastly, we present the baseline results for action classification with fine-grained annotations. As observed in application type classification and protocol classification tasks, Random Forest model is again the best performing baseline model with poor F1 score $0.2486$ and mAP score $0.2127$

\begin{figure}[t]
\subfigure{\includegraphics[height=1.65in,width=1.65in]{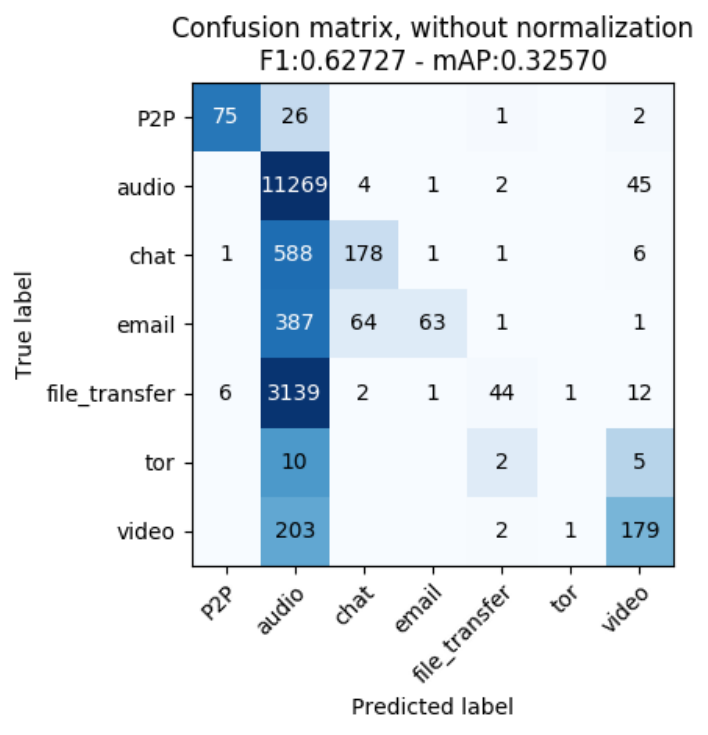}}
\subfigure{\includegraphics[height=1.65in,width=1.65in]{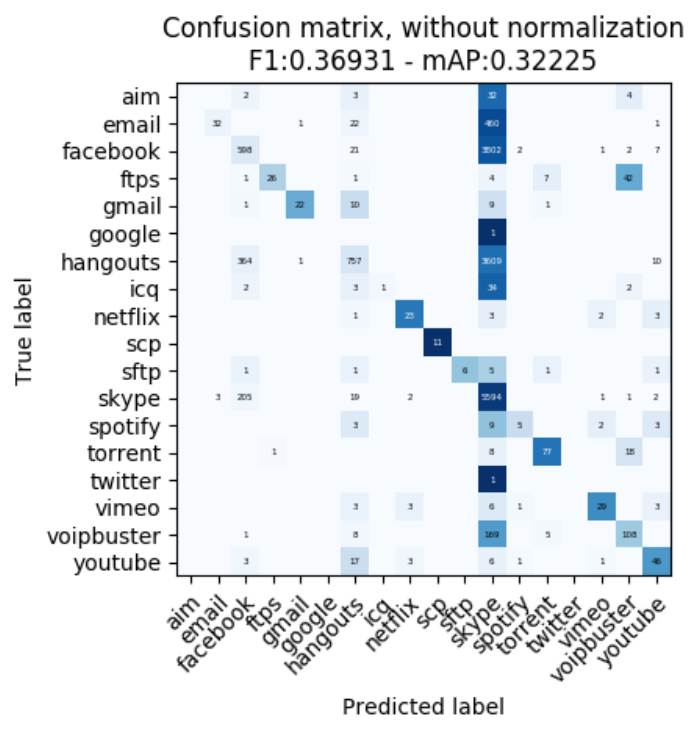}}
\subfigure{\includegraphics[height=2.4in,width=2.4in]{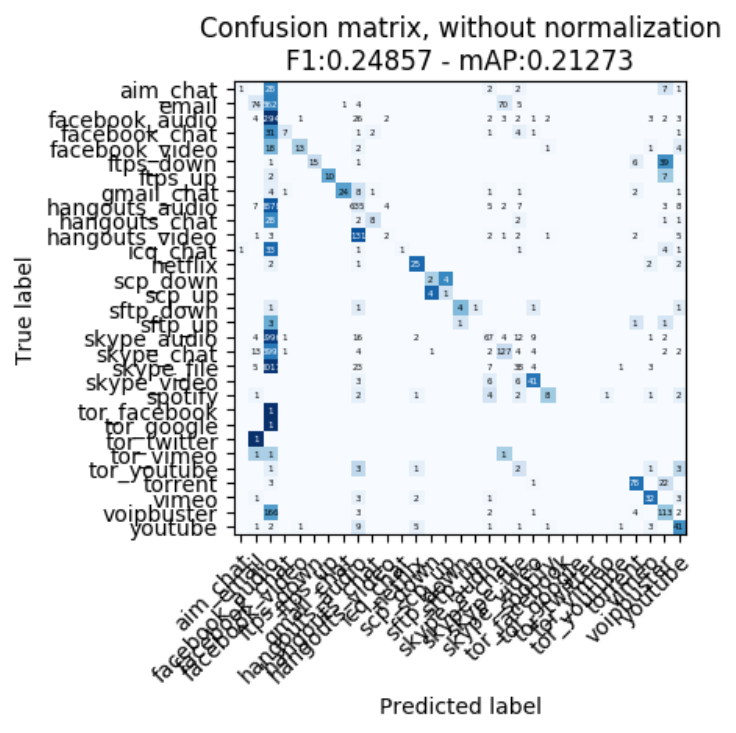}}
\caption{Confusion matrices of non-vpn2016 test-std results obtained by Random Forest model}
\label{results_non-vpn_rf}
\end{figure}

Figure \ref{results_non-vpn_rf} gives three confusion matrices for 3 level of annotations of non-vpn2016 dataset. We observe that all other classes are biased to mispredict a flow sample to \textit{audio} class. Likewise, \textit{skype} class for mid-level annotations and \textit{facebook\_audio} class for fine-grained annotations are the classes which bias the model to mispredict. The reason for such a bias would be due to the number of samples in the training set. These specified classes constitute the majority of the training sets and they are overwhelming the model causing the bias in prediction.

\section{NetML Challenge and Workshop}
We have set up an evaluation server for the participants to submit their prediction results for the test and challenge sets so that we can generate scores and list a leaderboard. We organize the first challenge and workshop to facilitate a systematic progress in network traffic analytics. The first instance of the workshop will be held at IJCAI 2020. We recommend that papers reporting results on NetML, CICIDS2017 and non-vpn2016 datasets described here should:

\begin{itemize}
    \item report results on test-std set displayed in the evaluation server leaderboard
    \item report TPR and FAR for binary classification
    \item report F1 and mAP for multi-class classification
    \item compare their results with baseline results presented in this paper
\end{itemize}

For further details and descriptions about the challenge, please visit the challenge page on github \url{https://www.github.com/ACANETS/NetML-Competition2020}

\section{Conclusion}
To enable data driven machine learning based research in network flow analytics, we introduce NetML datasets curated from open-sources for malware detection and network traffic classification. We release the flow features and different levels of annotations, aiming to present a common dataset for research community. In this paper, we characterize the features of each dataset and present baseline results for $7$ different scenarios using three commonly used machine learning models. We observe that Random Forest is the best performing model for malware detection tasks, but it lacks accuracy for traffic classification problems due to class imbalance.

Our results have inspired potential future work including, for example, using deep learning models to achieve a higher accuracy for the fine-grained classification tasks and implementing some measurements such as SMOTE \cite{SMOTE} to deal with class imbalance and hence obtain an unbiased prediction for traffic classification. We expect that the NetML datasets will promote machine learning based network traffic analytics and establish a common ground for researchers to propose and benchmark their approaches in this AI era.

\noindent{\bf ACKNOWLEDGMENT}
 
This work was sponsored in part by Intel Corporation. 


\bibliographystyle{ACM-Reference-Format}
\bibliography{reference}


\begin{thebibliography}{27}


\ifx \showCODEN    \undefined \def \showCODEN     #1{\unskip}     \fi
\ifx \showDOI      \undefined \def \showDOI       #1{#1}\fi
\ifx \showISBNx    \undefined \def \showISBNx     #1{\unskip}     \fi
\ifx \showISBNxiii \undefined \def \showISBNxiii  #1{\unskip}     \fi
\ifx \showISSN     \undefined \def \showISSN      #1{\unskip}     \fi
\ifx \showLCCN     \undefined \def \showLCCN      #1{\unskip}     \fi
\ifx \shownote     \undefined \def \shownote      #1{#1}          \fi
\ifx \showarticletitle \undefined \def \showarticletitle #1{#1}   \fi
\ifx \showURL      \undefined \def \showURL       {\relax}        \fi
\providecommand\bibfield[2]{#2}
\providecommand\bibinfo[2]{#2}
\providecommand\natexlab[1]{#1}
\providecommand\showeprint[2][]{arXiv:#2}

\bibitem[\protect\citeauthoryear{??}{sen}{2001}]%
        {senseval}
 \bibinfo{year}{2001}\natexlab{}.
\newblock \bibinfo{booktitle}{{\em SENSEVAL ’01: The Proceedings of the
  Second International Workshop on Evaluating Word Sense Disambiguation
  Systems}}. \bibinfo{publisher}{Association for Computational Linguistics},
  \bibinfo{address}{USA}.
\newblock


\bibitem[\protect\citeauthoryear{{Ahmad}, {Basheri}, {Iqbal}, and
  {Rahim}}{{Ahmad} et~al\mbox{.}}{2018}]%
        {Ahmad_2018_5ofLuNet}
\bibfield{author}{\bibinfo{person}{I. {Ahmad}}, \bibinfo{person}{M. {Basheri}},
  \bibinfo{person}{M.~J. {Iqbal}}, {and} \bibinfo{person}{A. {Rahim}}.}
  \bibinfo{year}{2018}\natexlab{}.
\newblock \showarticletitle{Performance Comparison of Support Vector Machine,
  Random Forest, and Extreme Learning Machine for Intrusion Detection}.
\newblock \bibinfo{journal}{{\em IEEE Access\/}}  \bibinfo{volume}{6}
  (\bibinfo{year}{2018}), \bibinfo{pages}{33789--33795}.
\newblock
\showISSN{2169-3536}
\showDOI{%
\url{https://doi.org/10.1109/ACCESS.2018.2841987}}


\bibitem[\protect\citeauthoryear{Anderson, Paul, and McGrew}{Anderson
  et~al\mbox{.}}{2016}]%
        {AndersonBlake_2016}
\bibfield{author}{\bibinfo{person}{Blake Anderson}, \bibinfo{person}{Subharthi
  Paul}, {and} \bibinfo{person}{David McGrew}.}
  \bibinfo{year}{2016}\natexlab{}.
\newblock \showarticletitle{Deciphering Malware's use of TLS (without
  Decryption)}.
\newblock \bibinfo{journal}{{\em Journal of Computer Virology and Hacking
  Techniques\/}} (\bibinfo{date}{07} \bibinfo{year}{2016}).
\newblock
\showDOI{%
\url{https://doi.org/10.1007/s11416-017-0306-6}}


\bibitem[\protect\citeauthoryear{Chawla, Bowyer, Hall, and Kegelmeyer}{Chawla
  et~al\mbox{.}}{2002}]%
        {SMOTE}
\bibfield{author}{\bibinfo{person}{Nitesh~V. Chawla}, \bibinfo{person}{Kevin~W.
  Bowyer}, \bibinfo{person}{Lawrence~O. Hall}, {and} \bibinfo{person}{W.~Philip
  Kegelmeyer}.} \bibinfo{year}{2002}\natexlab{}.
\newblock \showarticletitle{SMOTE: Synthetic Minority over-Sampling Technique}.
\newblock \bibinfo{journal}{{\em J. Artif. Int. Res.\/}} \bibinfo{volume}{16},
  \bibinfo{number}{1} (\bibinfo{date}{June} \bibinfo{year}{2002}),
  \bibinfo{pages}{321–357}.
\newblock
\showISSN{1076-9757}


\bibitem[\protect\citeauthoryear{{Conti}, {Mancini}, {Spolaor}, and
  {Verde}}{{Conti} et~al\mbox{.}}{2016}]%
        {Conti_2016_useractions}
\bibfield{author}{\bibinfo{person}{M. {Conti}}, \bibinfo{person}{L.~V.
  {Mancini}}, \bibinfo{person}{R. {Spolaor}}, {and} \bibinfo{person}{N.~V.
  {Verde}}.} \bibinfo{year}{2016}\natexlab{}.
\newblock \showarticletitle{Analyzing Android Encrypted Network Traffic to
  Identify User Actions}.
\newblock \bibinfo{journal}{{\em IEEE Transactions on Information Forensics and
  Security\/}} \bibinfo{volume}{11}, \bibinfo{number}{1} (\bibinfo{date}{Jan}
  \bibinfo{year}{2016}), \bibinfo{pages}{114--125}.
\newblock
\showISSN{1556-6021}
\showDOI{%
\url{https://doi.org/10.1109/TIFS.2015.2478741}}


\bibitem[\protect\citeauthoryear{Deng, Dong, Socher, Li, Li, and Fei-Fei}{Deng
  et~al\mbox{.}}{2009}]%
        {imagenet_cvpr09}
\bibfield{author}{\bibinfo{person}{J. Deng}, \bibinfo{person}{W. Dong},
  \bibinfo{person}{R. Socher}, \bibinfo{person}{L.-J. Li}, \bibinfo{person}{K.
  Li}, {and} \bibinfo{person}{L. Fei-Fei}.} \bibinfo{year}{2009}\natexlab{}.
\newblock \showarticletitle{{ImageNet: A Large-Scale Hierarchical Image
  Database}}. In \bibinfo{booktitle}{{\em CVPR09}}.
\newblock


\bibitem[\protect\citeauthoryear{{Ektefa}, {Memar}, {Sidi}, and
  {Affendey}}{{Ektefa} et~al\mbox{.}}{2010}]%
        {Ektefa_2010_datamining_ids}
\bibfield{author}{\bibinfo{person}{M. {Ektefa}}, \bibinfo{person}{S. {Memar}},
  \bibinfo{person}{F. {Sidi}}, {and} \bibinfo{person}{L.~S. {Affendey}}.}
  \bibinfo{year}{2010}\natexlab{}.
\newblock \showarticletitle{Intrusion detection using data mining techniques}.
  In \bibinfo{booktitle}{{\em 2010 International Conference on Information
  Retrieval Knowledge Management (CAMP)}}. \bibinfo{pages}{200--203}.
\newblock
\showISSN{null}
\showDOI{%
\url{https://doi.org/10.1109/INFRKM.2010.5466919}}


\bibitem[\protect\citeauthoryear{Farid, Nouria, and Zahidur~Rahman}{Farid
  et~al\mbox{.}}{2010}]%
        {Farid_2010_naive}
\bibfield{author}{\bibinfo{person}{Dewan Farid}, \bibinfo{person}{Harbi
  Nouria}, {and} \bibinfo{person}{Mohammad Zahidur~Rahman}.}
  \bibinfo{year}{2010}\natexlab{}.
\newblock \showarticletitle{Combining Naive Bayes and Decision Tree for
  Adaptive Intrusion Detection}.
\newblock \bibinfo{journal}{{\em International Journal of Network Security \&
  Its Applications\/}}  \bibinfo{volume}{2} (\bibinfo{date}{04}
  \bibinfo{year}{2010}).
\newblock
\showDOI{%
\url{https://doi.org/10.5121/ijnsa.2010.2202}}


\bibitem[\protect\citeauthoryear{Gao, Ma, Liu, Zhang, Ning, and Xu}{Gao
  et~al\mbox{.}}{2020}]%
        {Gao_2020_cicids2017_nsl_kdd}
\bibfield{author}{\bibinfo{person}{Minghui Gao}, \bibinfo{person}{Li Ma},
  \bibinfo{person}{Heng Liu}, \bibinfo{person}{Zhijun Zhang},
  \bibinfo{person}{Zhiyan Ning}, {and} \bibinfo{person}{Jian Xu}.}
  \bibinfo{year}{2020}\natexlab{}.
\newblock \showarticletitle{Malicious Network Traffic Detection Based on Deep
  Neural Networks and Association Analysis}.
\newblock \bibinfo{journal}{{\em Sensors\/}}  \bibinfo{volume}{20}
  (\bibinfo{date}{03} \bibinfo{year}{2020}), \bibinfo{pages}{1452}.
\newblock
\showDOI{%
\url{https://doi.org/10.3390/s20051452}}


\bibitem[\protect\citeauthoryear{Habibi~Lashkari, Draper~Gil, Mamun, and
  Ghorbani}{Habibi~Lashkari et~al\mbox{.}}{2016}]%
        {iscx_vpn2016}
\bibfield{author}{\bibinfo{person}{Arash Habibi~Lashkari},
  \bibinfo{person}{Gerard Draper~Gil}, \bibinfo{person}{Mohammad Mamun}, {and}
  \bibinfo{person}{Ali Ghorbani}.} \bibinfo{year}{2016}\natexlab{}.
\newblock \showarticletitle{Characterization of Encrypted and VPN Traffic Using
  Time-Related Features}.
\newblock
\showDOI{%
\url{https://doi.org/10.5220/0005740704070414}}


\bibitem[\protect\citeauthoryear{Habibi~Lashkari, Draper~Gil, Mamun, and
  Ghorbani}{Habibi~Lashkari et~al\mbox{.}}{2017}]%
        {flowmeter}
\bibfield{author}{\bibinfo{person}{Arash Habibi~Lashkari},
  \bibinfo{person}{Gerard Draper~Gil}, \bibinfo{person}{Mohammad Mamun}, {and}
  \bibinfo{person}{Ali Ghorbani}.} \bibinfo{year}{2017}\natexlab{}.
\newblock \showarticletitle{Characterization of Tor Traffic using Time based
  Features}. \bibinfo{pages}{253--262}.
\newblock
\showDOI{%
\url{https://doi.org/10.5220/0006105602530262}}


\bibitem[\protect\citeauthoryear{{Hu}, {Hu}, and {Maybank}}{{Hu}
  et~al\mbox{.}}{2008}]%
        {Hu_2008_adaboost}
\bibfield{author}{\bibinfo{person}{W. {Hu}}, \bibinfo{person}{W. {Hu}}, {and}
  \bibinfo{person}{S. {Maybank}}.} \bibinfo{year}{2008}\natexlab{}.
\newblock \showarticletitle{AdaBoost-Based Algorithm for Network Intrusion
  Detection}.
\newblock \bibinfo{journal}{{\em IEEE Transactions on Systems, Man, and
  Cybernetics, Part B (Cybernetics)\/}} \bibinfo{volume}{38},
  \bibinfo{number}{2} (\bibinfo{date}{April} \bibinfo{year}{2008}),
  \bibinfo{pages}{577--583}.
\newblock
\showISSN{1941-0492}
\showDOI{%
\url{https://doi.org/10.1109/TSMCB.2007.914695}}


\bibitem[\protect\citeauthoryear{{KDD Cup 1999 Data}}{{KDD Cup 1999
  Data}}{1999}]%
        {KDDcup99}
\bibfield{author}{\bibinfo{person}{{KDD Cup 1999 Data}}.}
  \bibinfo{year}{1999}\natexlab{}.
\newblock   (\bibinfo{year}{1999}).
\newblock
\showURL{%
\url{http://kdd.ics.uci.edu/databases/kddcup99/kddcup99.html}}
\newblock
\shownote{[Online; accessed 14-March-2020].}


\bibitem[\protect\citeauthoryear{Lin, Maire, Belongie, Bourdev, Girshick, Hays,
  Perona, Ramanan, Zitnick, and Dollár}{Lin et~al\mbox{.}}{2014}]%
        {lin2014microsoft_coco}
\bibfield{author}{\bibinfo{person}{Tsung-Yi Lin}, \bibinfo{person}{Michael
  Maire}, \bibinfo{person}{Serge Belongie}, \bibinfo{person}{Lubomir Bourdev},
  \bibinfo{person}{Ross Girshick}, \bibinfo{person}{James Hays},
  \bibinfo{person}{Pietro Perona}, \bibinfo{person}{Deva Ramanan},
  \bibinfo{person}{C.~Lawrence Zitnick}, {and} \bibinfo{person}{Piotr
  Dollár}.} \bibinfo{year}{2014}\natexlab{}.
\newblock \bibinfo{title}{Microsoft COCO: Common Objects in Context}.
\newblock   (\bibinfo{year}{2014}).
\newblock
\showeprint[arxiv]{cs.CV/1405.0312}


\bibitem[\protect\citeauthoryear{Moore, Zuev, and Crogan}{Moore
  et~al\mbox{.}}{2013}]%
        {Moore_2013_DiscriminatorsFU}
\bibfield{author}{\bibinfo{person}{Andrew~W. Moore}, \bibinfo{person}{Denis
  Zuev}, {and} \bibinfo{person}{Michael~L. Crogan}.}
  \bibinfo{year}{2013}\natexlab{}.
\newblock \showarticletitle{Discriminators for use in flow-based
  classification}.
\newblock


\bibitem[\protect\citeauthoryear{{Panda}, {Abraham}, and {Patra}}{{Panda}
  et~al\mbox{.}}{2010}]%
        {Panda_2010_twolevel}
\bibfield{author}{\bibinfo{person}{M. {Panda}}, \bibinfo{person}{A. {Abraham}},
  {and} \bibinfo{person}{M.~R. {Patra}}.} \bibinfo{year}{2010}\natexlab{}.
\newblock \showarticletitle{Discriminative multinomial Naïve Bayes for network
  intrusion detection}. In \bibinfo{booktitle}{{\em 2010 Sixth International
  Conference on Information Assurance and Security}}. \bibinfo{pages}{5--10}.
\newblock
\showISSN{null}
\showDOI{%
\url{https://doi.org/10.1109/ISIAS.2010.5604193}}


\bibitem[\protect\citeauthoryear{Panda, Abraham, and Patra}{Panda
  et~al\mbox{.}}{2012}]%
        {Panda_2012_twolevel}
\bibfield{author}{\bibinfo{person}{Mrutyunjaya Panda}, \bibinfo{person}{Ajith
  Abraham}, {and} \bibinfo{person}{Manas~Ranjan Patra}.}
  \bibinfo{year}{2012}\natexlab{}.
\newblock \showarticletitle{A Hybrid Intelligent Approach for Network Intrusion
  Detection}.
\newblock \bibinfo{journal}{{\em Procedia Engineering\/}}  \bibinfo{volume}{30}
  (\bibinfo{year}{2012}), \bibinfo{pages}{1 -- 9}.
\newblock
\showISSN{1877-7058}
\showDOI{%
\url{https://doi.org/10.1016/j.proeng.2012.01.827}}
\newblock
\shownote{International Conference on Communication Technology and System
  Design 2011.}


\bibitem[\protect\citeauthoryear{Qin, Lei, Bai, and Zhang}{Qin
  et~al\mbox{.}}{2019}]%
        {netai_19}
\bibfield{author}{\bibinfo{person}{Meng Qin}, \bibinfo{person}{Kai Lei},
  \bibinfo{person}{Bo Bai}, {and} \bibinfo{person}{Gong Zhang}.}
  \bibinfo{year}{2019}\natexlab{}.
\newblock \showarticletitle{Towards a Profiling View for Unsupervised Traffic
  Classification by Exploring the Statistic Features and Link Patterns}. In
  \bibinfo{booktitle}{{\em Proceedings of the 2019 Workshop on Network Meets AI
  \& ML}} {\em (\bibinfo{series}{NetAI’19})}. \bibinfo{publisher}{Association
  for Computing Machinery}, \bibinfo{address}{New York, NY, USA},
  \bibinfo{pages}{50–56}.
\newblock
\showISBNx{9781450368728}
\showDOI{%
\url{https://doi.org/10.1145/3341216.3342213}}


\bibitem[\protect\citeauthoryear{Sharafaldin, Habibi~Lashkari, and
  Ghorbani}{Sharafaldin et~al\mbox{.}}{2018}]%
        {CICIDS2017}
\bibfield{author}{\bibinfo{person}{Iman Sharafaldin}, \bibinfo{person}{Arash
  Habibi~Lashkari}, {and} \bibinfo{person}{Ali Ghorbani}.}
  \bibinfo{year}{2018}\natexlab{}.
\newblock \showarticletitle{Toward Generating a New Intrusion Detection Dataset
  and Intrusion Traffic Characterization}. \bibinfo{pages}{108--116}.
\newblock
\showDOI{%
\url{https://doi.org/10.5220/0006639801080116}}


\bibitem[\protect\citeauthoryear{{Stratosphere}}{{Stratosphere}}{2015}]%
        {stratosphere}
\bibfield{author}{\bibinfo{person}{{Stratosphere}}.}
  \bibinfo{year}{2015}\natexlab{}.
\newblock \bibinfo{title}{Stratosphere Laboratory Datasets}.
\newblock   (\bibinfo{year}{2015}).
\newblock
\showURL{%
\url{https://www.stratosphereips.org/datasets-overview}}
\newblock
\shownote{[Online; accessed 12-March-2020].}


\bibitem[\protect\citeauthoryear{Tavallaee, Bagheri, Lu, and
  Ghorbani}{Tavallaee et~al\mbox{.}}{2009}]%
        {nsl_kdd}
\bibfield{author}{\bibinfo{person}{Mahbod Tavallaee}, \bibinfo{person}{Ebrahim
  Bagheri}, \bibinfo{person}{Wei Lu}, {and} \bibinfo{person}{Ali Ghorbani}.}
  \bibinfo{year}{2009}\natexlab{}.
\newblock \showarticletitle{A detailed analysis of the KDD CUP 99 data set}.
\newblock \bibinfo{journal}{{\em IEEE Symposium. Computational Intelligence for
  Security and Defense Applications, CISDA\/}}  \bibinfo{volume}{2}
  (\bibinfo{date}{07} \bibinfo{year}{2009}).
\newblock
\showDOI{%
\url{https://doi.org/10.1109/CISDA.2009.5356528}}


\bibitem[\protect\citeauthoryear{{Taylor}, {Spolaor}, {Conti}, and
  {Martinovic}}{{Taylor} et~al\mbox{.}}{2016}]%
        {Taylor_2016_eta}
\bibfield{author}{\bibinfo{person}{V.~F. {Taylor}}, \bibinfo{person}{R.
  {Spolaor}}, \bibinfo{person}{M. {Conti}}, {and} \bibinfo{person}{I.
  {Martinovic}}.} \bibinfo{year}{2016}\natexlab{}.
\newblock \showarticletitle{AppScanner: Automatic Fingerprinting of Smartphone
  Apps from Encrypted Network Traffic}. In \bibinfo{booktitle}{{\em 2016 IEEE
  European Symposium on Security and Privacy (EuroS P)}}.
  \bibinfo{pages}{439--454}.
\newblock
\showISSN{null}
\showDOI{%
\url{https://doi.org/10.1109/EuroSP.2016.40}}


\bibitem[\protect\citeauthoryear{{Wang}, {Yang}, and {Ren}}{{Wang}
  et~al\mbox{.}}{2009}]%
        {Wang_2009_df}
\bibfield{author}{\bibinfo{person}{J. {Wang}}, \bibinfo{person}{Q. {Yang}},
  {and} \bibinfo{person}{D. {Ren}}.} \bibinfo{year}{2009}\natexlab{}.
\newblock \showarticletitle{An Intrusion Detection Algorithm Based on Decision
  Tree Technology}. In \bibinfo{booktitle}{{\em 2009 Asia-Pacific Conference on
  Information Processing}}, Vol.~\bibinfo{volume}{2}.
  \bibinfo{pages}{333--335}.
\newblock
\showISSN{null}
\showDOI{%
\url{https://doi.org/10.1109/APCIP.2009.218}}


\bibitem[\protect\citeauthoryear{{Wang}, {Ye}, {Chen}, and {Qian}}{{Wang}
  et~al\mbox{.}}{2018}]%
        {Wang_2018_datanet}
\bibfield{author}{\bibinfo{person}{P. {Wang}}, \bibinfo{person}{F. {Ye}},
  \bibinfo{person}{X. {Chen}}, {and} \bibinfo{person}{Y. {Qian}}.}
  \bibinfo{year}{2018}\natexlab{}.
\newblock \showarticletitle{Datanet: Deep Learning Based Encrypted Network
  Traffic Classification in SDN Home Gateway}.
\newblock \bibinfo{journal}{{\em IEEE Access\/}}  \bibinfo{volume}{6}
  (\bibinfo{year}{2018}), \bibinfo{pages}{55380--55391}.
\newblock
\showISSN{2169-3536}
\showDOI{%
\url{https://doi.org/10.1109/ACCESS.2018.2872430}}


\bibitem[\protect\citeauthoryear{{Yamansavascilar}, {Guvensan}, {Yavuz}, and
  {Karsligil}}{{Yamansavascilar} et~al\mbox{.}}{2017}]%
        {Baris_2017_iscx}
\bibfield{author}{\bibinfo{person}{B. {Yamansavascilar}},
  \bibinfo{person}{M.~A. {Guvensan}}, \bibinfo{person}{A.~G. {Yavuz}}, {and}
  \bibinfo{person}{M.~E. {Karsligil}}.} \bibinfo{year}{2017}\natexlab{}.
\newblock \showarticletitle{Application identification via network traffic
  classification}. In \bibinfo{booktitle}{{\em 2017 International Conference on
  Computing, Networking and Communications (ICNC)}}. \bibinfo{pages}{843--848}.
\newblock
\showISSN{null}
\showDOI{%
\url{https://doi.org/10.1109/ICCNC.2017.7876241}}


\bibitem[\protect\citeauthoryear{{Yao}, {Liu}, {Zhang}, {Wu}, {Jiang}, and
  {Yu}}{{Yao} et~al\mbox{.}}{2019}]%
        {Yao_2019_attentionLSTM}
\bibfield{author}{\bibinfo{person}{H. {Yao}}, \bibinfo{person}{C. {Liu}},
  \bibinfo{person}{P. {Zhang}}, \bibinfo{person}{S. {Wu}}, \bibinfo{person}{C.
  {Jiang}}, {and} \bibinfo{person}{S. {Yu}}.} \bibinfo{year}{2019}\natexlab{}.
\newblock \showarticletitle{Identification of Encrypted Traffic Through
  Attention Mechanism Based Long Short Term Memory}.
\newblock \bibinfo{journal}{{\em IEEE Transactions on Big Data\/}}
  (\bibinfo{year}{2019}), \bibinfo{pages}{1--1}.
\newblock
\showISSN{2372-2096}
\showDOI{%
\url{https://doi.org/10.1109/TBDATA.2019.2940675}}


\bibitem[\protect\citeauthoryear{{Zhang}, {Zulkernine}, and {Haque}}{{Zhang}
  et~al\mbox{.}}{2008}]%
        {Zhang_2008_7ofLuNet}
\bibfield{author}{\bibinfo{person}{J. {Zhang}}, \bibinfo{person}{M.
  {Zulkernine}}, {and} \bibinfo{person}{A. {Haque}}.}
  \bibinfo{year}{2008}\natexlab{}.
\newblock \showarticletitle{Random-Forests-Based Network Intrusion Detection
  Systems}.
\newblock \bibinfo{journal}{{\em IEEE Transactions on Systems, Man, and
  Cybernetics, Part C (Applications and Reviews)\/}} \bibinfo{volume}{38},
  \bibinfo{number}{5} (\bibinfo{date}{Sep.} \bibinfo{year}{2008}),
  \bibinfo{pages}{649--659}.
\newblock
\showISSN{1558-2442}
\showDOI{%
\url{https://doi.org/10.1109/TSMCC.2008.923876}}


\end{thebibliography}

\appendix
\section{Appendix}
\begin{figure}[t]
\subfigure{\includegraphics[height=1.2in,width=1.65in]{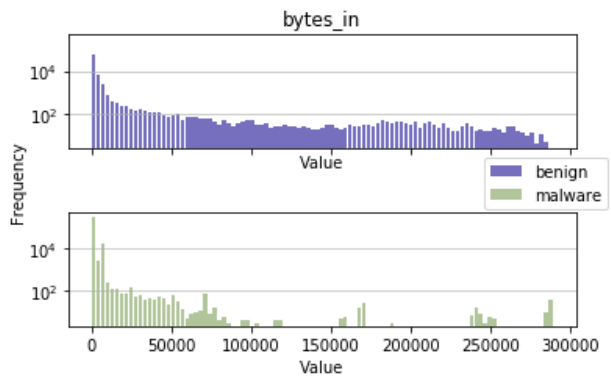}}
\subfigure{\includegraphics[height=1.2in,width=1.65in]{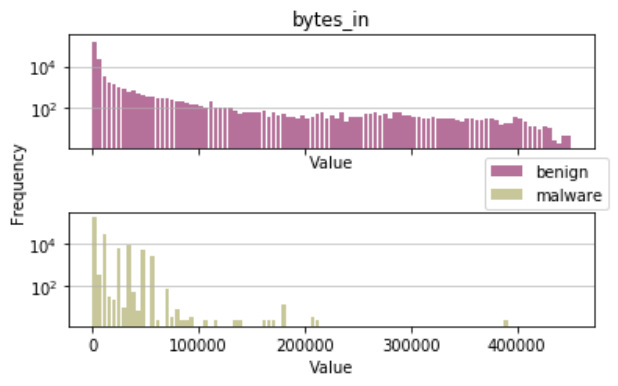}}
\subfigure{\includegraphics[height=1.2in,width=1.65in]{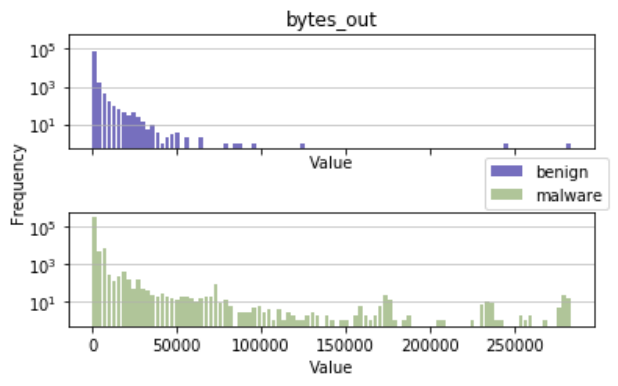}}
\subfigure{\includegraphics[height=1.2in,width=1.65in]{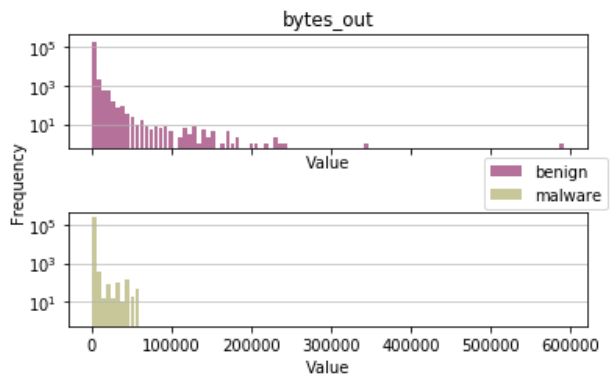}}
\subfigure{\includegraphics[height=1.2in,width=1.65in]{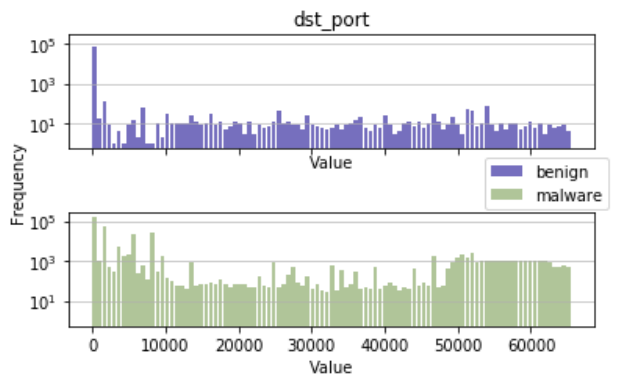}}
\subfigure{\includegraphics[height=1.2in,width=1.65in]{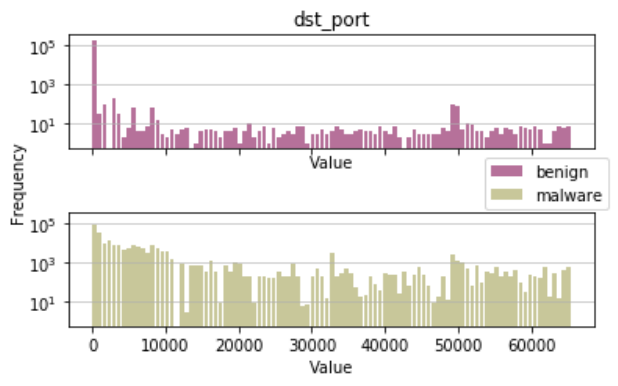}}
\subfigure{\includegraphics[height=1.2in,width=1.65in]{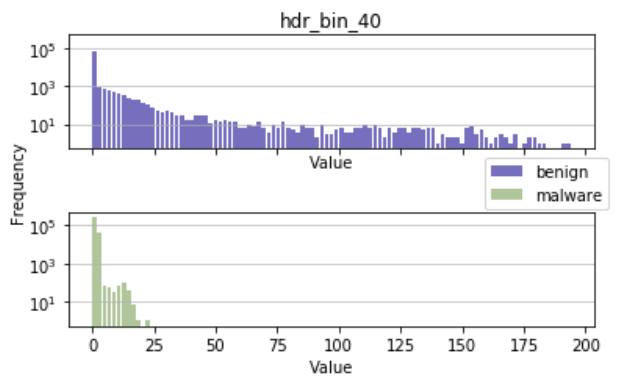}}
\subfigure{\includegraphics[height=1.2in,width=1.65in]{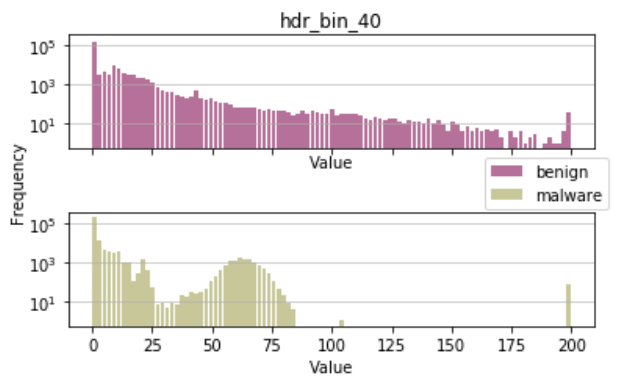}}
\subfigure{\includegraphics[height=1.2in,width=1.65in]{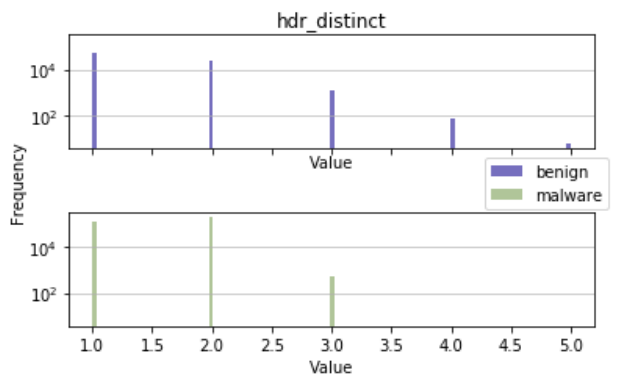}}
\subfigure{\includegraphics[height=1.2in,width=1.65in]{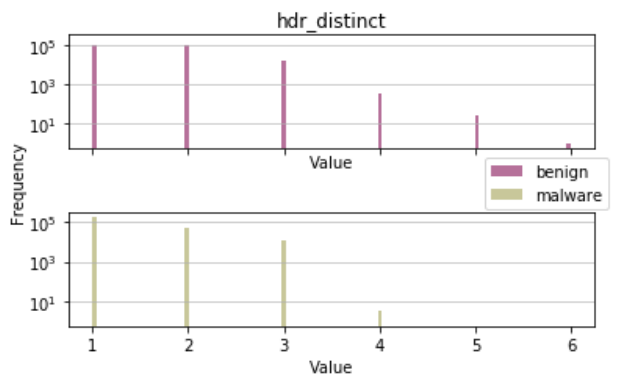}}
\subfigure{\includegraphics[height=1.2in,width=1.65in]{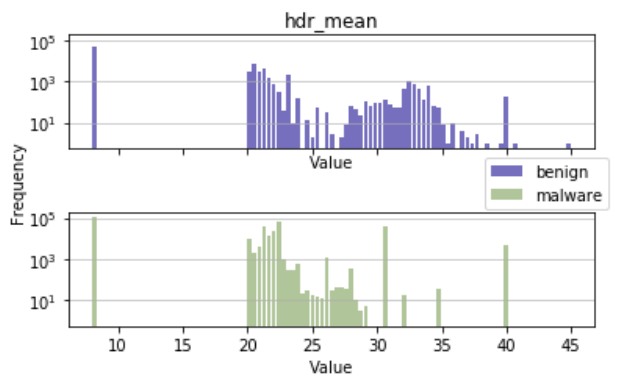}}
\subfigure{\includegraphics[height=1.2in,width=1.65in]{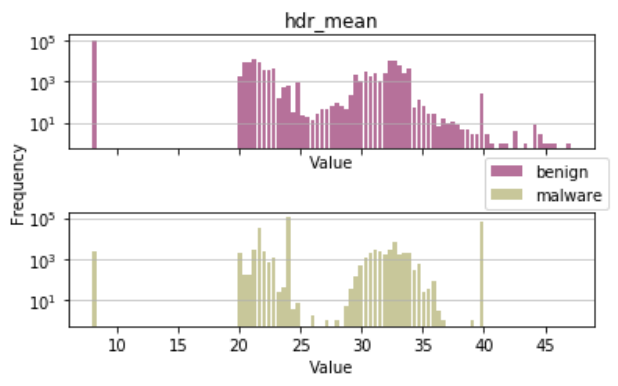}}
\caption{Metadata single value features - 1 \\ (Left: NetML dataset , Right: CICIDS2017 dataset)}
\label{metadata_single_1_malware}
\end{figure}

\begin{figure}[t]
\subfigure{\includegraphics[height=1.2in,width=2.5in]{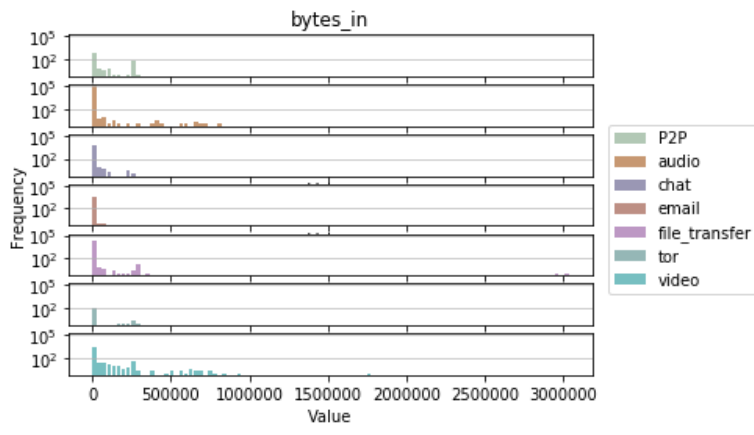}}
\subfigure{\includegraphics[height=1.2in,width=2.65in]{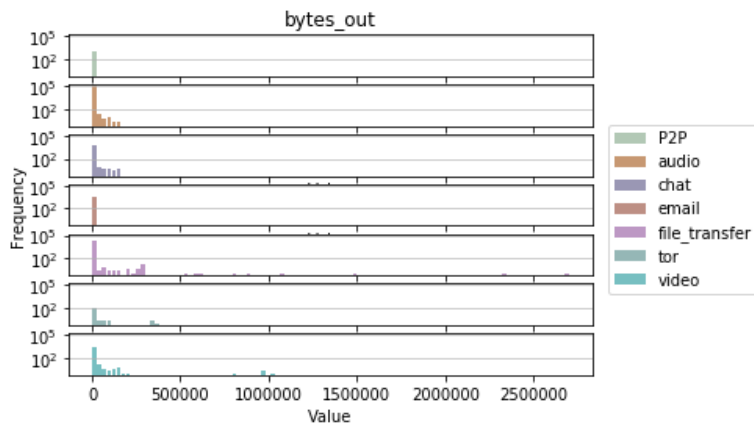}}
\subfigure{\includegraphics[height=1.2in,width=2.5in]{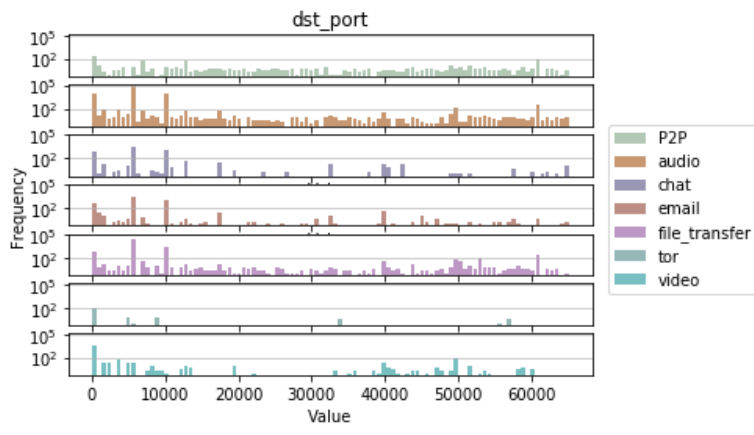}}
\subfigure{\includegraphics[height=1.2in,width=2.5in]{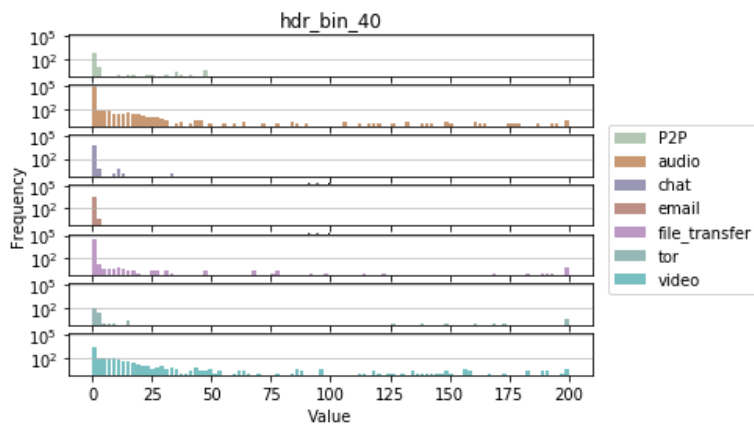}}
\subfigure{\includegraphics[height=1.2in,width=2.5in]{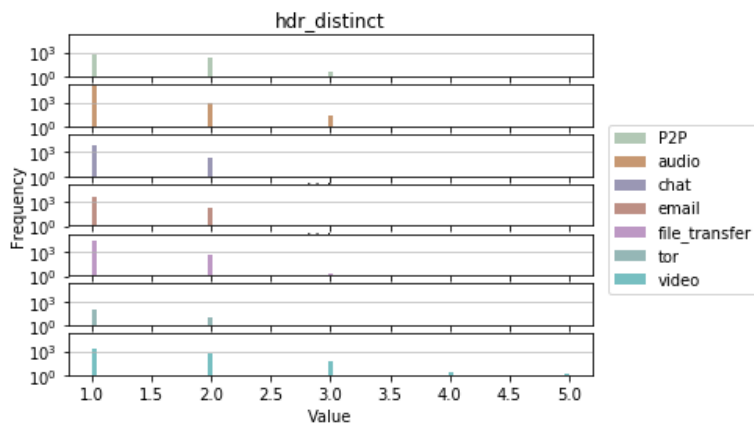}}
\subfigure{\includegraphics[height=1.2in,width=2.5in]{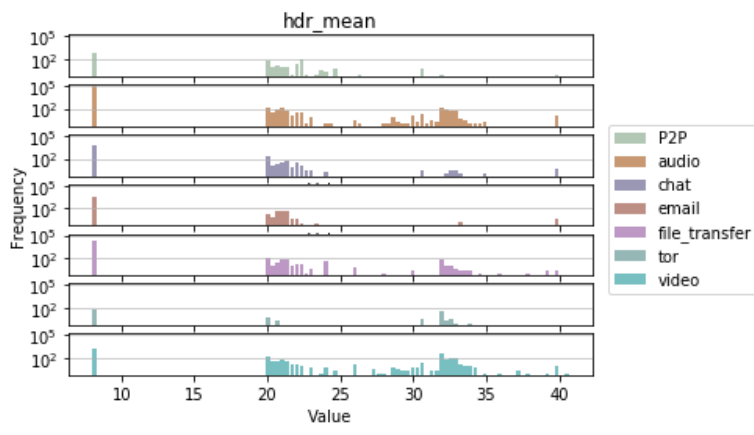}}
\caption{Metadata single value features - 1 \\ (non-vpn2016 dataset)}
\label{metadata_single_1_nonvpn}
\end{figure}

\begin{figure}[t]
\subfigure{\includegraphics[height=1.2in,width=1.65in]{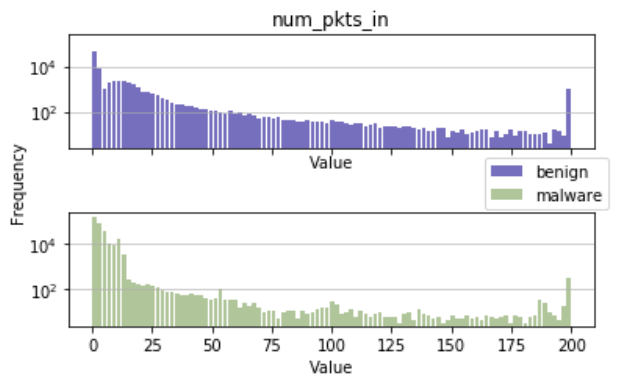}}
\subfigure{\includegraphics[height=1.2in,width=1.65in]{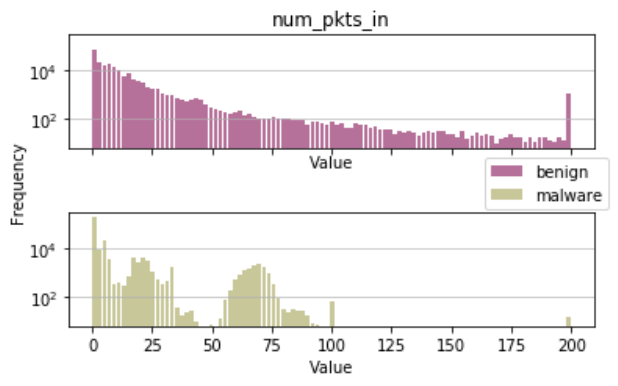}}
\subfigure{\includegraphics[height=1.2in,width=1.65in]{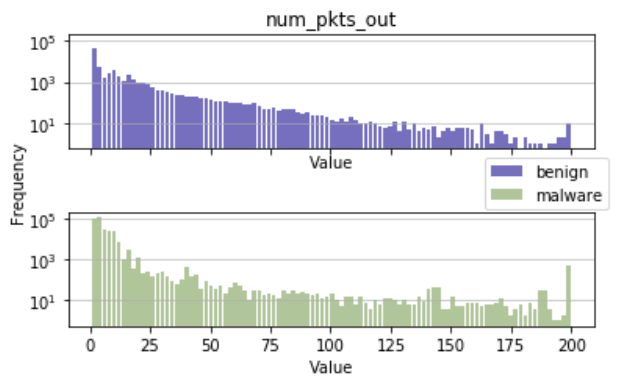}}
\subfigure{\includegraphics[height=1.2in,width=1.65in]{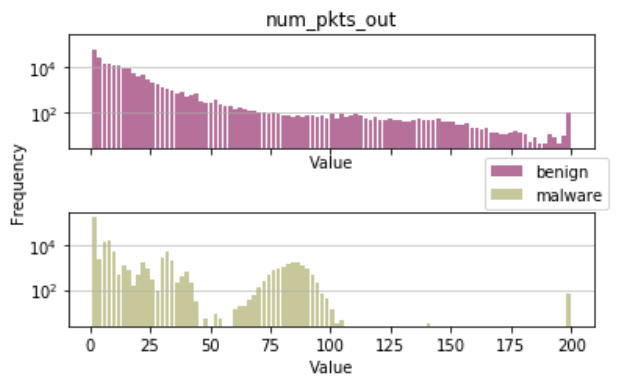}}
\subfigure{\includegraphics[height=1.2in,width=1.65in]{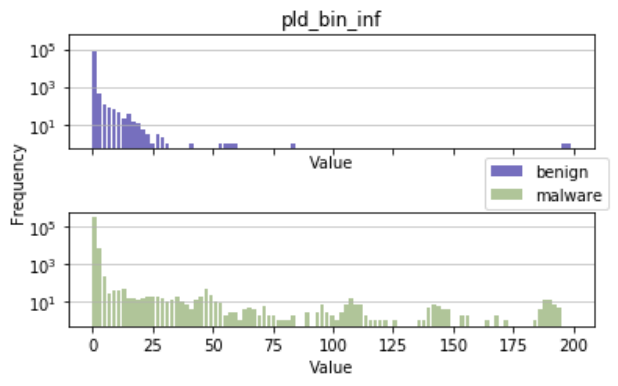}}
\subfigure{\includegraphics[height=1.2in,width=1.65in]{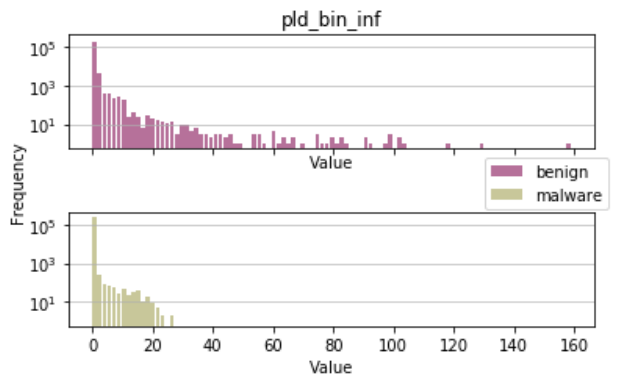}}
\subfigure{\includegraphics[height=1.2in,width=1.65in]{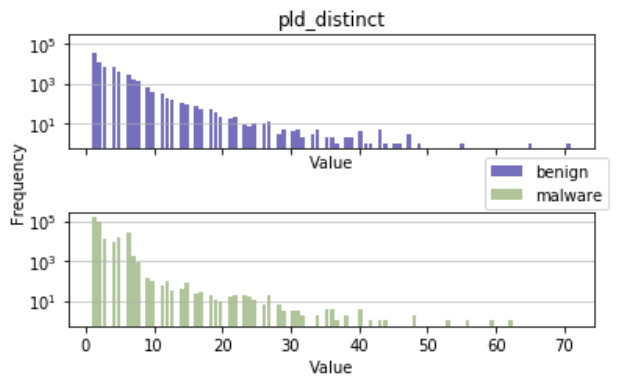}}
\subfigure{\includegraphics[height=1.2in,width=1.65in]{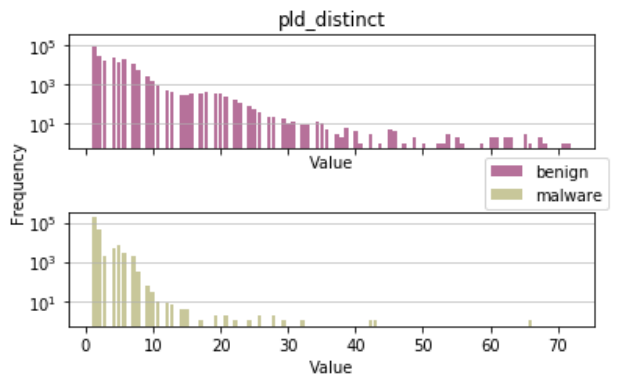}}
\subfigure{\includegraphics[height=1.2in,width=1.65in]{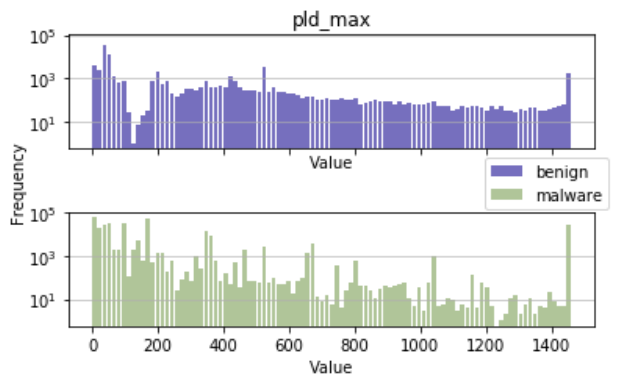}}
\subfigure{\includegraphics[height=1.2in,width=1.65in]{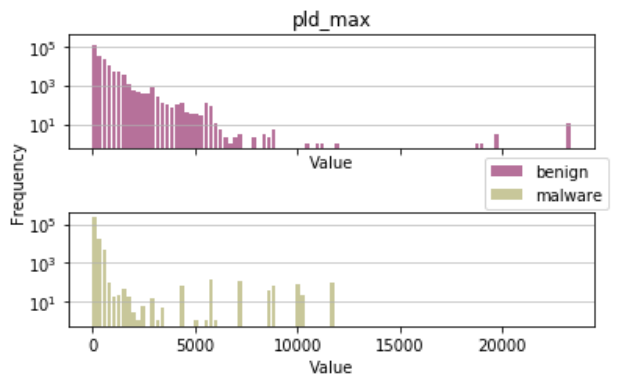}}
\subfigure{\includegraphics[height=1.2in,width=1.65in]{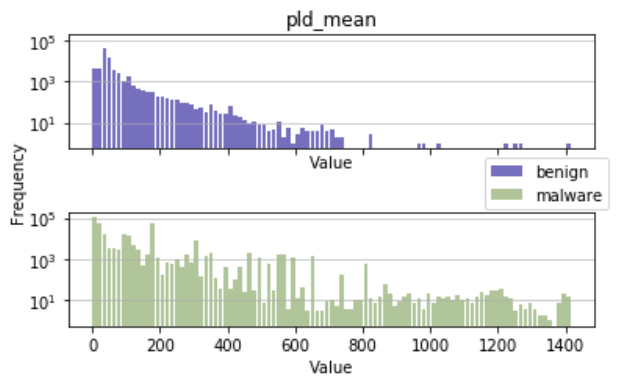}}
\subfigure{\includegraphics[height=1.2in,width=1.65in]{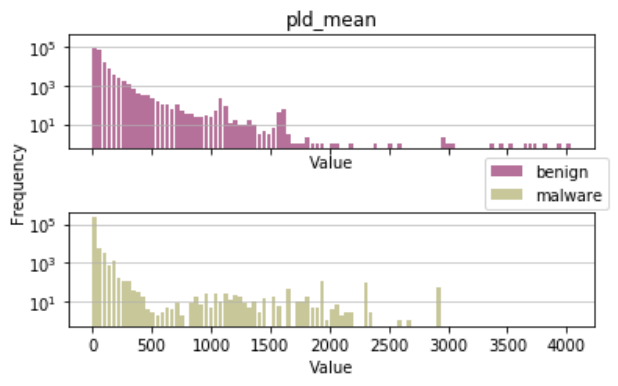}}
\caption{Metadata single value features - 2 \\ (Left: NetML dataset , Right: CICIDS2017 dataset)}
\label{metadata_single_2_malware}
\end{figure}

\begin{figure}[t]
\subfigure{\includegraphics[height=1.2in,width=2.5in]{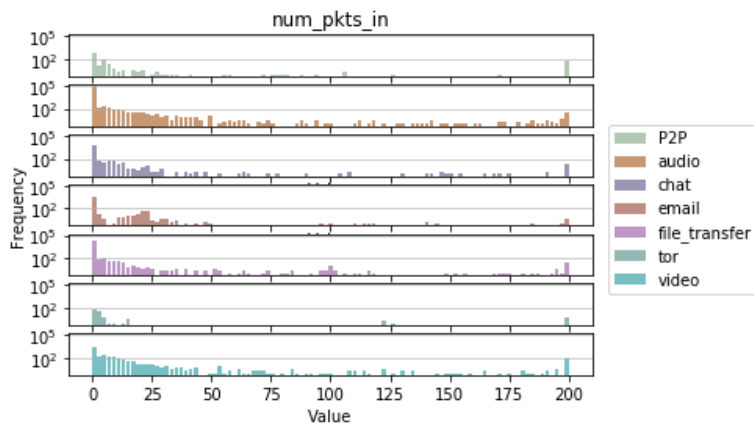}}
\subfigure{\includegraphics[height=1.2in,width=2.5in]{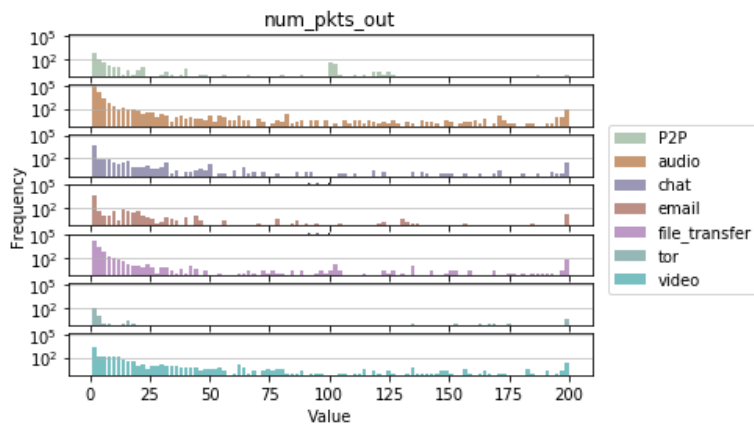}}
\subfigure{\includegraphics[height=1.2in,width=2.5in]{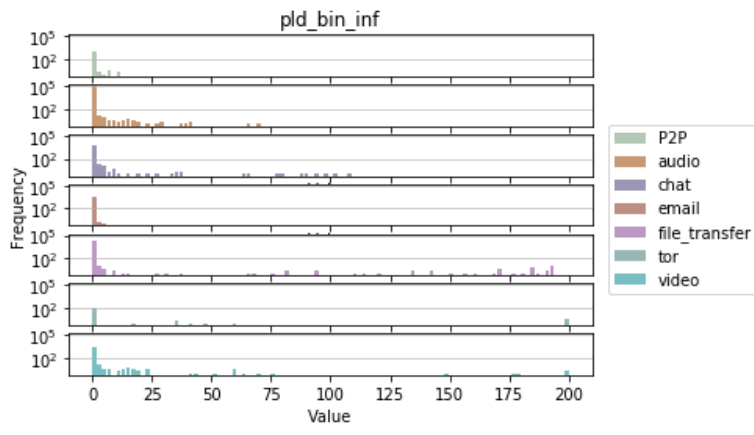}}
\subfigure{\includegraphics[height=1.2in,width=2.5in]{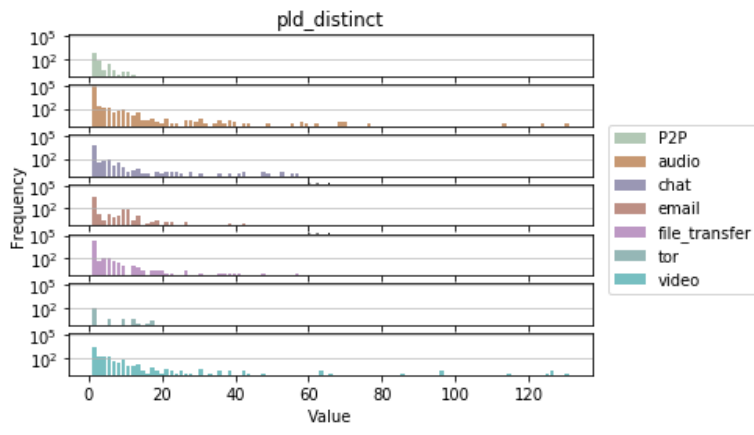}}
\subfigure{\includegraphics[height=1.2in,width=2.5in]{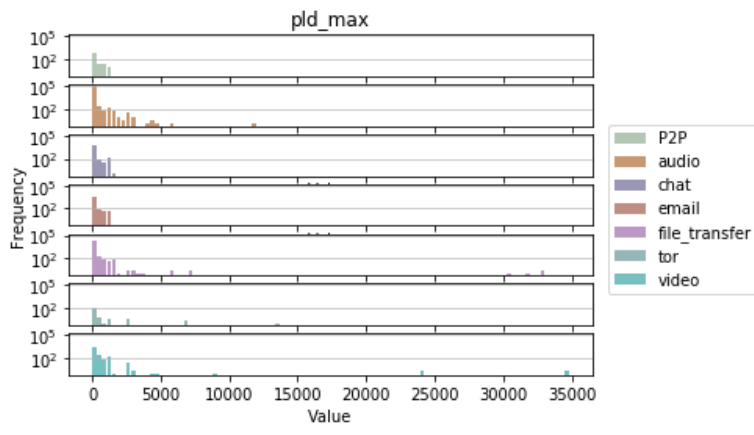}}
\subfigure{\includegraphics[height=1.2in,width=2.5in]{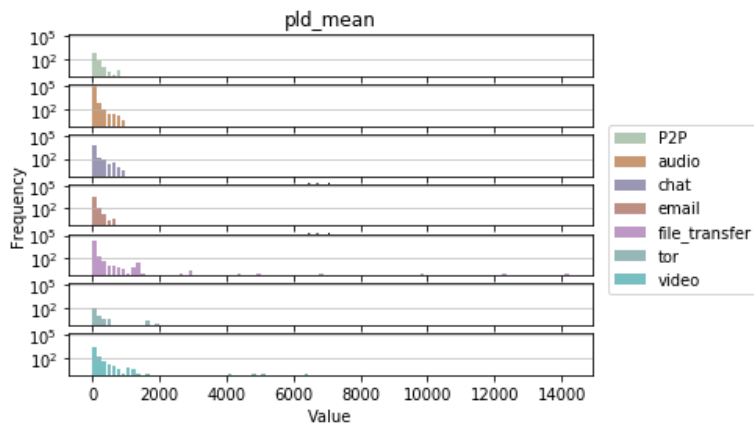}}
\caption{Metadata single value features - 2 \\ (non-vpn2016 dataset)}
\label{metadata_single_2_nonvpn}
\end{figure}

\begin{figure}[t]
\subfigure{\includegraphics[height=1.2in,width=1.65in]{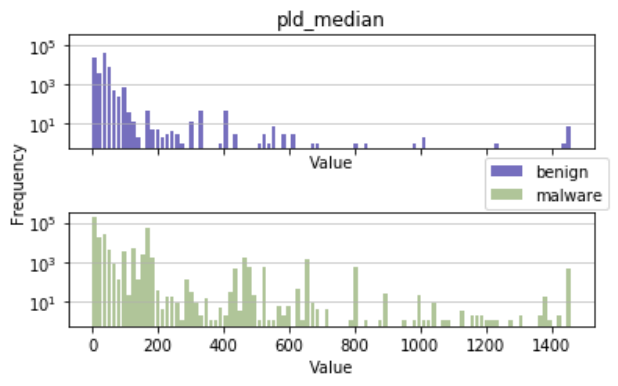}}
\subfigure{\includegraphics[height=1.2in,width=1.65in]{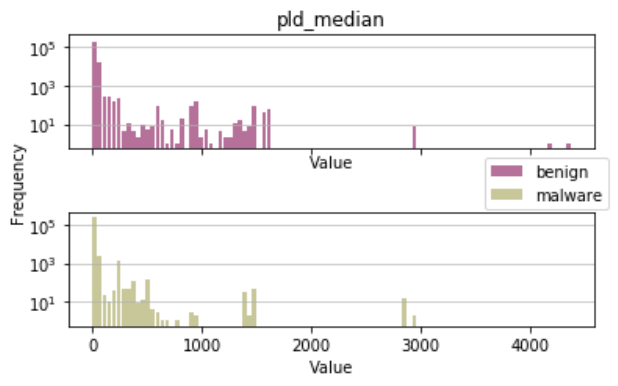}}
\subfigure{\includegraphics[height=1.2in,width=1.65in]{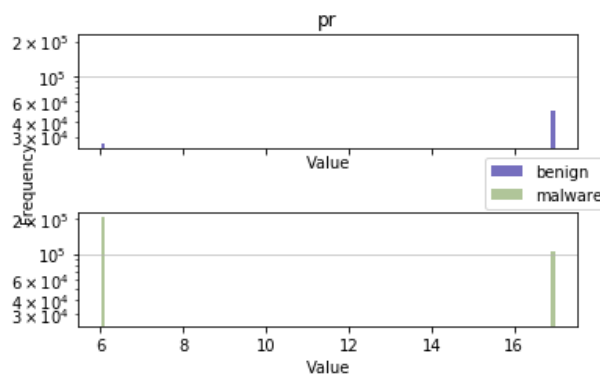}}
\subfigure{\includegraphics[height=1.2in,width=1.65in]{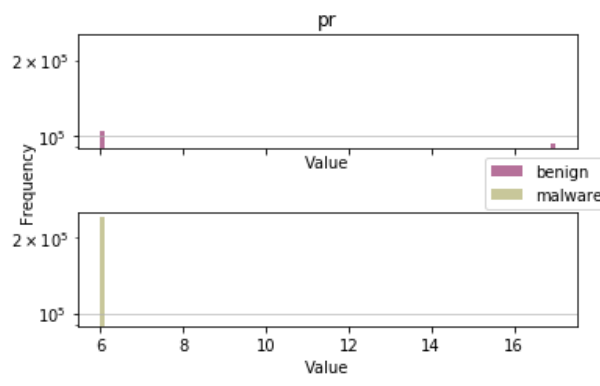}}
\subfigure{\includegraphics[height=1.2in,width=1.65in]{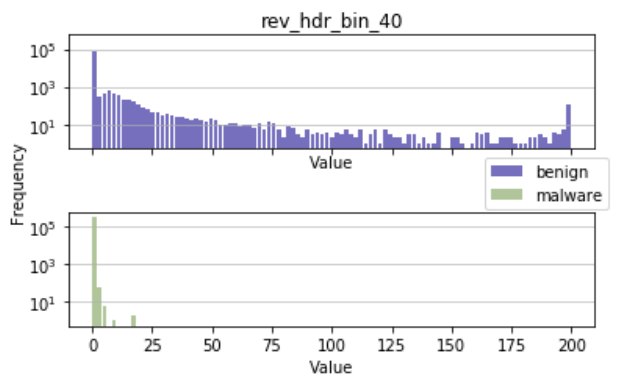}}
\subfigure{\includegraphics[height=1.2in,width=1.65in]{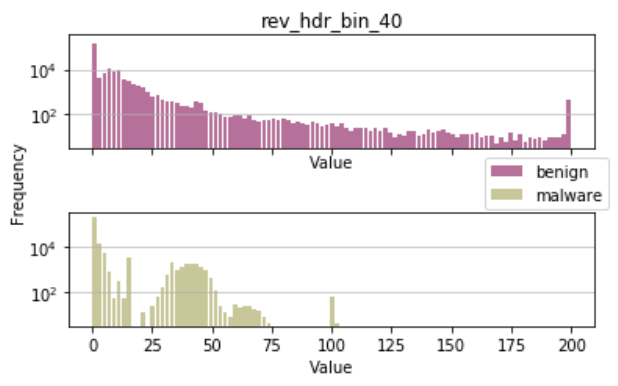}}
\subfigure{\includegraphics[height=1.2in,width=1.65in]{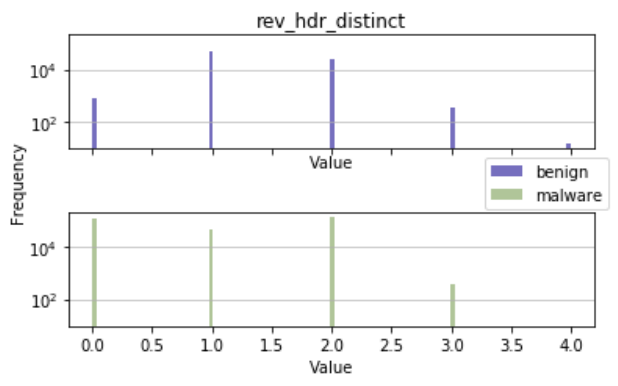}}
\subfigure{\includegraphics[height=1.2in,width=1.65in]{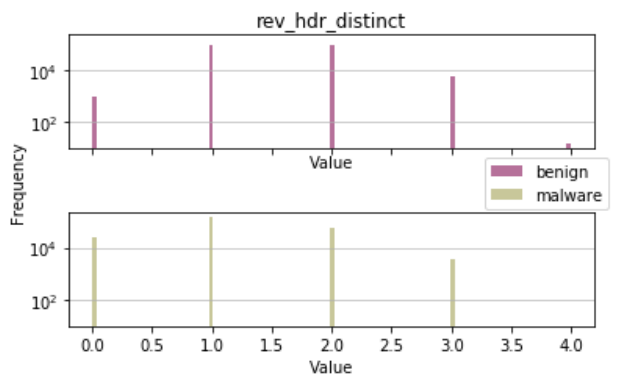}}
\subfigure{\includegraphics[height=1.2in,width=1.65in]{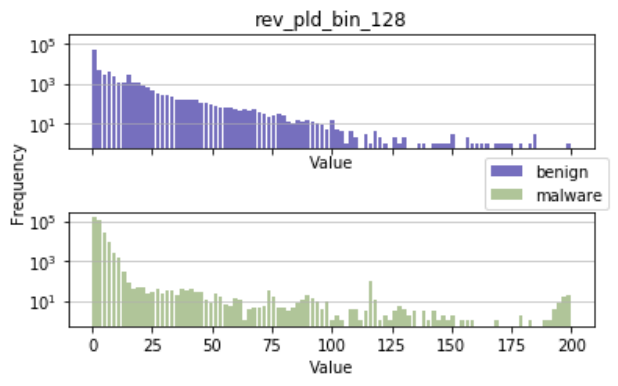}}
\subfigure{\includegraphics[height=1.2in,width=1.65in]{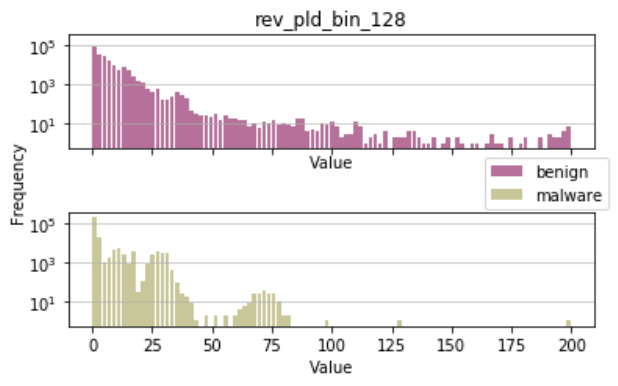}}
\subfigure{\includegraphics[height=1.2in,width=1.65in]{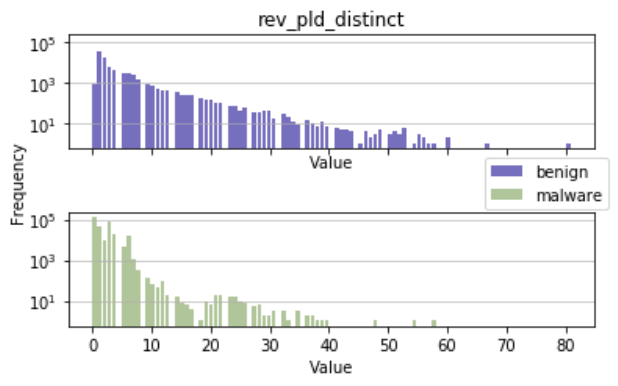}}
\subfigure{\includegraphics[height=1.2in,width=1.65in]{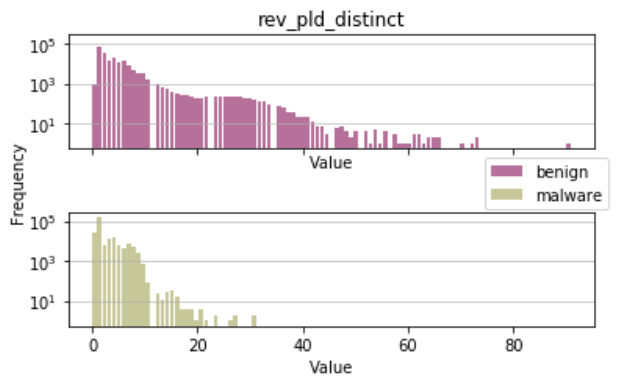}}
\caption{Metadata single value features - 3 \\ (Left: NetML dataset , Right: CICIDS2017 dataset)}
\label{metadata_single_3_malware}
\end{figure}

\begin{figure}[t]
\subfigure{\includegraphics[height=1.2in,width=2.5in]{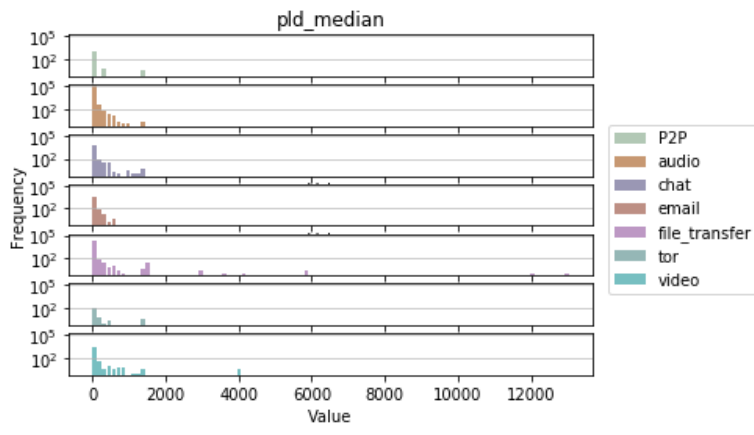}}
\subfigure{\includegraphics[height=1.2in,width=2.5in]{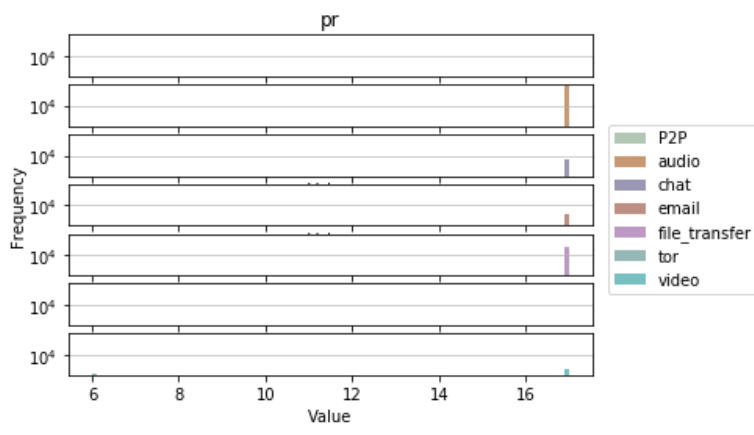}}
\subfigure{\includegraphics[height=1.2in,width=2.5in]{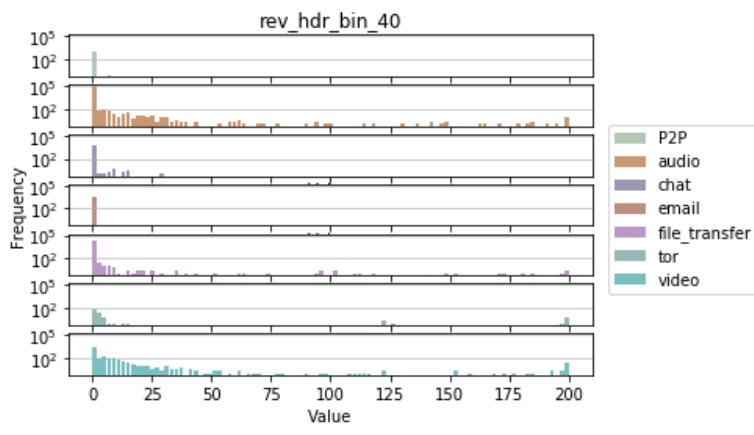}}
\subfigure{\includegraphics[height=1.2in,width=2.5in]{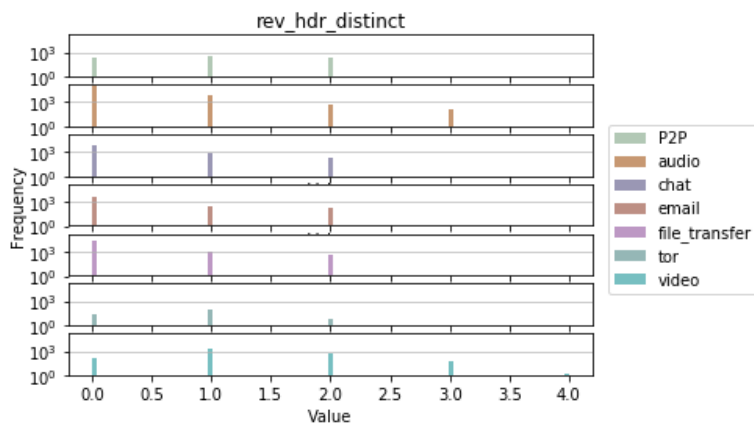}}
\subfigure{\includegraphics[height=1.2in,width=2.5in]{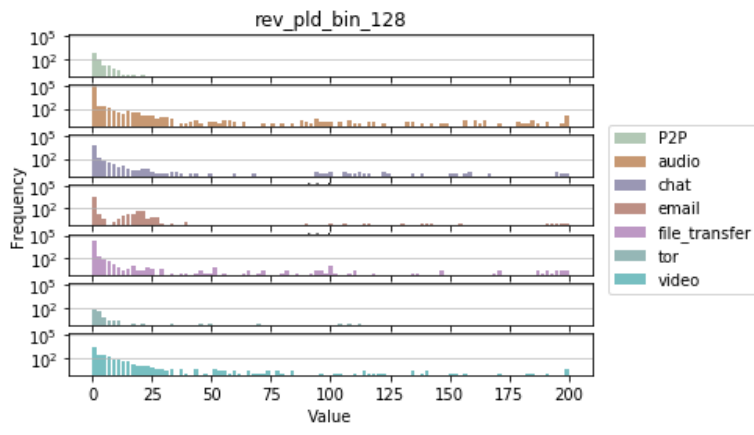}}
\subfigure{\includegraphics[height=1.2in,width=2.5in]{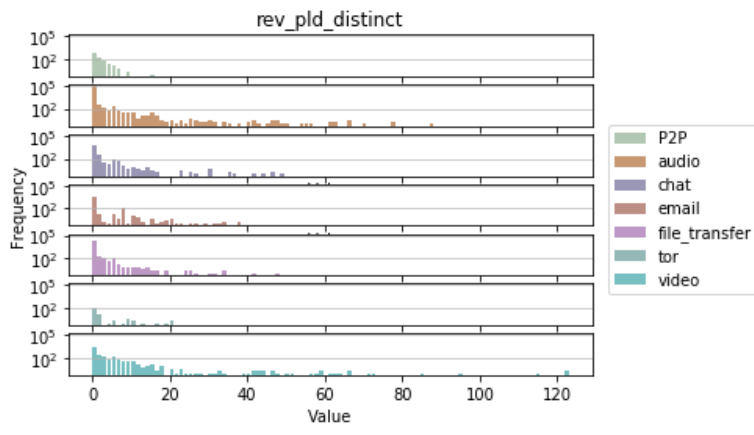}}
\caption{Metadata single value features - 3 \\ (non-vpn2016 dataset)}
\label{metadata_single_3_nonvpn}
\end{figure}

\begin{figure}[t]
\subfigure{\includegraphics[height=1.2in,width=1.65in]{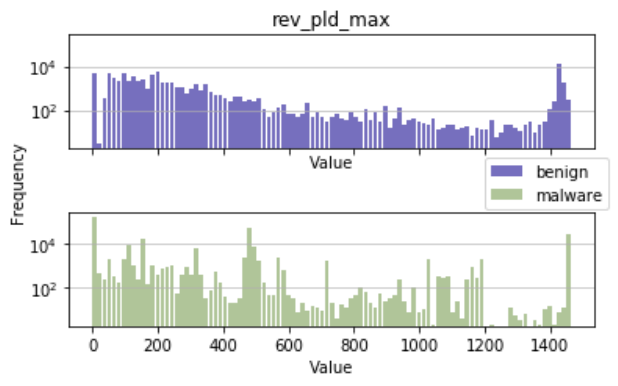}}
\subfigure{\includegraphics[height=1.2in,width=1.65in]{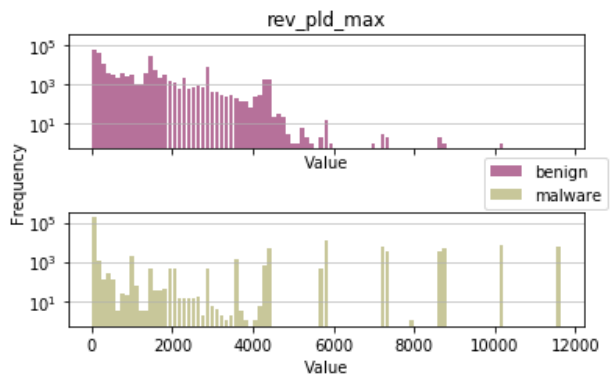}}
\subfigure{\includegraphics[height=1.2in,width=1.65in]{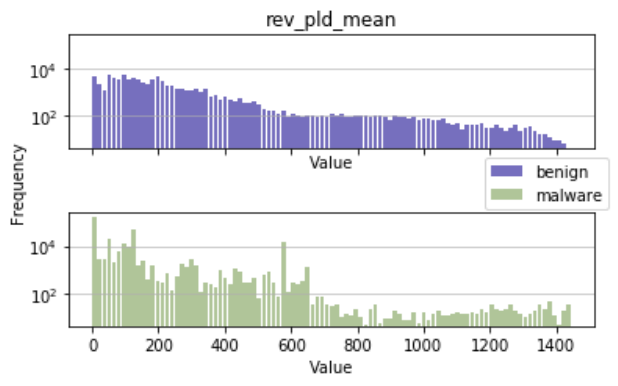}}
\subfigure{\includegraphics[height=1.2in,width=1.65in]{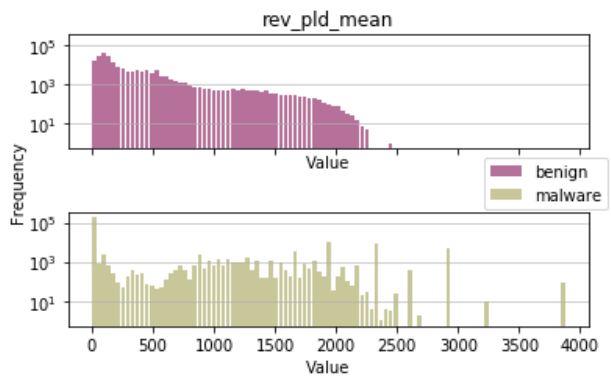}}
\subfigure{\includegraphics[height=1.2in,width=1.65in]{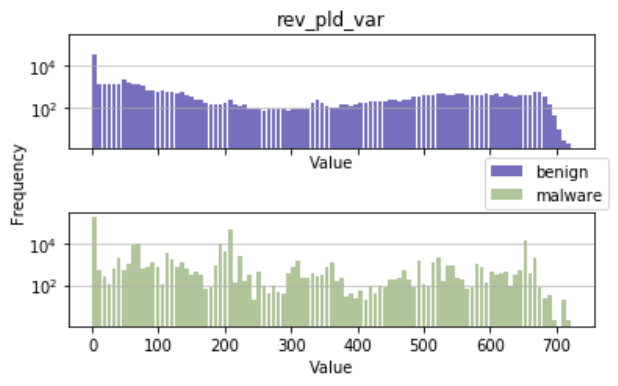}}
\subfigure{\includegraphics[height=1.2in,width=1.65in]{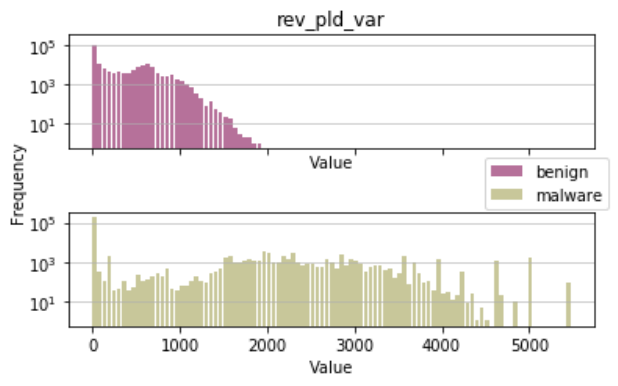}}
\subfigure{\includegraphics[height=1.2in,width=1.65in]{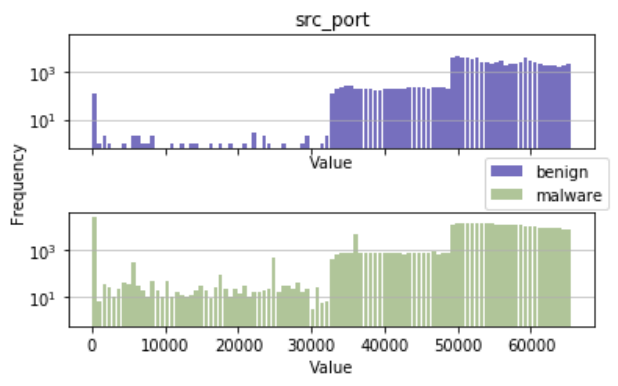}}
\subfigure{\includegraphics[height=1.2in,width=1.65in]{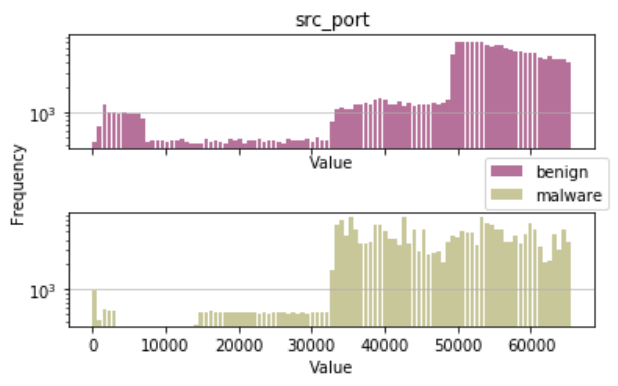}}
\subfigure{\includegraphics[height=1.2in,width=1.65in]{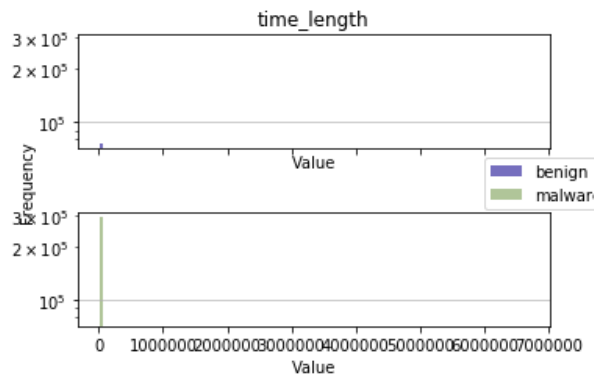}}
\subfigure{\includegraphics[height=1.2in,width=1.65in]{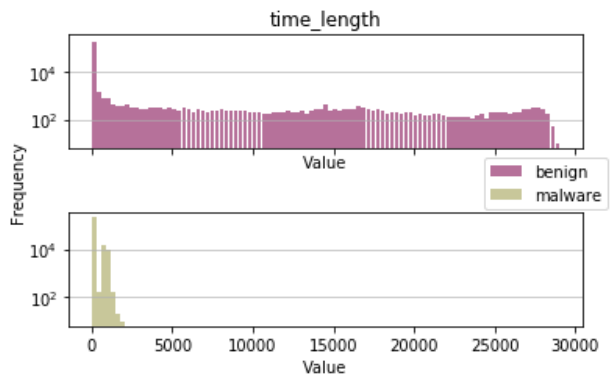}}
\caption{Metadata single value features - 4 \\ (Left: NetML dataset , Right: CICIDS2017 dataset)}
\label{metadata_single_4_malware}
\end{figure}

\begin{figure}[t]
\subfigure{\includegraphics[height=1.2in,width=2.5in]{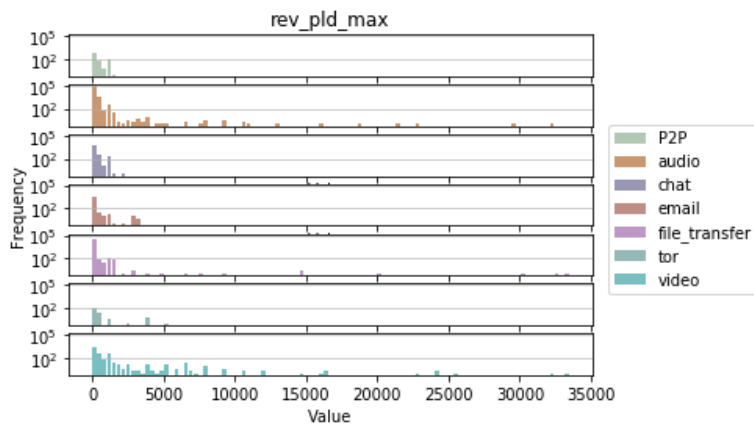}}
\subfigure{\includegraphics[height=1.2in,width=2.5in]{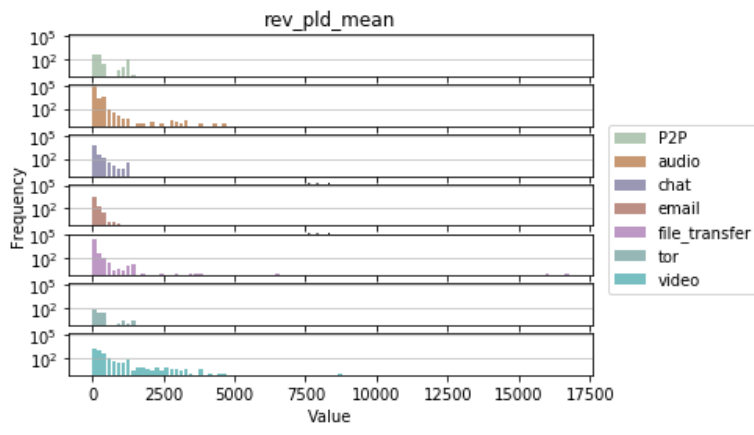}}
\subfigure{\includegraphics[height=1.2in,width=2.5in]{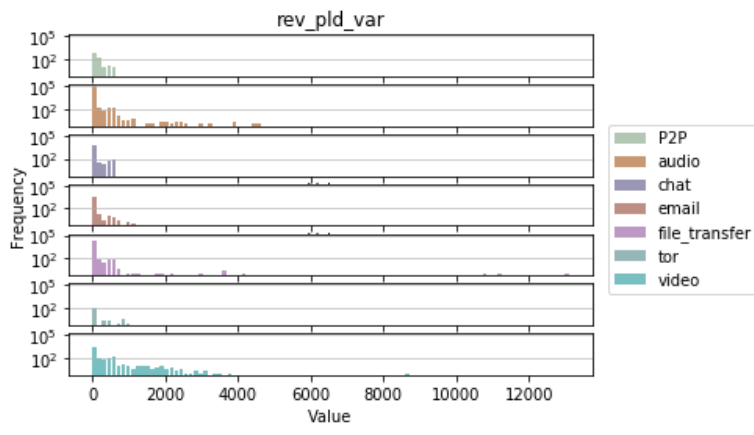}}
\subfigure{\includegraphics[height=1.2in,width=2.5in]{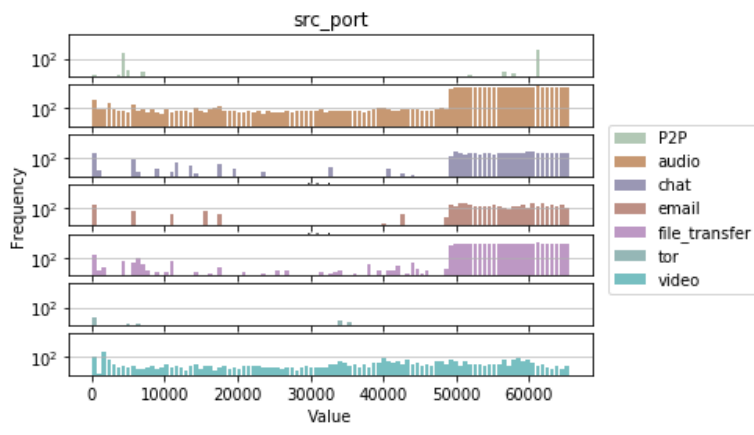}}
\subfigure{\includegraphics[height=1.2in,width=2.5in]{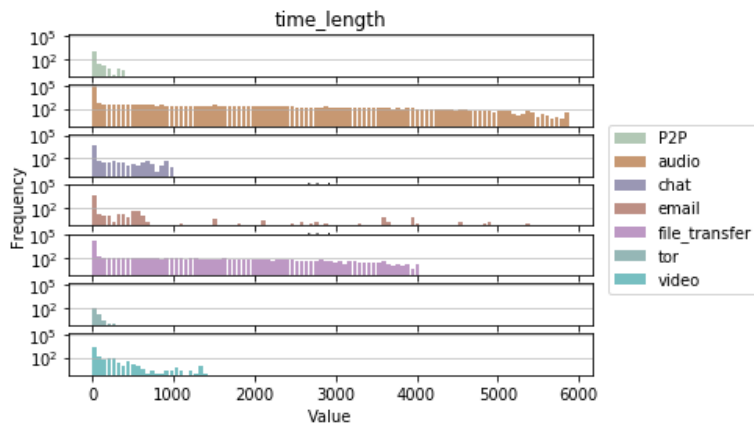}}
\caption{Metadata single value features - 4 \\ (non-vpn2016 dataset)}
\label{metadata_single_4_nonvpn}
\end{figure}

\begin{figure}[t]
\subfigure{\includegraphics[height=1.2in,width=1.65in]{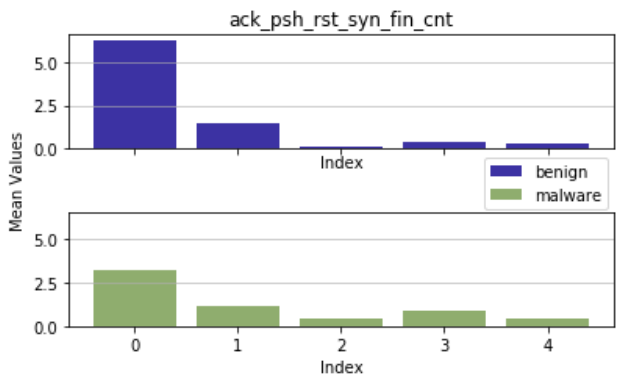}}
\subfigure{\includegraphics[height=1.2in,width=1.65in]{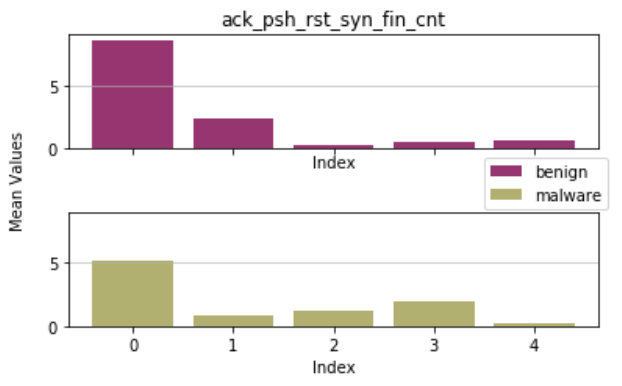}}
\subfigure{\includegraphics[height=1.2in,width=1.65in]{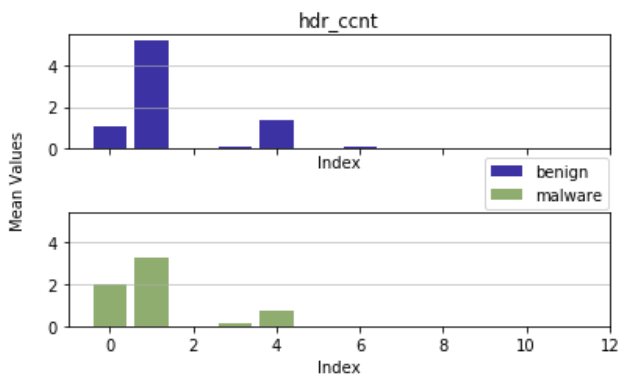}}
\subfigure{\includegraphics[height=1.2in,width=1.65in]{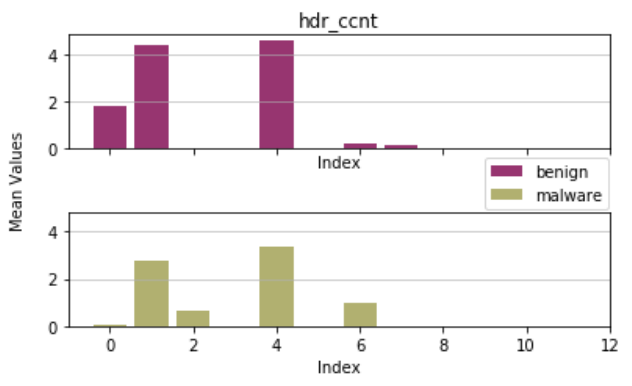}}
\subfigure{\includegraphics[height=1.2in,width=1.65in]{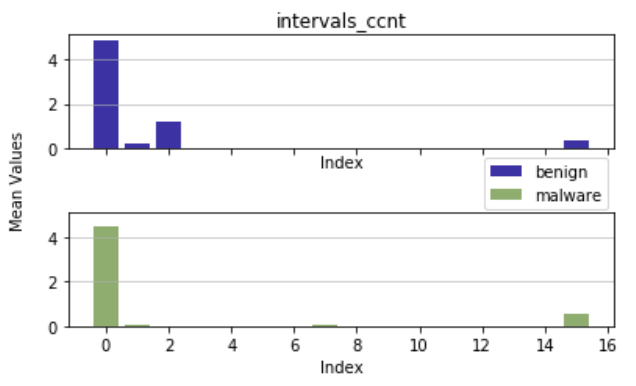}}
\subfigure{\includegraphics[height=1.2in,width=1.65in]{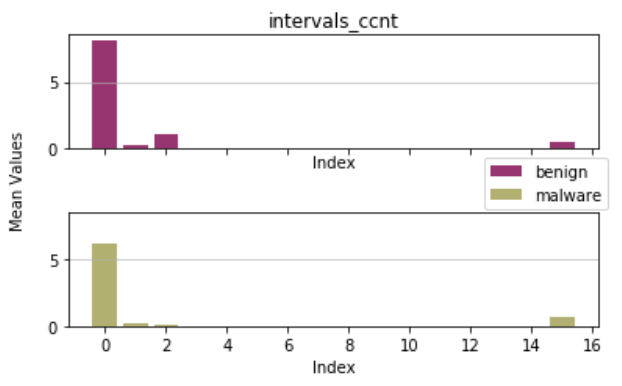}}
\subfigure{\includegraphics[height=1.2in,width=1.65in]{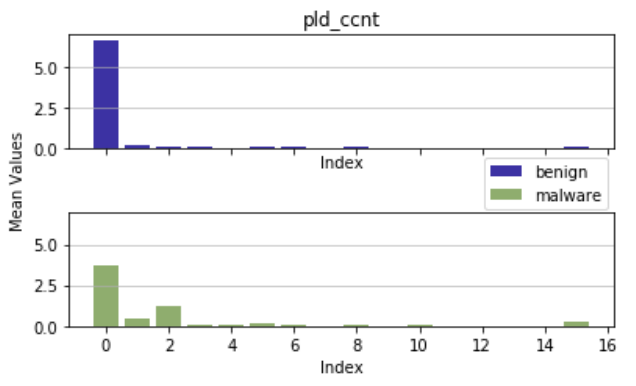}}
\subfigure{\includegraphics[height=1.2in,width=1.65in]{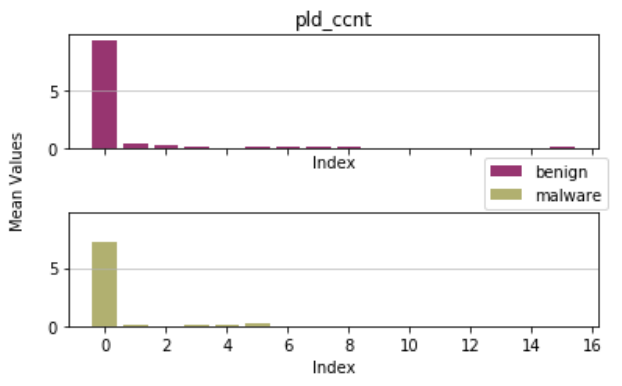}}
\caption{Metadata histogram-like Features - 1 \\ (Left: NetML dataset , Right: CICIDS2017 dataset)}
\label{metadata_arrays_1_malware}
\end{figure}

\begin{figure}[t]
\subfigure{\includegraphics[height=1.2in,width=2.5in]{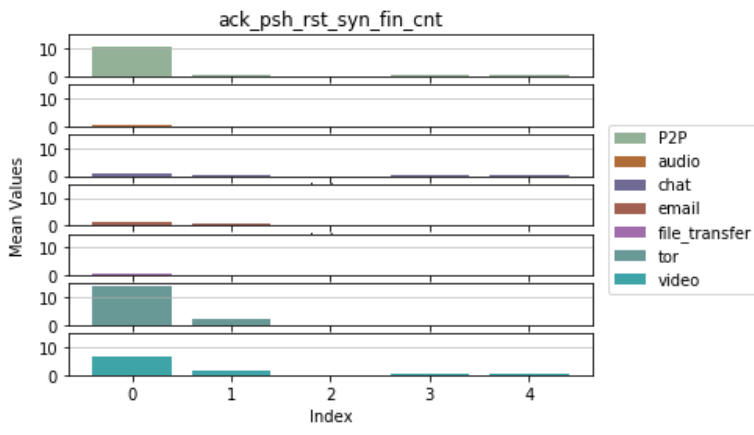}}
\subfigure{\includegraphics[height=1.2in,width=2.5in]{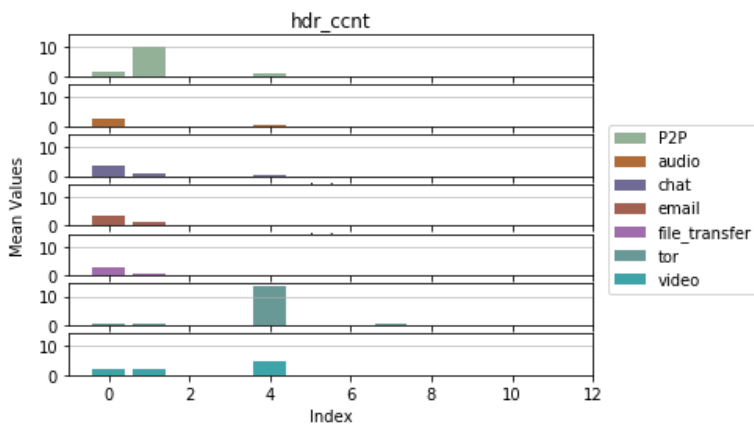}}
\subfigure{\includegraphics[height=1.2in,width=2.5in]{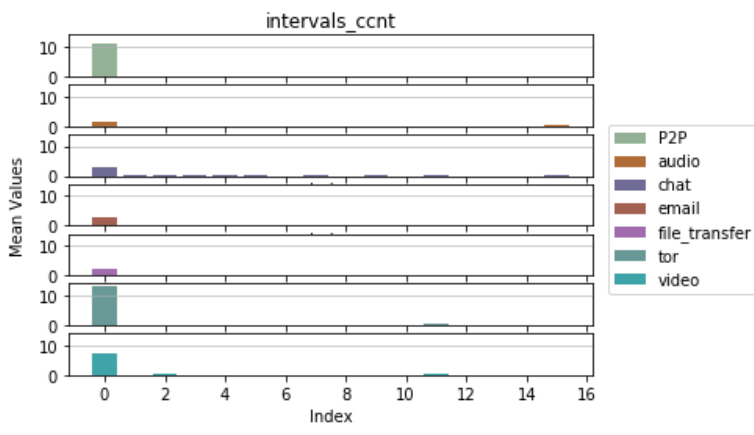}}
\subfigure{\includegraphics[height=1.2in,width=2.5in]{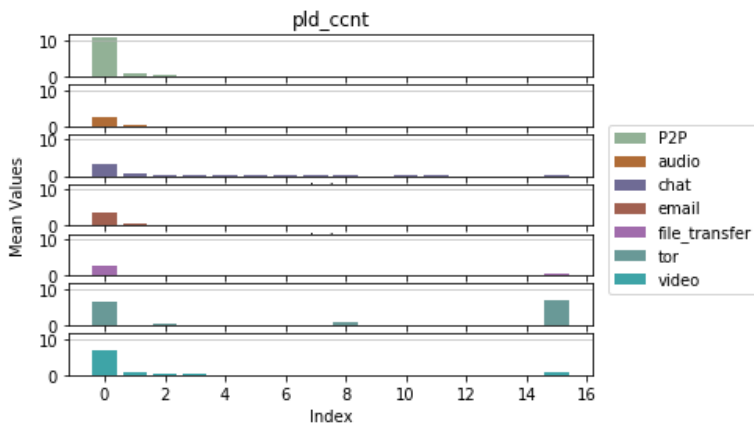}}
\caption{Metadata histogram-like Features - 1 \\ (non-vpn2016 dataset)}
\label{metadata_arrays_1_nonvpn}
\end{figure}

\begin{figure}[t]
\subfigure{\includegraphics[height=1.2in,width=1.65in]{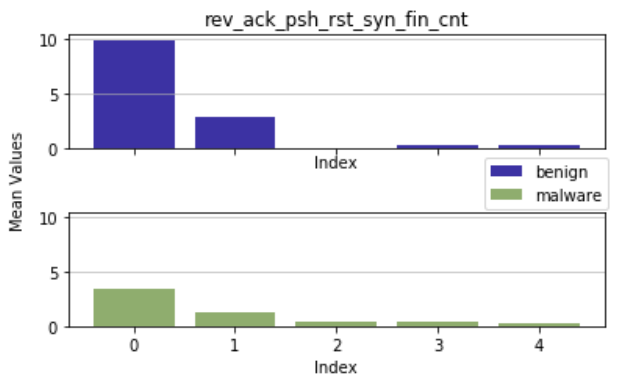}}
\subfigure{\includegraphics[height=1.2in,width=1.65in]{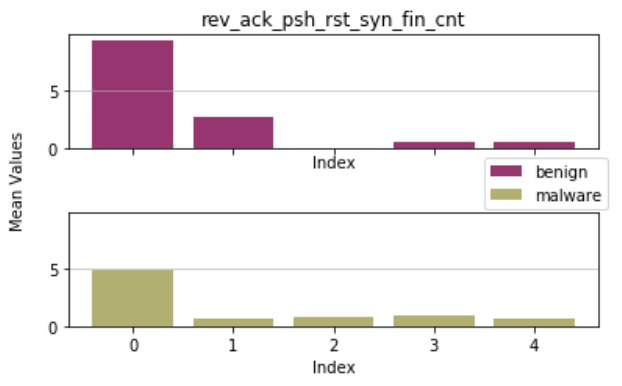}}
\subfigure{\includegraphics[height=1.2in,width=1.65in]{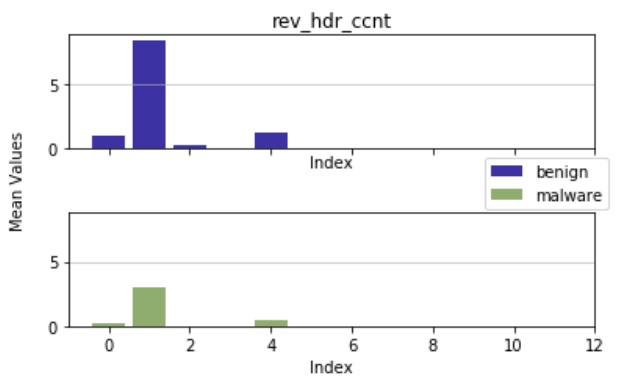}}
\subfigure{\includegraphics[height=1.2in,width=1.65in]{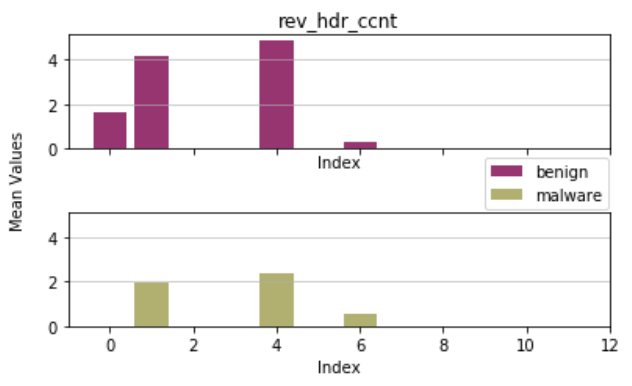}}
\subfigure{\includegraphics[height=1.2in,width=1.65in]{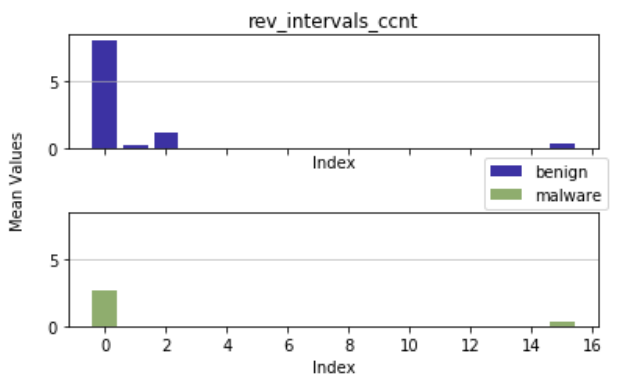}}
\subfigure{\includegraphics[height=1.2in,width=1.65in]{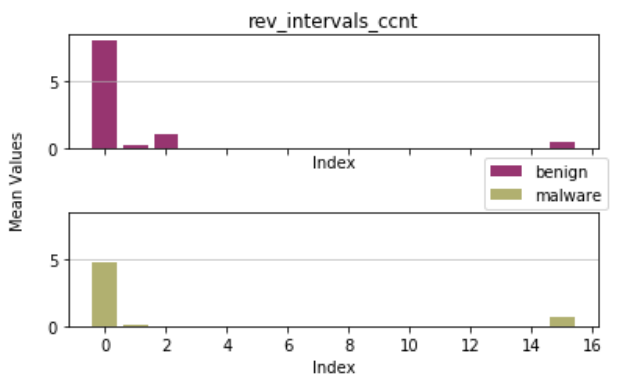}}
\subfigure{\includegraphics[height=1.2in,width=1.65in]{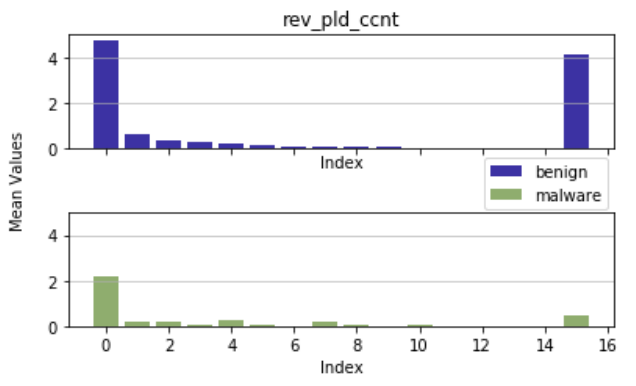}}
\subfigure{\includegraphics[height=1.2in,width=1.65in]{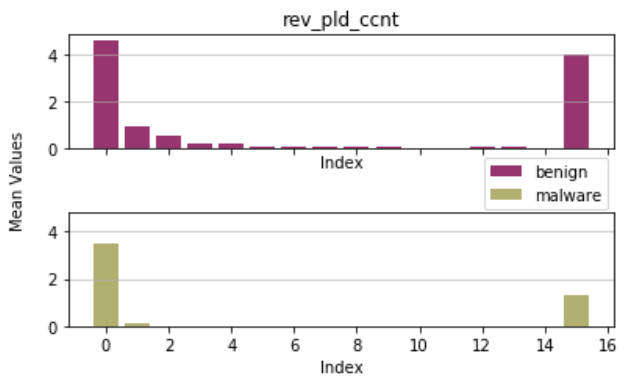}}
\caption{Metadata histogram-like Features - 2 \\ (Left: NetML dataset , Right: CICIDS2017 dataset)}
\label{metadata_arrays_2_malware}
\end{figure}

\begin{figure}[t]
\subfigure{\includegraphics[height=1.2in,width=2.5in]{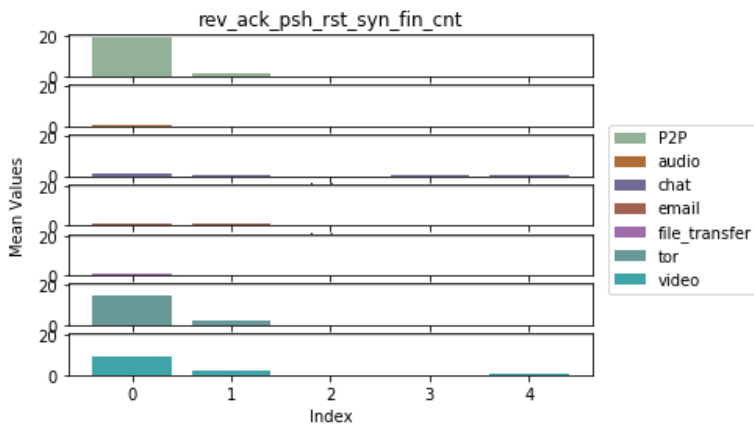}}
\subfigure{\includegraphics[height=1.2in,width=2.5in]{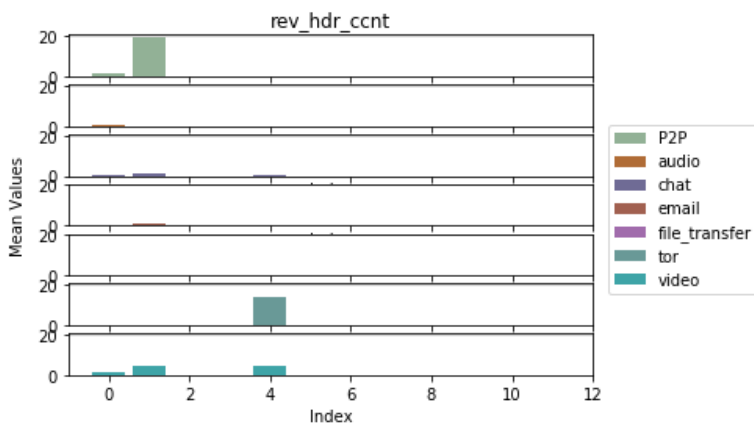}}
\subfigure{\includegraphics[height=1.2in,width=2.5in]{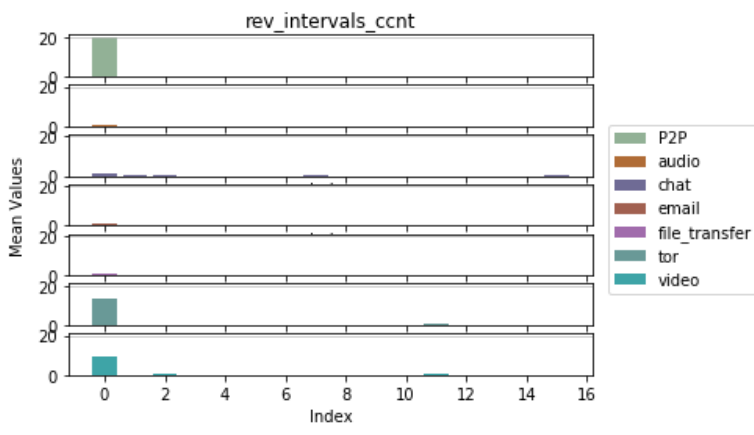}}
\subfigure{\includegraphics[height=1.2in,width=2.5in]{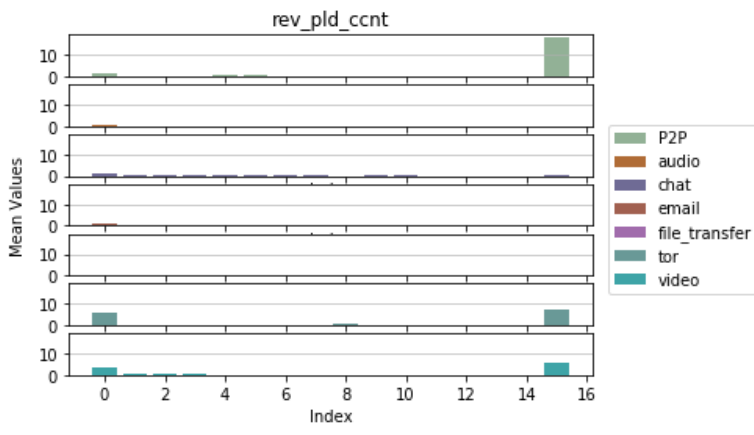}}
\caption{Metadata histogram-like Features - 2 \\ (non-vpn2016 dataset)}
\label{metadata_arrays_2_nonvpn}
\end{figure}

\begin{figure}[t]
\subfigure{\includegraphics[height=1.2in,width=1.65in]{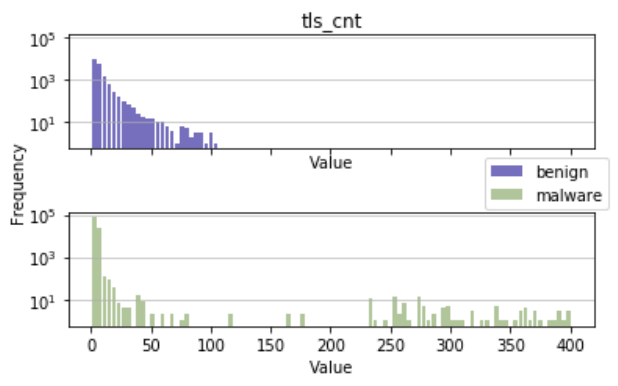}}
\subfigure{\includegraphics[height=1.2in,width=1.65in]{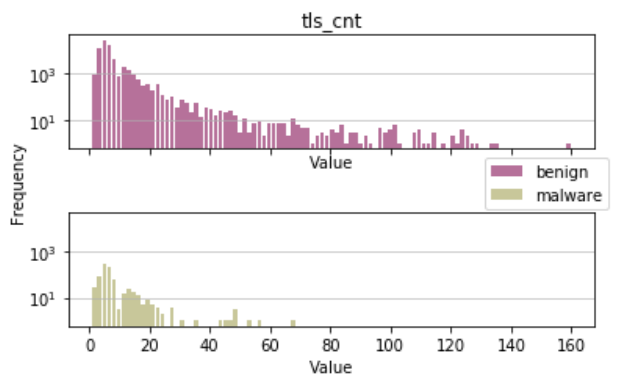}}
\subfigure{\includegraphics[height=1.2in,width=1.65in]{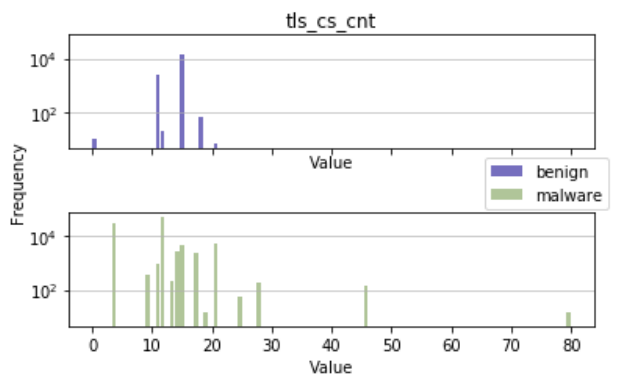}}
\subfigure{\includegraphics[height=1.2in,width=1.65in]{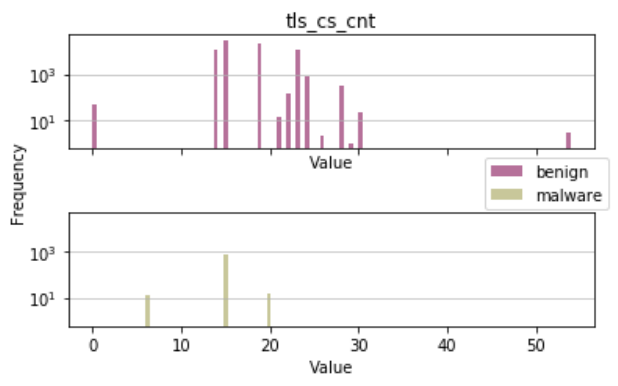}}
\subfigure{\includegraphics[height=1.2in,width=1.65in]{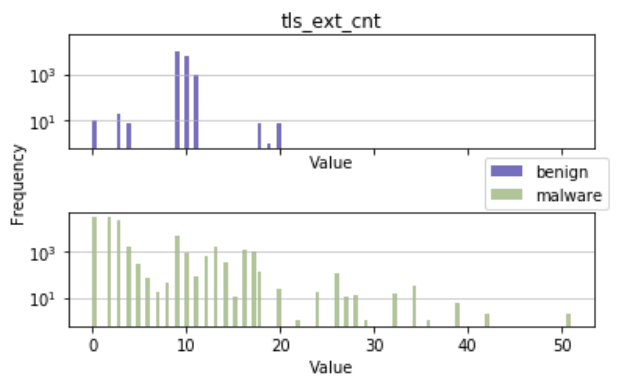}}
\subfigure{\includegraphics[height=1.2in,width=1.65in]{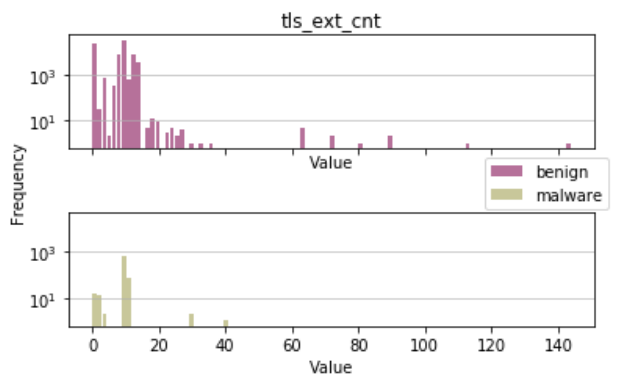}}
\subfigure{\includegraphics[height=1.2in,width=1.65in]{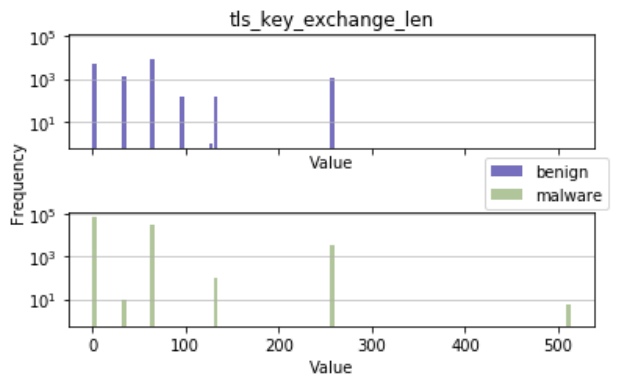}}
\subfigure{\includegraphics[height=1.2in,width=1.65in]{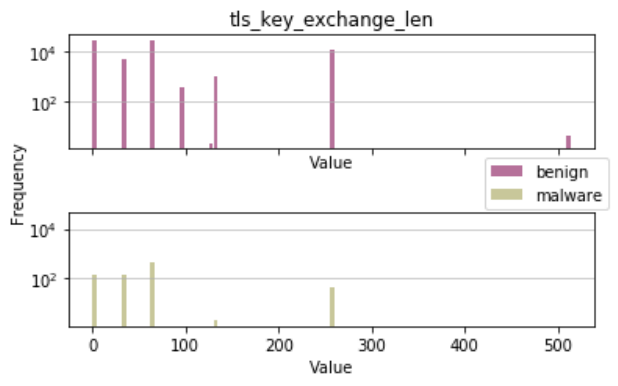}}
\caption{TLS Features offered/supported by client \\ (Left: NetML dataset , Right: CICIDS2017 dataset)}
\label{tls_client_single_malware}
\end{figure}

\begin{figure}[t]
\subfigure{\includegraphics[height=1.2in,width=2.5in]{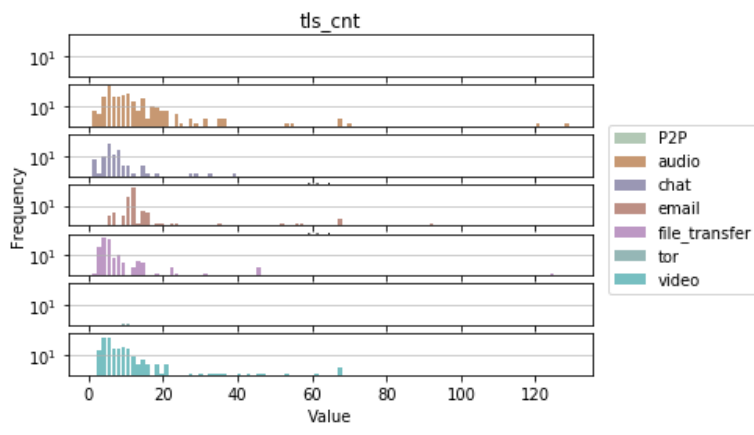}}
\subfigure{\includegraphics[height=1.2in,width=2.5in]{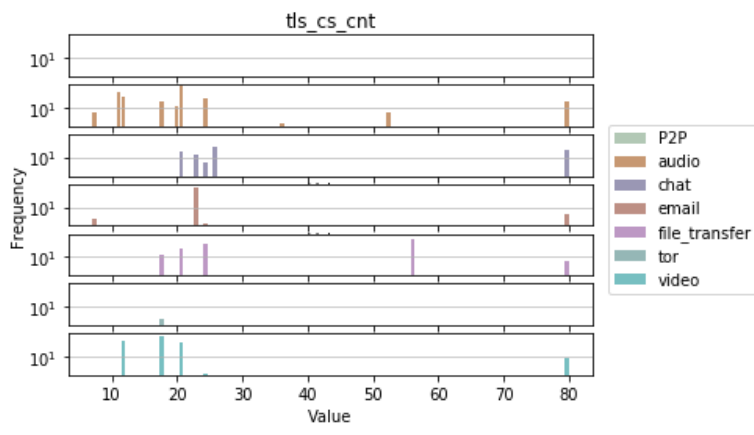}}
\subfigure{\includegraphics[height=1.2in,width=2.5in]{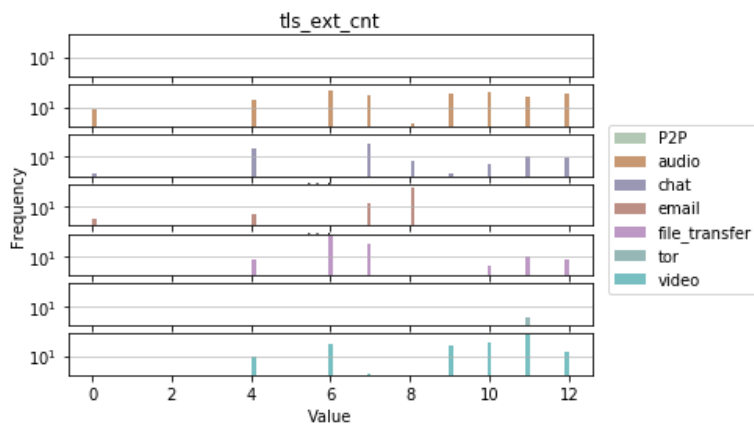}}
\subfigure{\includegraphics[height=1.2in,width=2.5in]{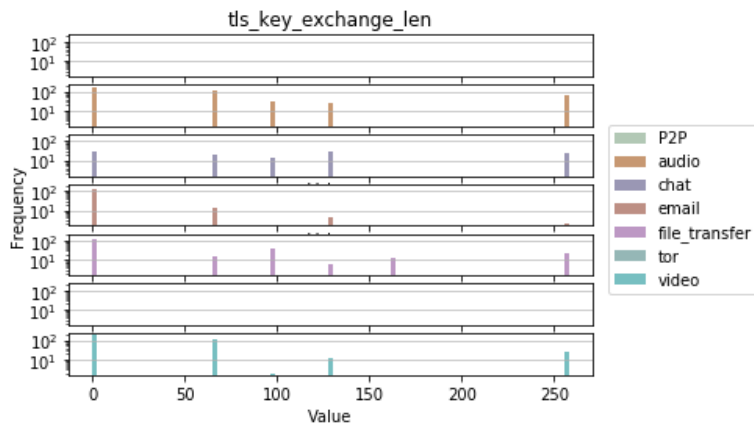}}
\caption{TLS Features offered/supported by client \\ (non-vpn2016 dataset)}
\label{tls_client_single_nonvpn}
\end{figure}

\begin{figure}[t]
\subfigure{\includegraphics[height=1.2in,width=1.65in]{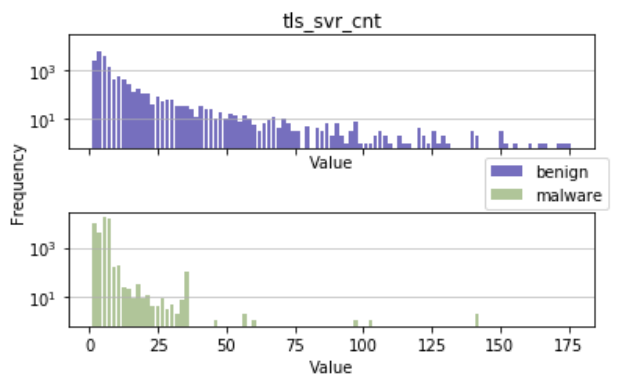}}
\subfigure{\includegraphics[height=1.2in,width=1.65in]{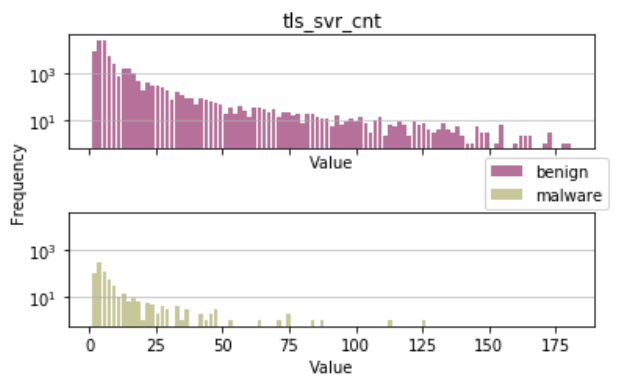}}
\subfigure{\includegraphics[height=1.2in,width=1.65in]{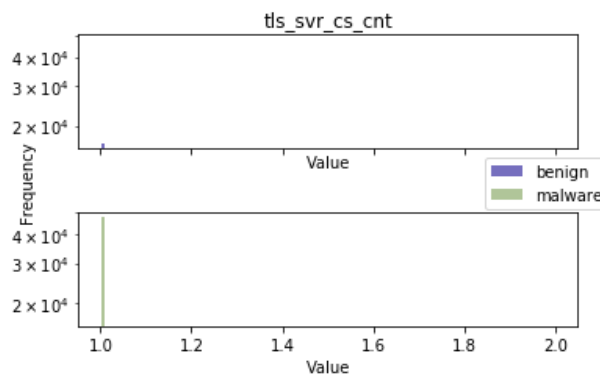}}
\subfigure{\includegraphics[height=1.2in,width=1.65in]{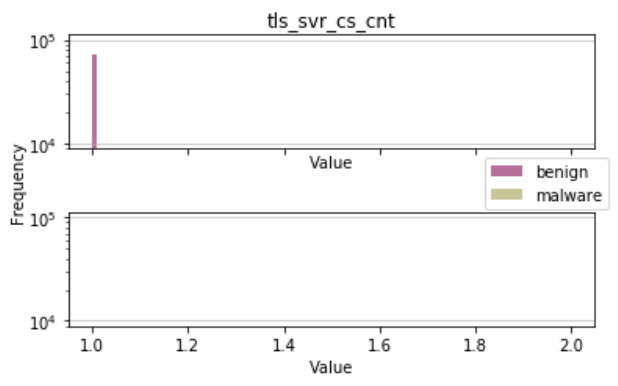}}
\subfigure{\includegraphics[height=1.2in,width=1.65in]{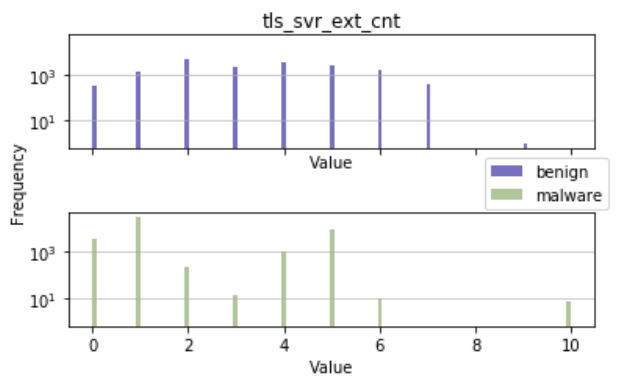}}
\subfigure{\includegraphics[height=1.2in,width=1.65in]{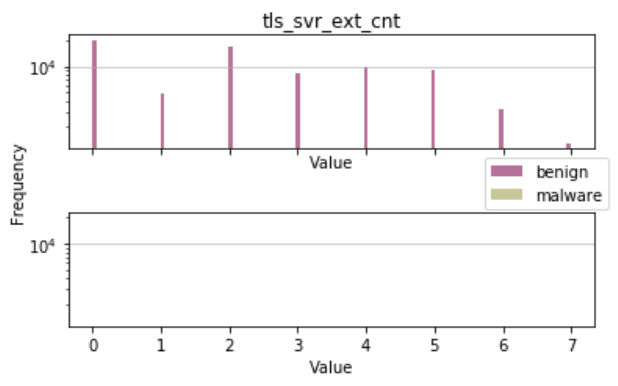}}
\subfigure{\includegraphics[height=1.2in,width=1.65in]{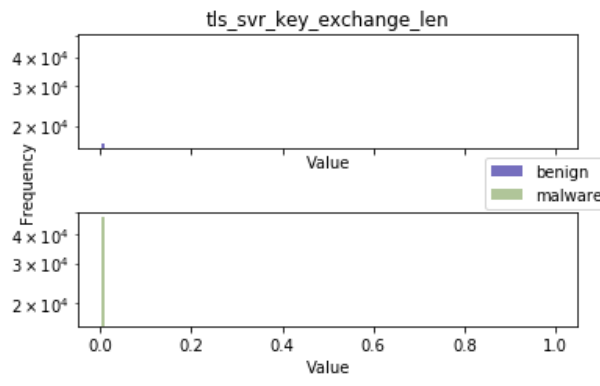}}
\subfigure{\includegraphics[height=1.2in,width=1.65in]{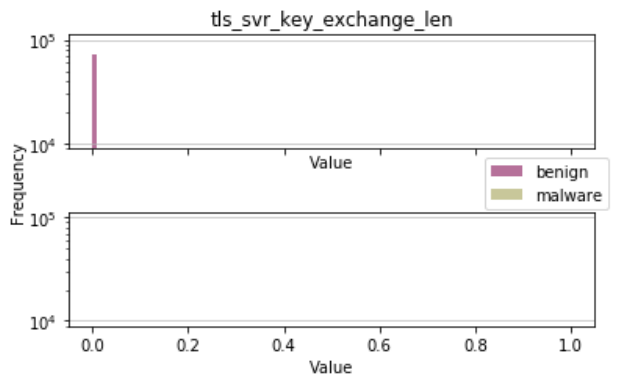}}
\caption{TLS Features offered/supported by server \\ (Left: NetML dataset , Right: CICIDS2017 dataset)}
\label{tls_server_single_malware}
\end{figure}

\begin{figure}[t]
\subfigure{\includegraphics[height=1.2in,width=2.5in]{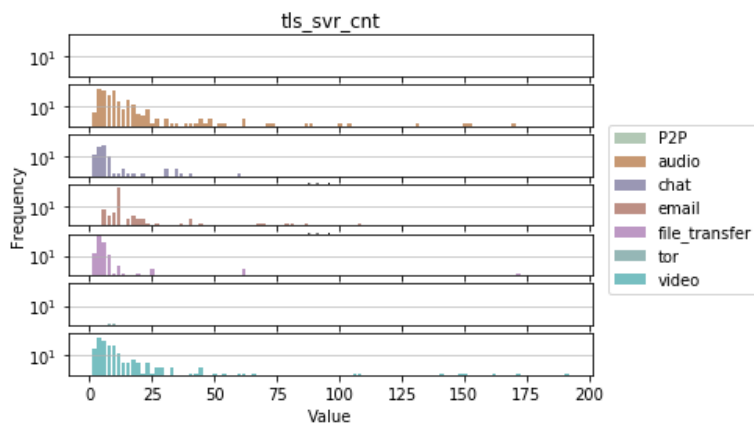}}
\subfigure{\includegraphics[height=1.2in,width=2.5in]{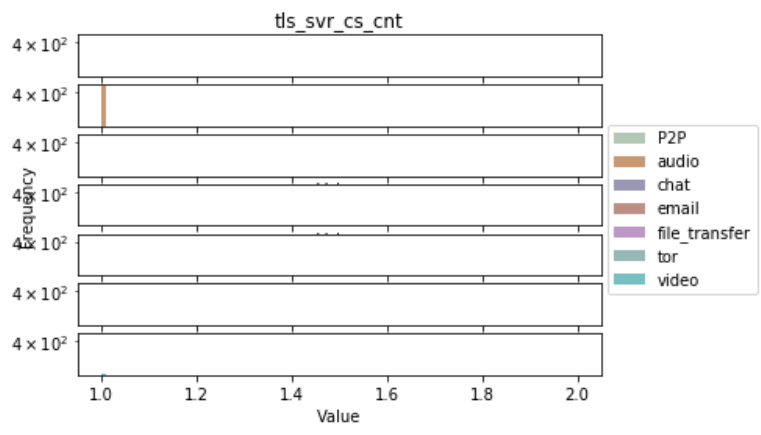}}
\subfigure{\includegraphics[height=1.2in,width=2.5in]{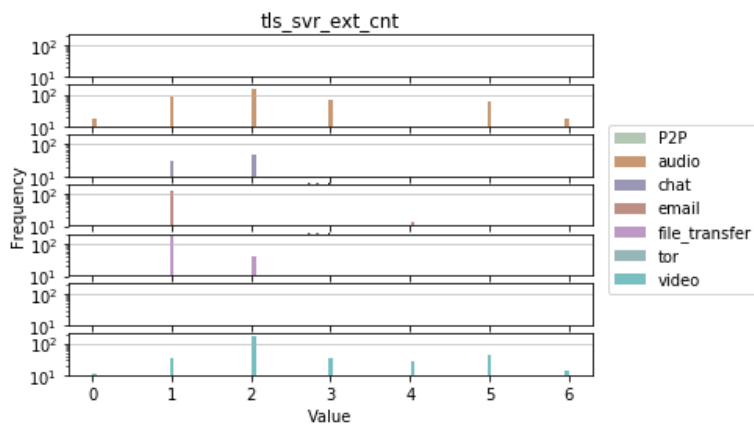}}
\subfigure{\includegraphics[height=1.2in,width=2.5in]{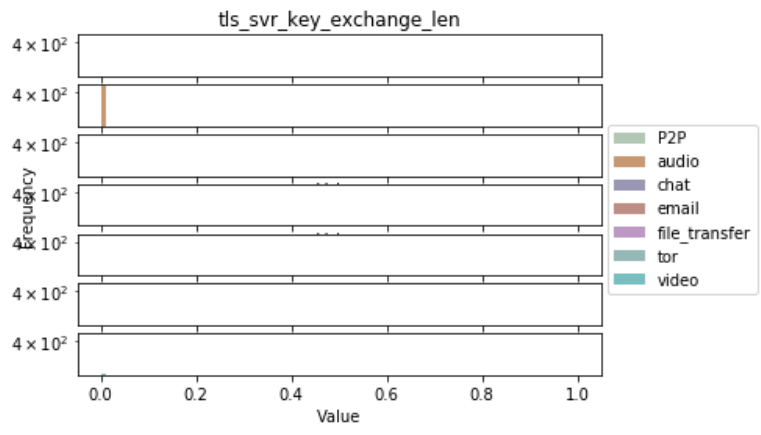}}
\caption{TLS Features offered/supported by server \\ (non-vpn2016 dataset)}
\label{tls_server_single_nonvpn}
\end{figure}

\begin{figure}[t]
\subfigure{\includegraphics[height=1.2in,width=1.65in]{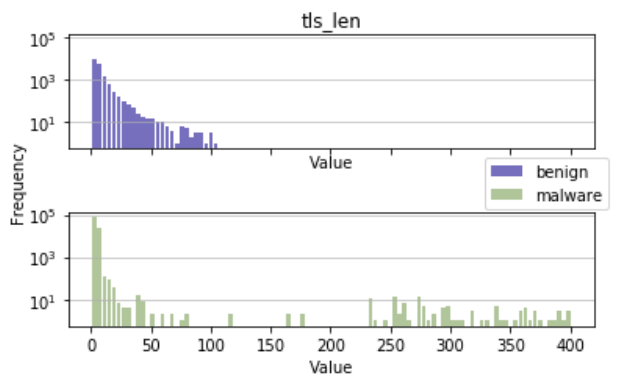}}
\subfigure{\includegraphics[height=1.2in,width=1.65in]{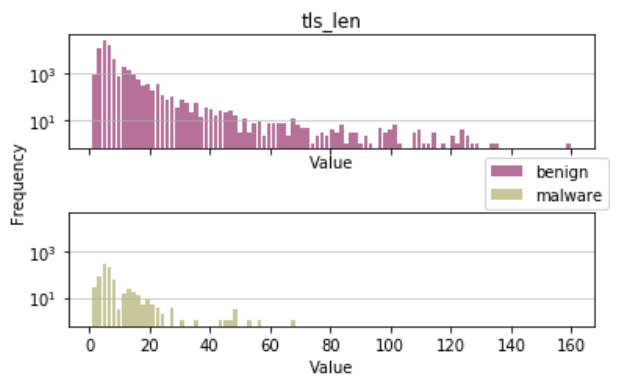}}
\subfigure{\includegraphics[height=1.2in,width=1.65in]{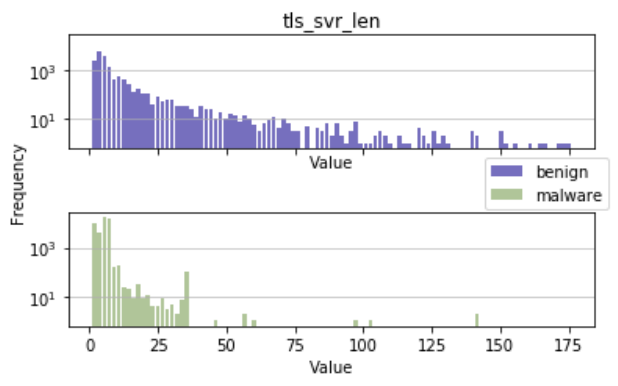}}
\subfigure{\includegraphics[height=1.2in,width=1.65in]{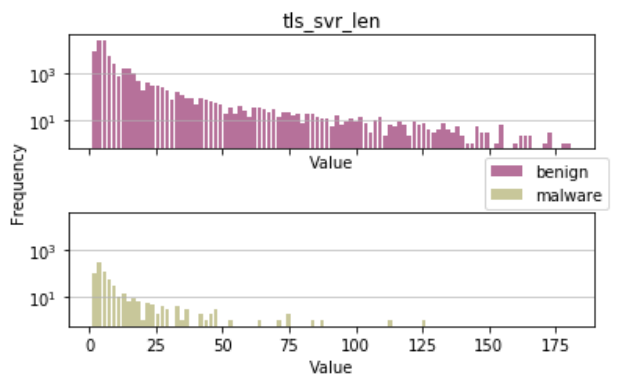}}
\subfigure{\includegraphics[height=1.2in,width=1.65in]{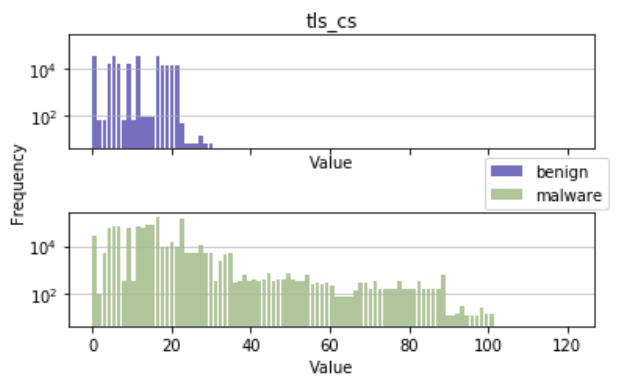}}
\subfigure{\includegraphics[height=1.2in,width=1.65in]{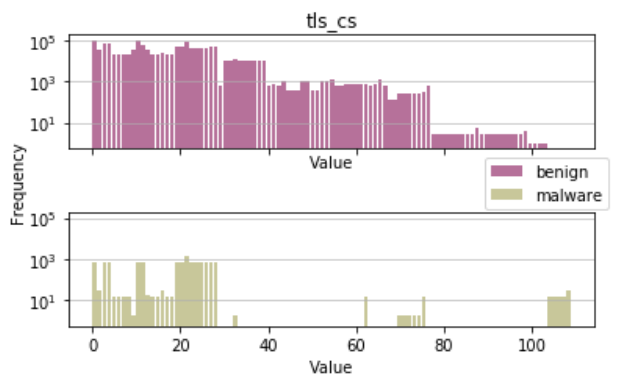}}
\subfigure{\includegraphics[height=1.2in,width=1.65in]{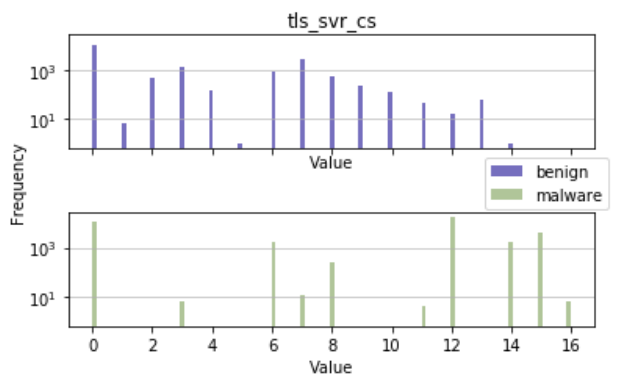}}
\subfigure{\includegraphics[height=1.2in,width=1.65in]{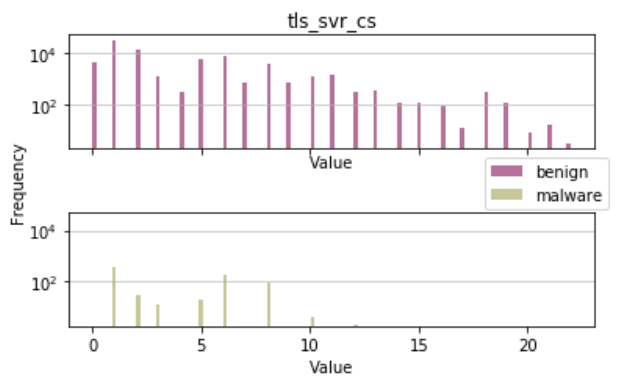}}
\subfigure{\includegraphics[height=1.2in,width=1.65in]{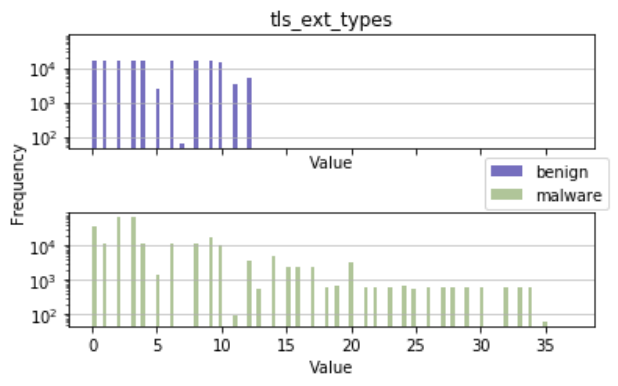}}
\subfigure{\includegraphics[height=1.2in,width=1.65in]{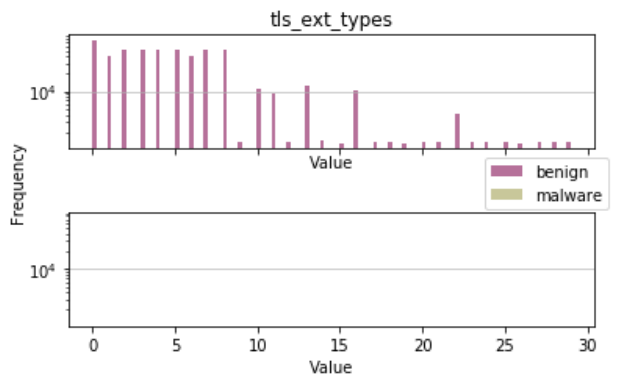}}
\subfigure{\includegraphics[height=1.2in,width=1.65in]{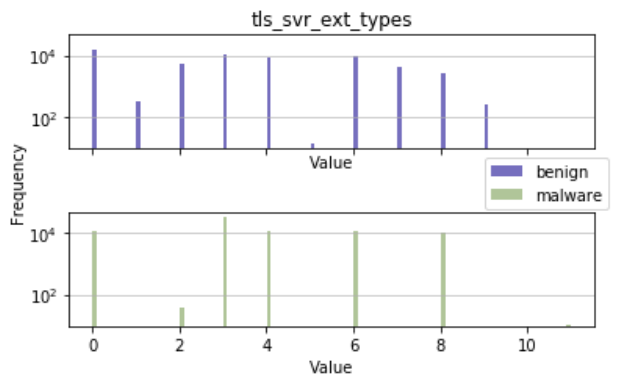}}
\subfigure{\includegraphics[height=1.2in,width=1.65in]{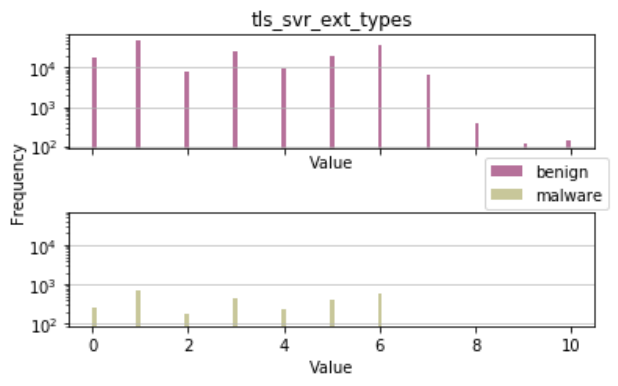}}
\caption{TLS histogram-like features \\ (Left: NetML dataset , Right: CICIDS2017 dataset)}
\label{tls_arrays_malware}
\end{figure}

\begin{figure}[t]
\subfigure{\includegraphics[height=1.2in,width=2.5in]{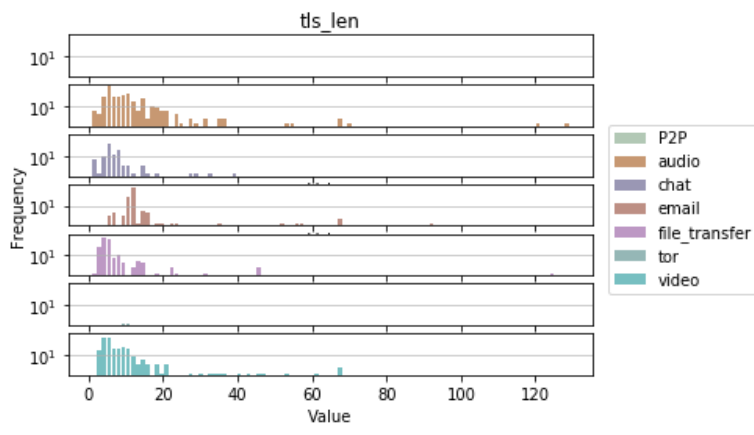}}
\subfigure{\includegraphics[height=1.2in,width=2.5in]{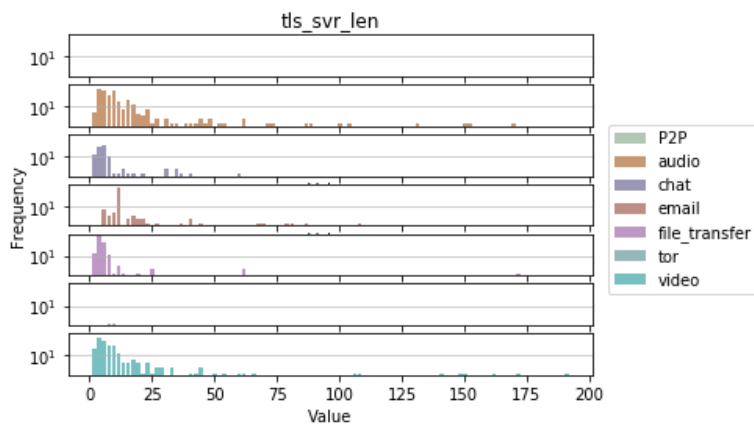}}
\subfigure{\includegraphics[height=1.2in,width=2.5in]{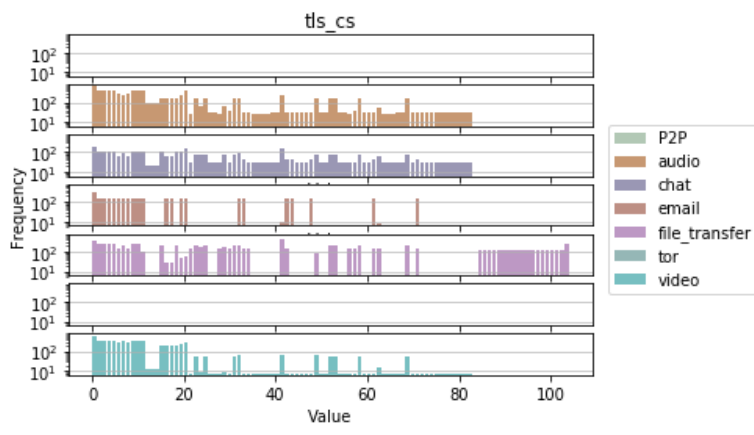}}
\subfigure{\includegraphics[height=1.2in,width=2.5in]{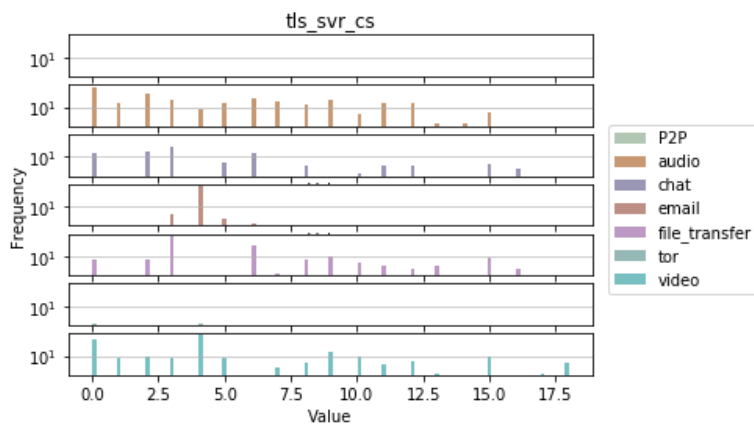}}
\subfigure{\includegraphics[height=1.2in,width=2.5in]{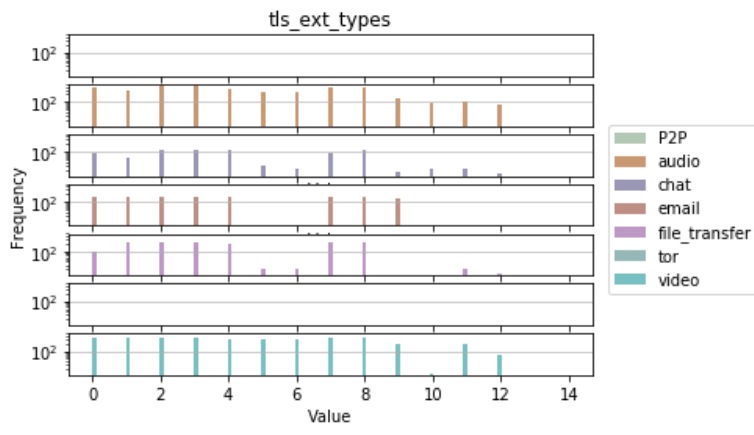}}
\subfigure{\includegraphics[height=1.2in,width=2.5in]{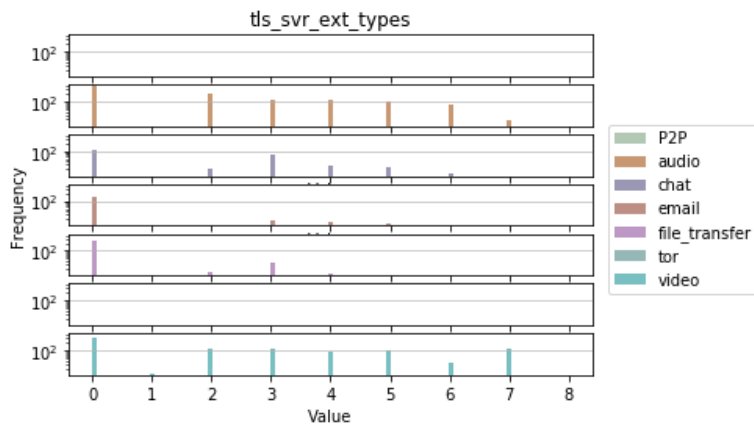}}
\caption{TLS histogram-like features \\ (non-vpn2016 dataset)}
\label{tls_arrays_nonvpn}
\end{figure}

\begin{figure}[t]
\subfigure{\includegraphics[height=1.2in,width=1.65in]{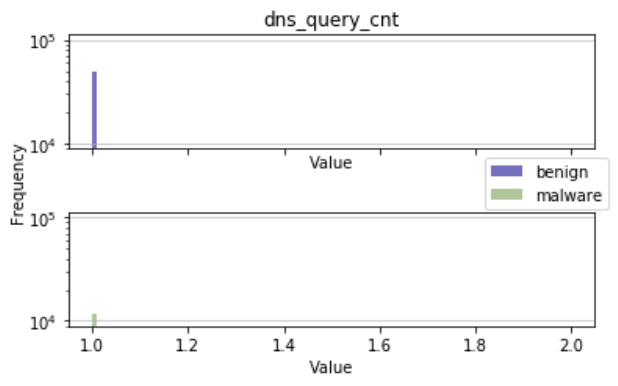}}
\subfigure{\includegraphics[height=1.2in,width=1.65in]{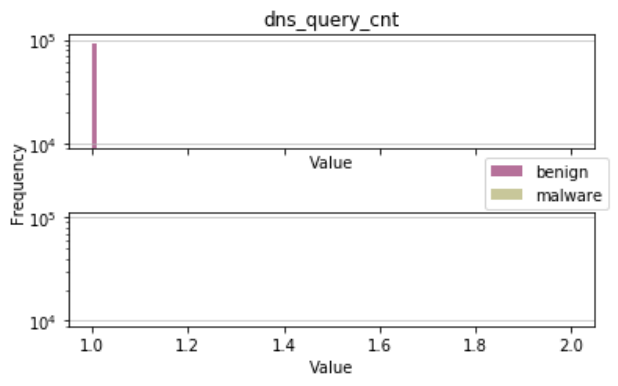}}
\subfigure{\includegraphics[height=1.2in,width=1.65in]{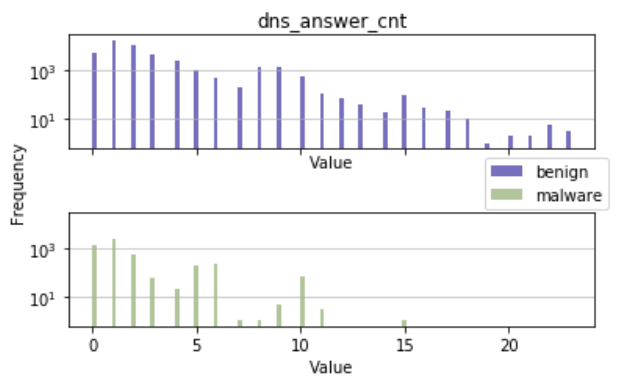}}
\subfigure{\includegraphics[height=1.2in,width=1.65in]{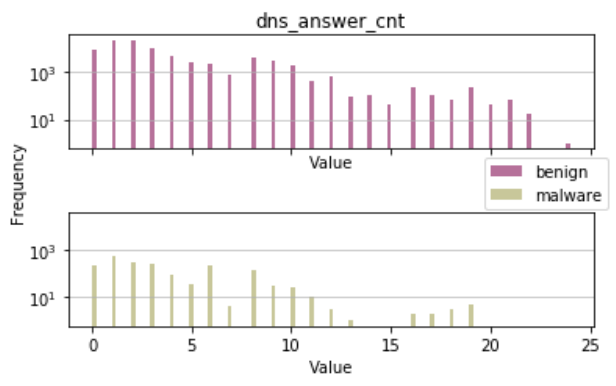}}
\caption{DNS single-valued features \\ (Left: NetML dataset , Right: CICIDS2017 dataset)}
\label{dns_single_malware}
\vspace{4in}
\end{figure}

\begin{figure}[t]
\subfigure{\includegraphics[height=1.2in,width=2.5in]{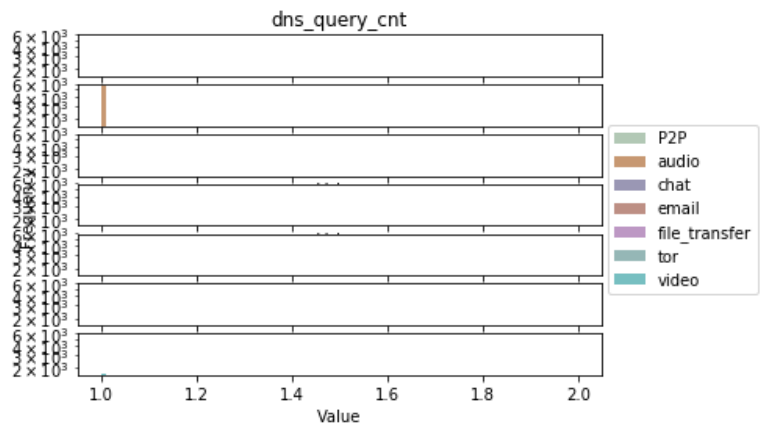}}
\subfigure{\includegraphics[height=1.2in,width=2.5in]{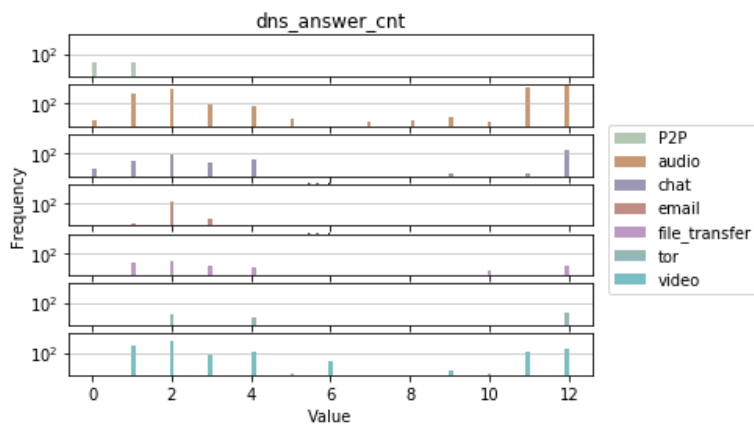}}
\caption{DNS single-valued features \\ (non-vpn2016 dataset)}
\label{dns_single_nonvpn}
\end{figure}

\begin{figure}[t]
\subfigure{\includegraphics[height=1.2in,width=1.65in]{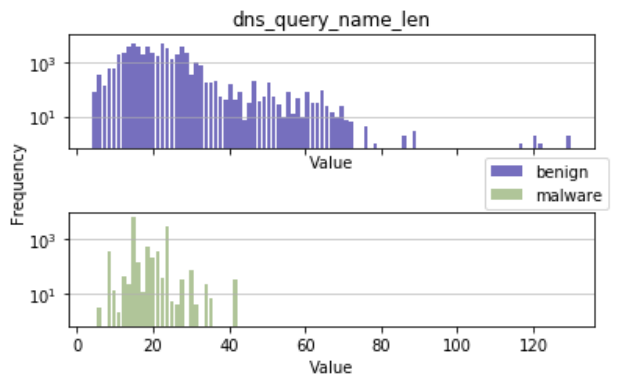}}
\subfigure{\includegraphics[height=1.2in,width=1.65in]{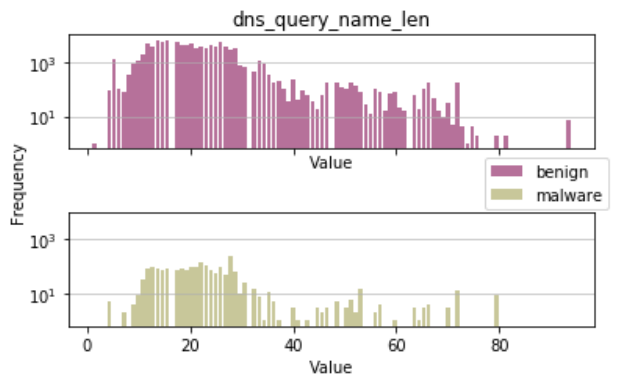}}
\subfigure{\includegraphics[height=1.2in,width=1.65in]{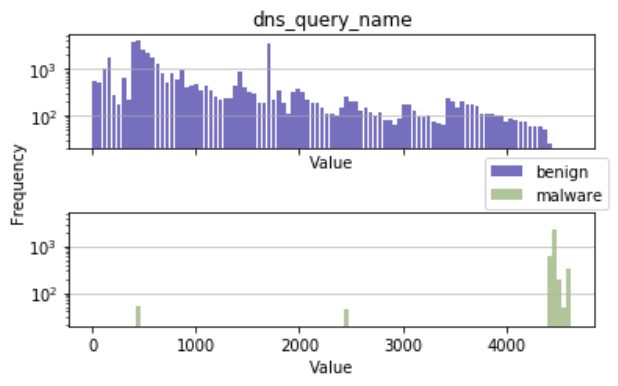}}
\subfigure{\includegraphics[height=1.2in,width=1.65in]{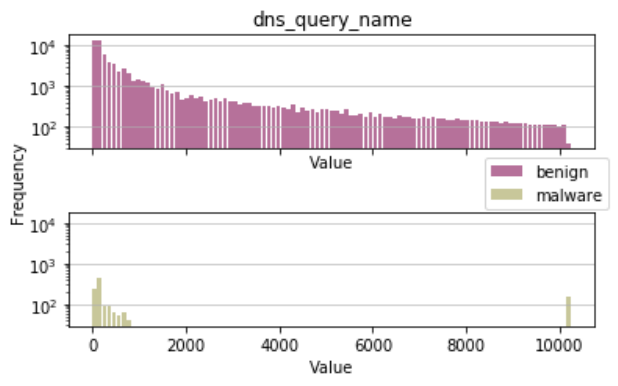}}
\subfigure{\includegraphics[height=1.2in,width=1.65in]{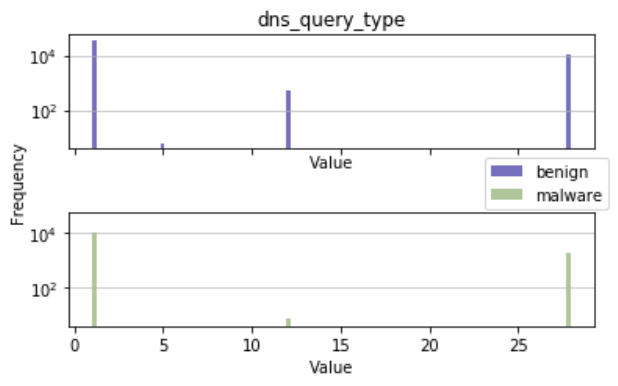}}
\subfigure{\includegraphics[height=1.2in,width=1.65in]{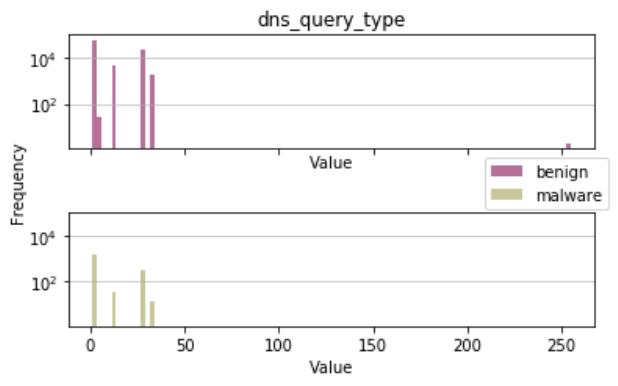}}
\subfigure{\includegraphics[height=1.2in,width=1.65in]{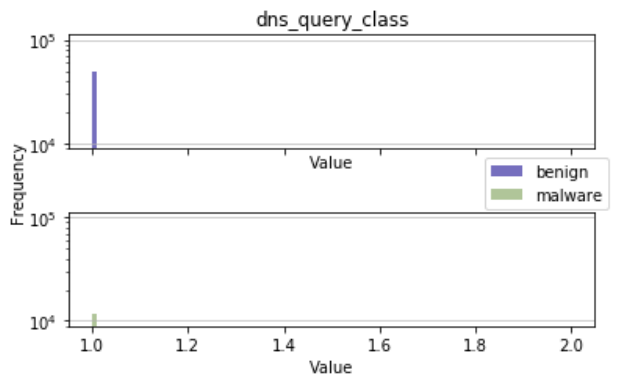}}
\subfigure{\includegraphics[height=1.2in,width=1.65in]{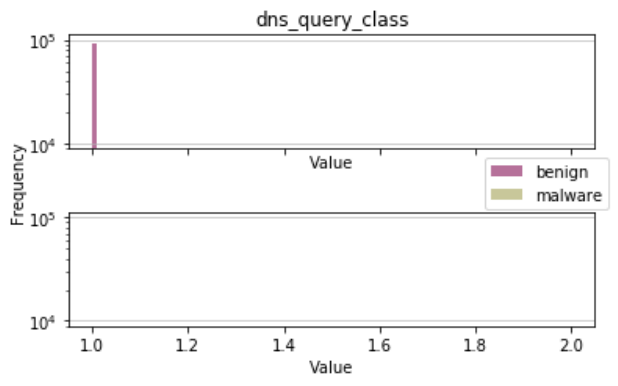}}
\subfigure{\includegraphics[height=1.2in,width=1.65in]{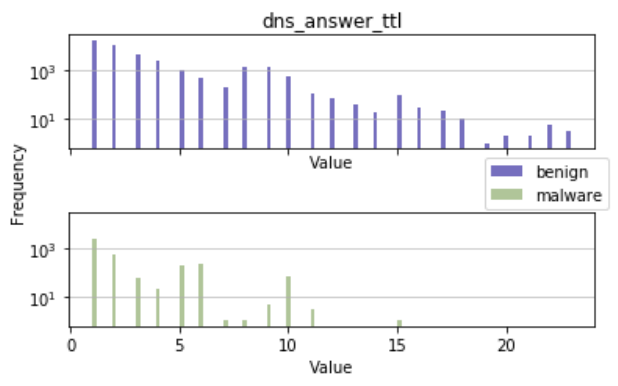}}
\subfigure{\includegraphics[height=1.2in,width=1.65in]{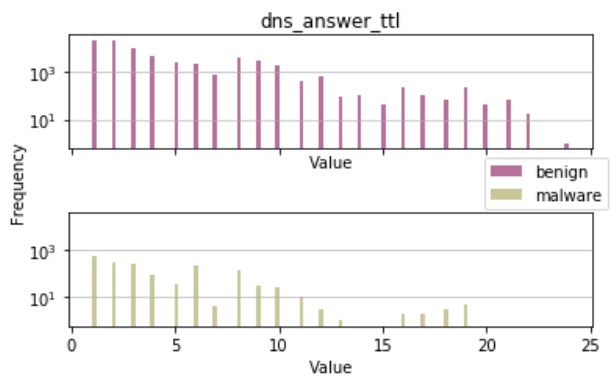}}
\subfigure{\includegraphics[height=1.2in,width=1.65in]{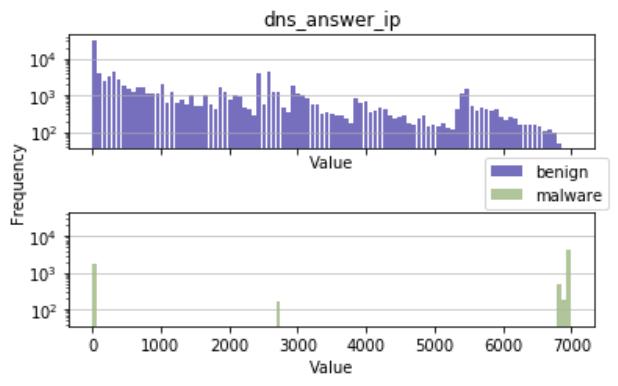}}
\subfigure{\includegraphics[height=1.2in,width=1.65in]{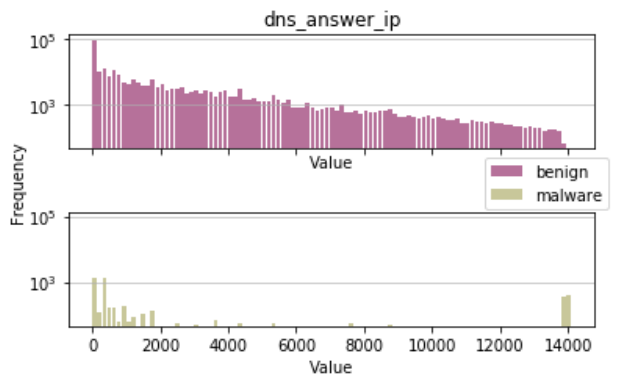}}
\caption{DNS array-like features \\ (Left: NetML dataset , Right: CICIDS2017 dataset)}
\label{dns_arrays_malware}
\end{figure}

\begin{figure}[t]
\subfigure{\includegraphics[height=1.2in,width=2.5in]{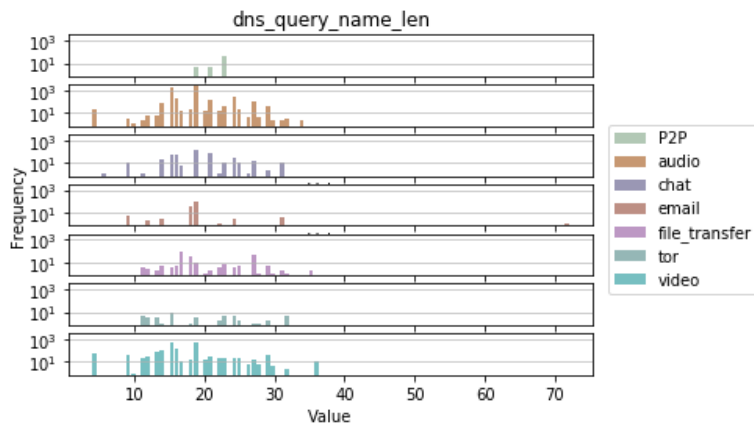}}
\subfigure{\includegraphics[height=1.2in,width=2.5in]{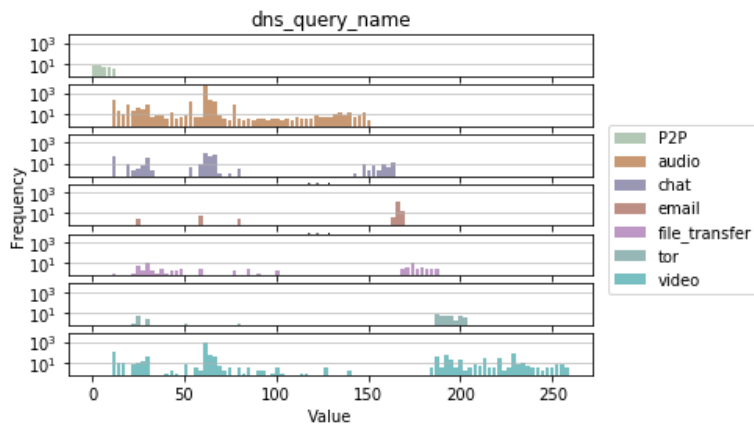}}
\subfigure{\includegraphics[height=1.2in,width=2.5in]{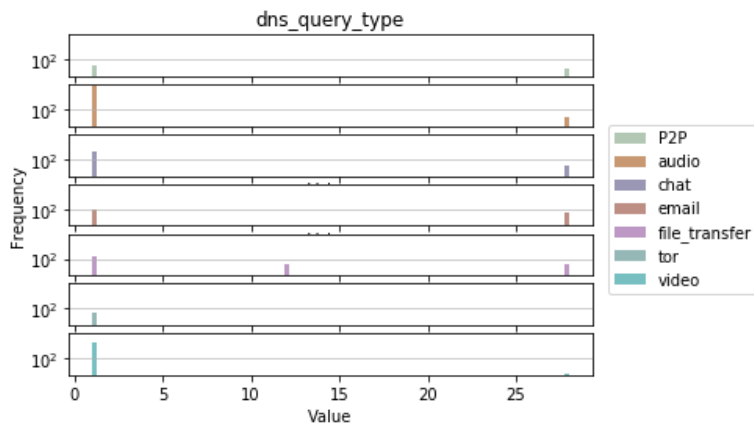}}
\subfigure{\includegraphics[height=1.2in,width=2.5in]{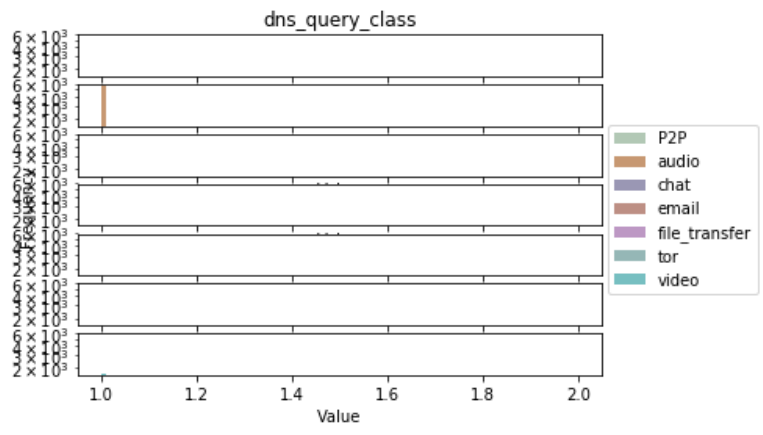}}
\subfigure{\includegraphics[height=1.2in,width=2.5in]{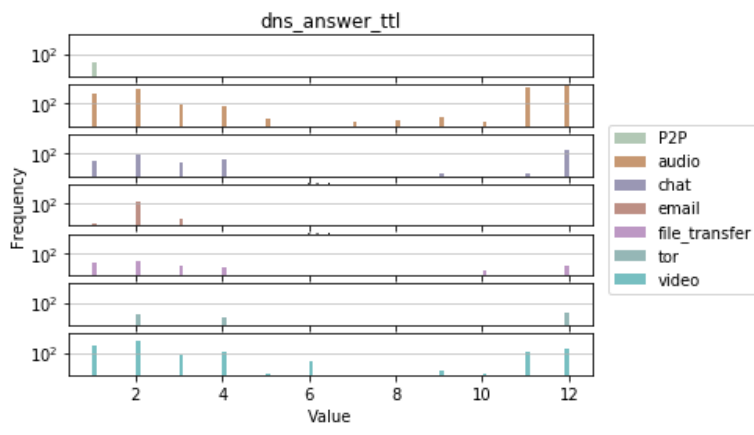}}
\subfigure{\includegraphics[height=1.2in,width=2.5in]{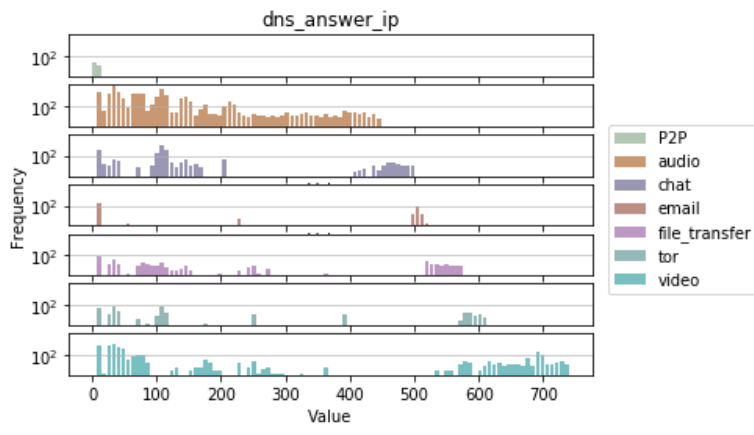}}
\caption{DNS array-like features \\ (non-vpn2016 dataset)}
\label{dns_arrays_nonvpn}
\end{figure}

\begin{figure}[t]
\subfigure{\includegraphics[height=1.2in,width=1.65in]{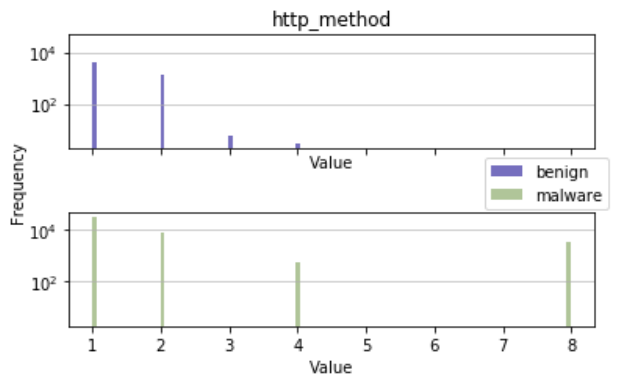}}
\subfigure{\includegraphics[height=1.2in,width=1.65in]{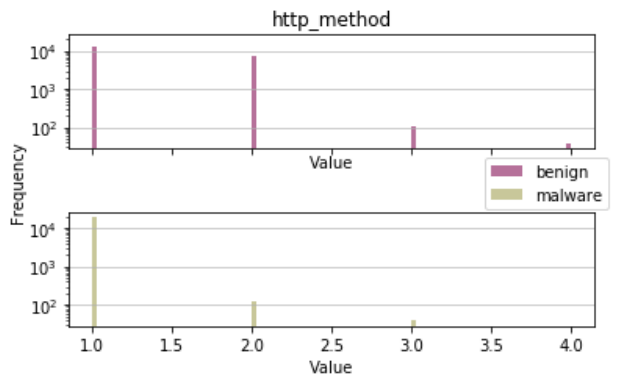}}
\subfigure{\includegraphics[height=1.2in,width=1.65in]{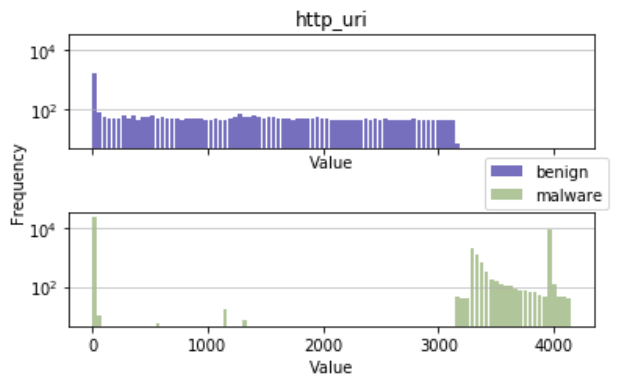}}
\subfigure{\includegraphics[height=1.2in,width=1.65in]{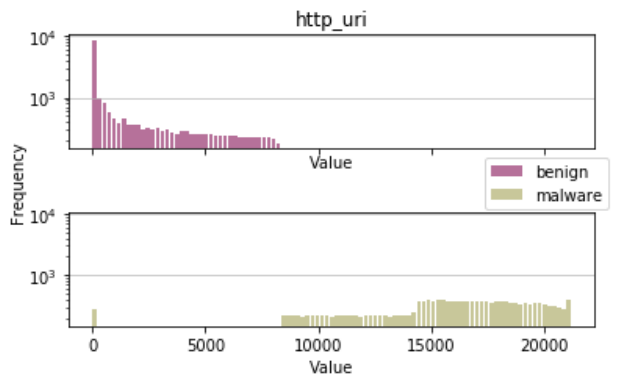}}
\subfigure{\includegraphics[height=1.2in,width=1.65in]{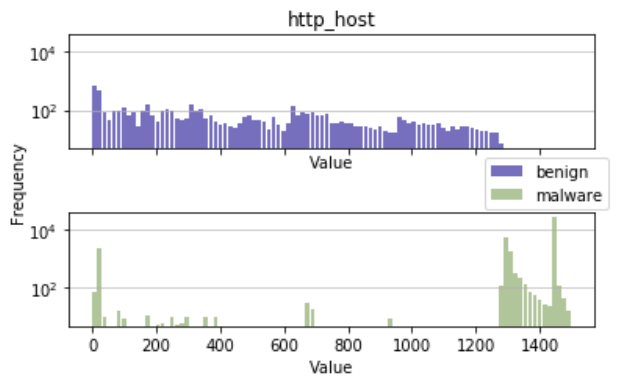}}
\subfigure{\includegraphics[height=1.2in,width=1.65in]{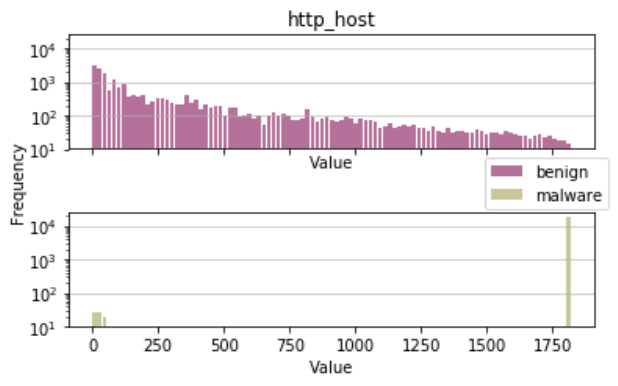}}
\subfigure{\includegraphics[height=1.2in,width=1.65in]{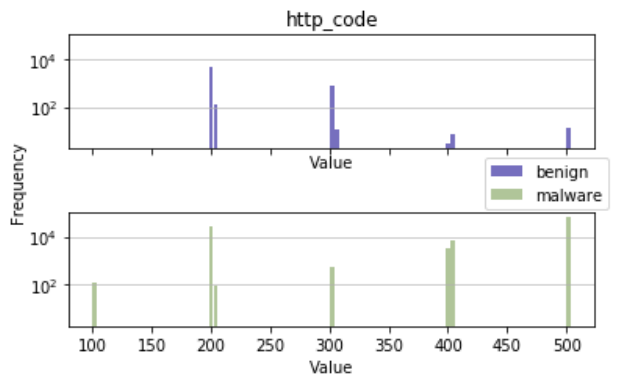}}
\subfigure{\includegraphics[height=1.2in,width=1.65in]{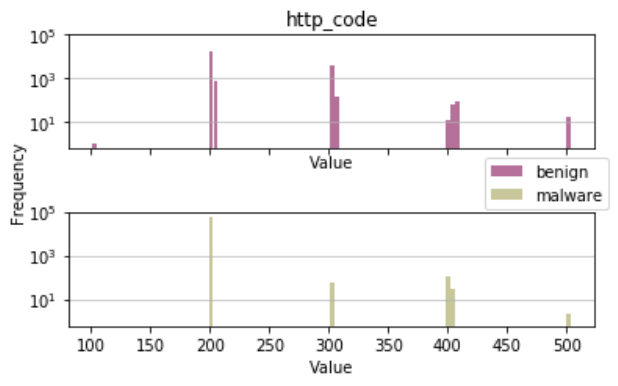}}
\subfigure{\includegraphics[height=1.2in,width=1.65in]{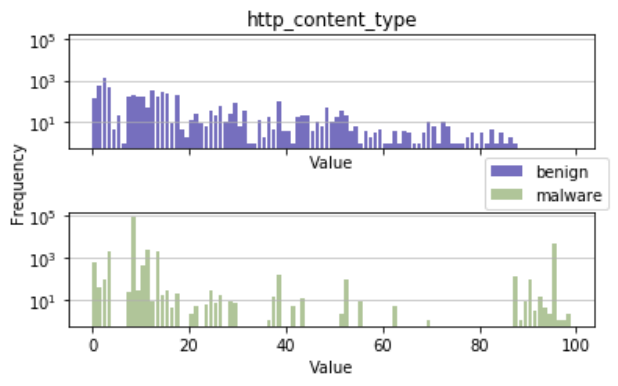}}
\subfigure{\includegraphics[height=1.2in,width=1.65in]{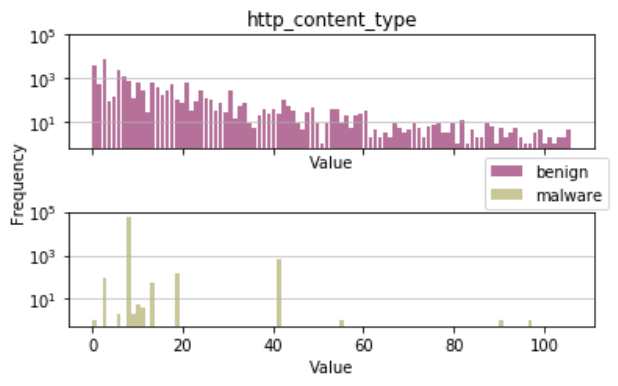}}
\subfigure{\includegraphics[height=1.2in,width=1.65in]{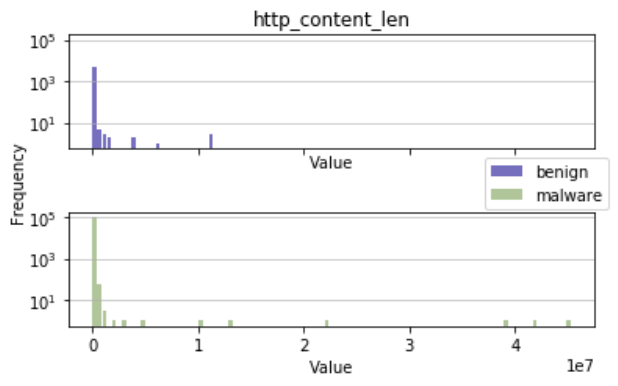}}
\subfigure{\includegraphics[height=1.2in,width=1.65in]{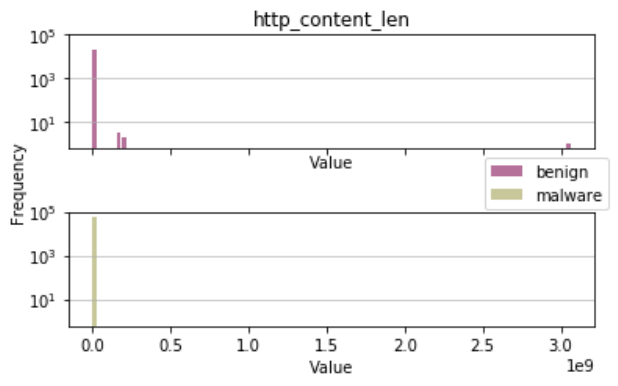}}
\caption{HTTP features \\ (Left: NetML dataset , Right: CICIDS2017 dataset)}
\label{http_all_malware}
\end{figure}

\begin{figure}[t]
\subfigure{\includegraphics[height=1.2in,width=2.5in]{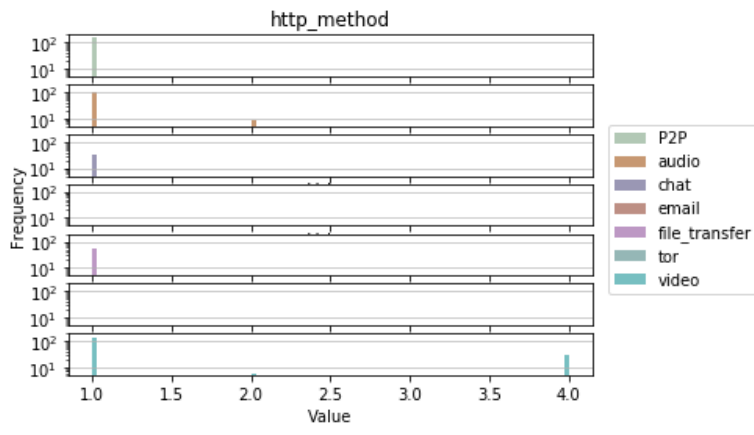}}
\subfigure{\includegraphics[height=1.2in,width=2.5in]{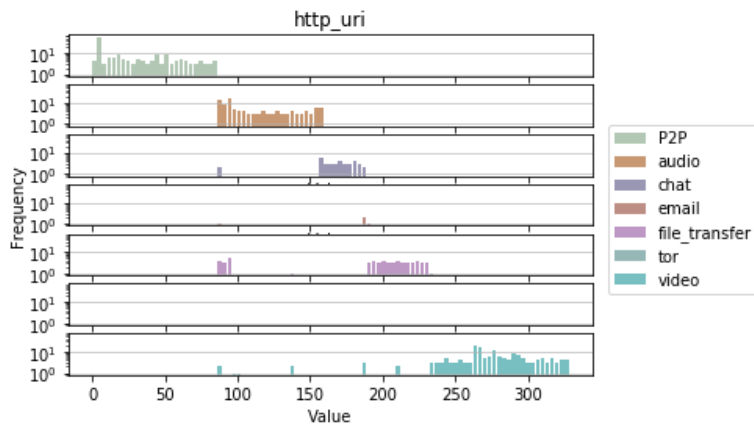}}
\subfigure{\includegraphics[height=1.2in,width=2.5in]{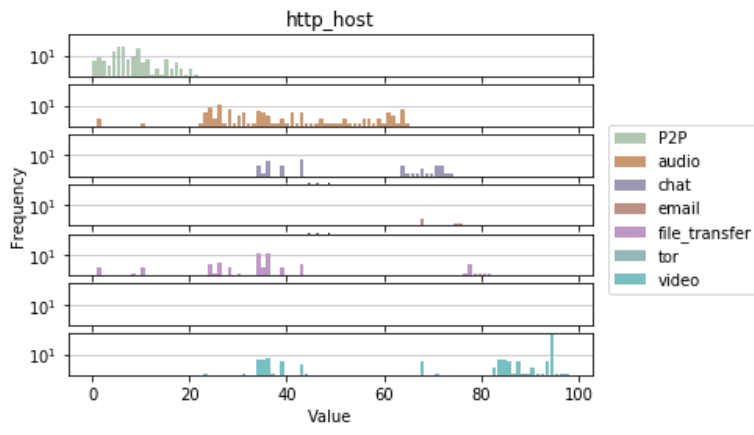}}
\subfigure{\includegraphics[height=1.2in,width=2.5in]{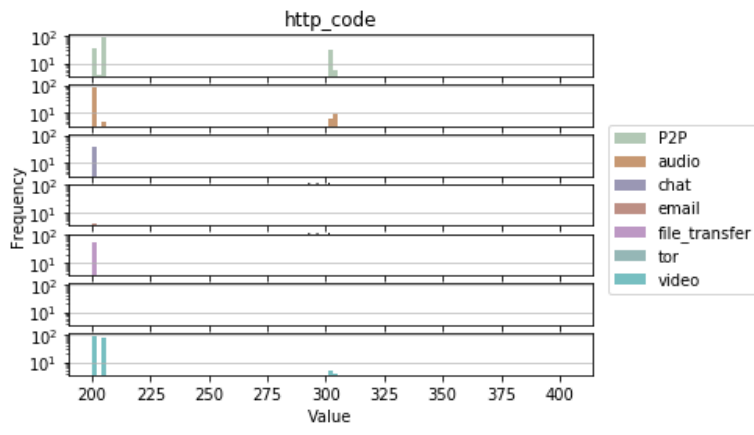}}
\subfigure{\includegraphics[height=1.2in,width=2.5in]{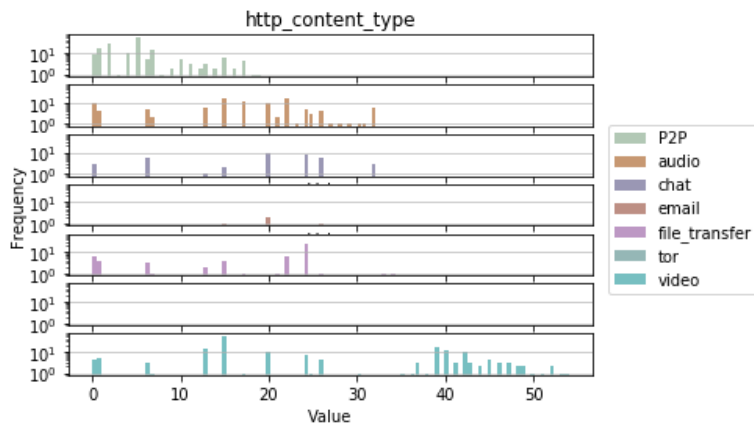}}
\subfigure{\includegraphics[height=1.2in,width=2.5in]{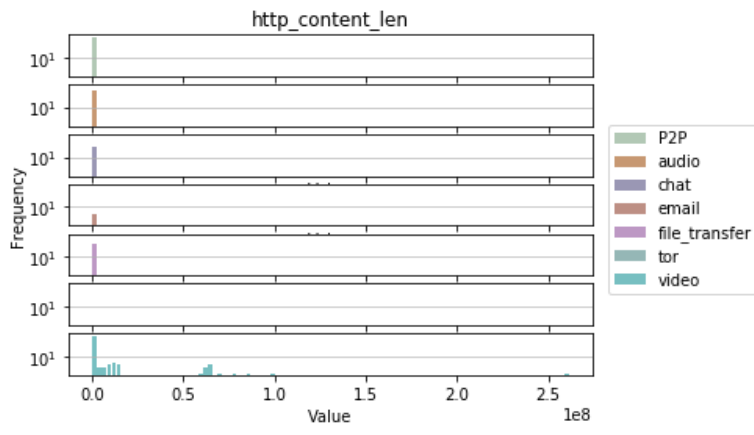}}
\caption{HTTP features \\ (non-vpn2016 dataset)}
\label{http_all_nonvpn}
\end{figure}

\begin{figure}[t]
\subfigure{\includegraphics[height=1.2in,width=1.65in]{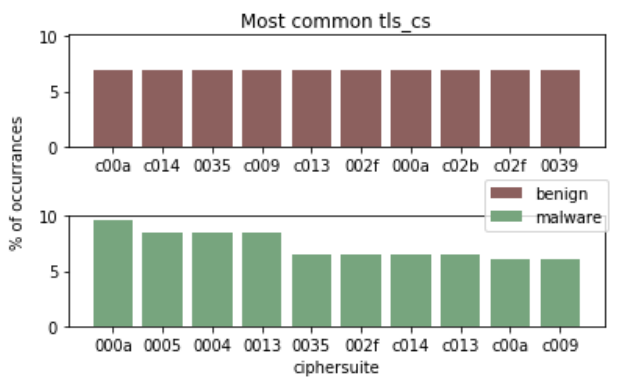}}
\subfigure{\includegraphics[height=1.2in,width=1.65in]{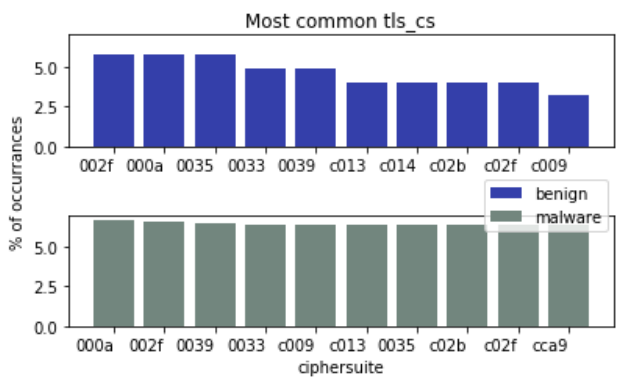}}
\subfigure{\includegraphics[height=1.2in,width=1.65in]{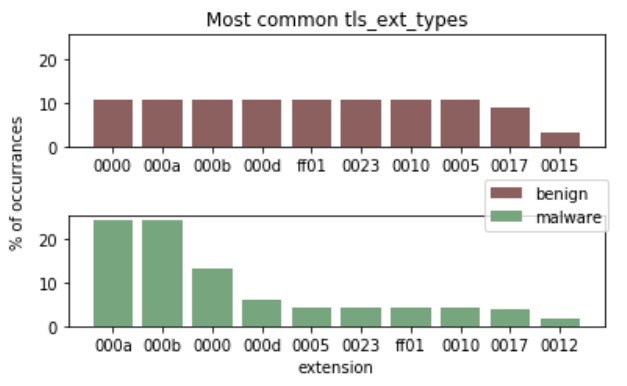}}
\subfigure{\includegraphics[height=1.2in,width=1.65in]{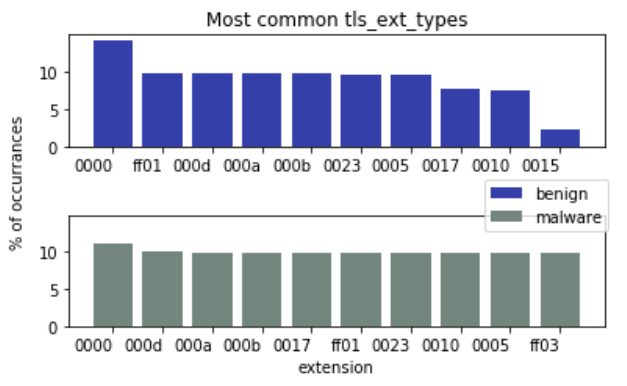}}
\subfigure{\includegraphics[height=1.2in,width=1.65in]{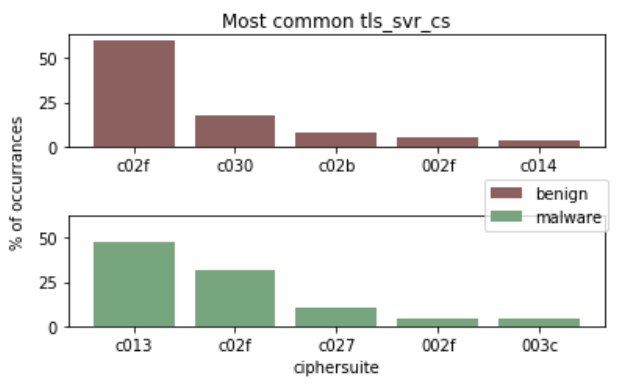}}
\subfigure{\includegraphics[height=1.2in,width=1.65in]{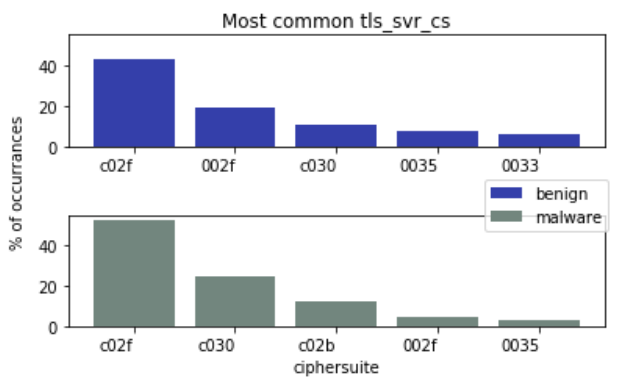}}
\subfigure{\includegraphics[height=1.2in,width=1.65in]{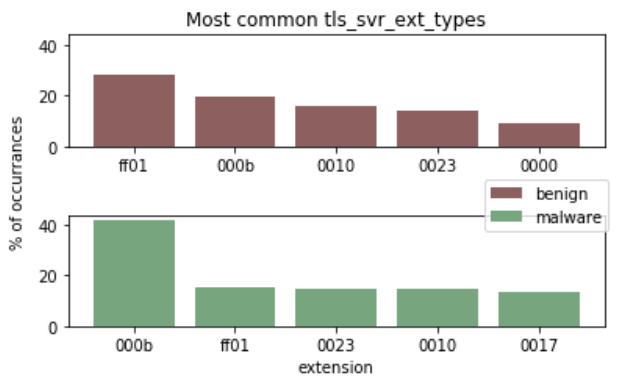}}
\subfigure{\includegraphics[height=1.2in,width=1.65in]{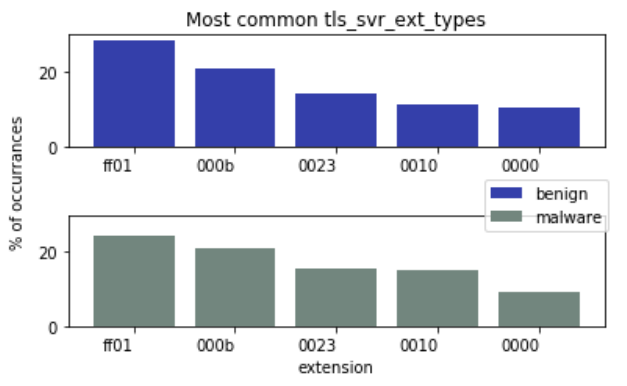}}
\caption{Most common TLS ciphersuites and extention types \\ (Left: NetML dataset , Right: CICIDS2017 dataset)}
\label{common_tls_malware}
\end{figure}

\begin{figure}[t]
\subfigure{\includegraphics[height=1.2in,width=1.65in]{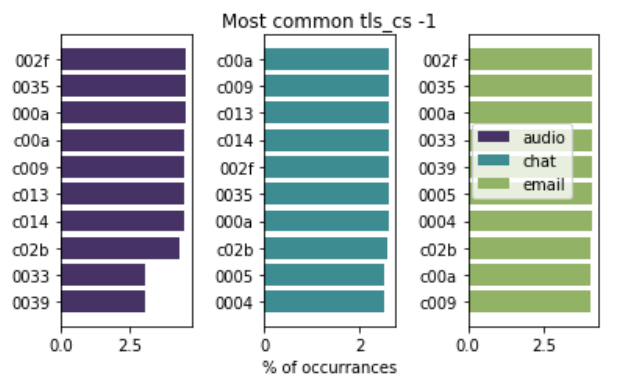}}
\subfigure{\includegraphics[height=1.2in,width=1.65in]{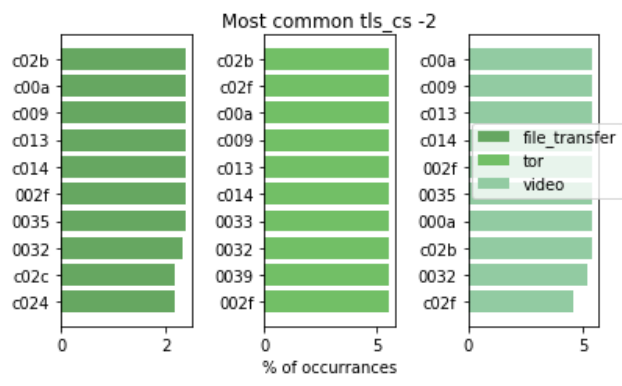}}
\subfigure{\includegraphics[height=1.2in,width=1.65in]{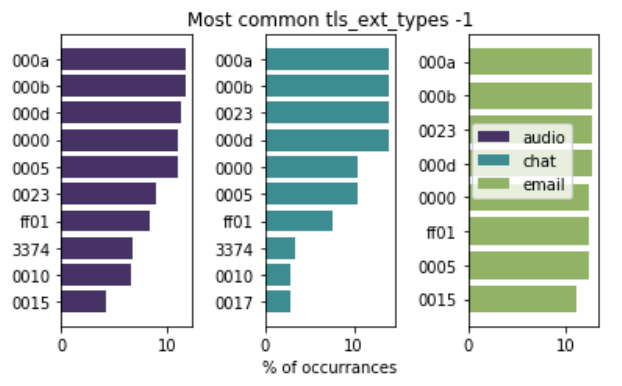}}
\subfigure{\includegraphics[height=1.2in,width=1.65in]{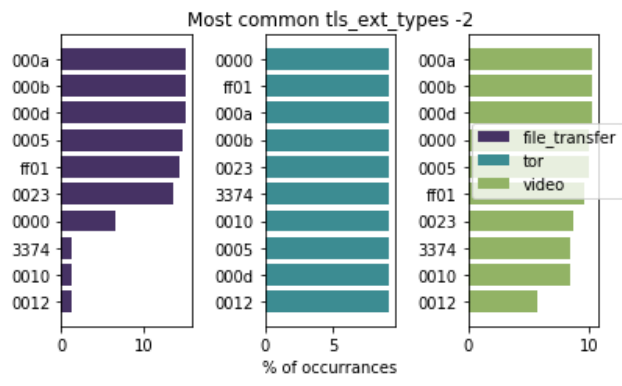}}
\subfigure{\includegraphics[height=1.2in,width=1.65in]{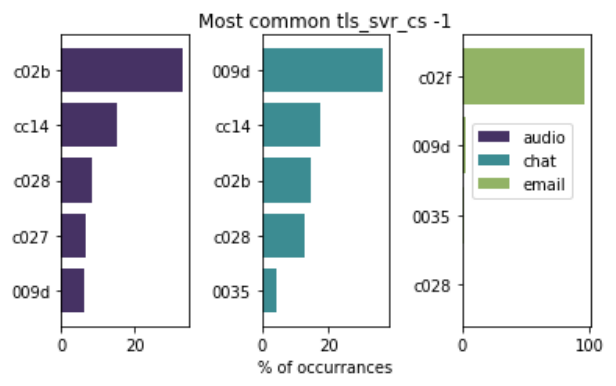}}
\subfigure{\includegraphics[height=1.2in,width=1.65in]{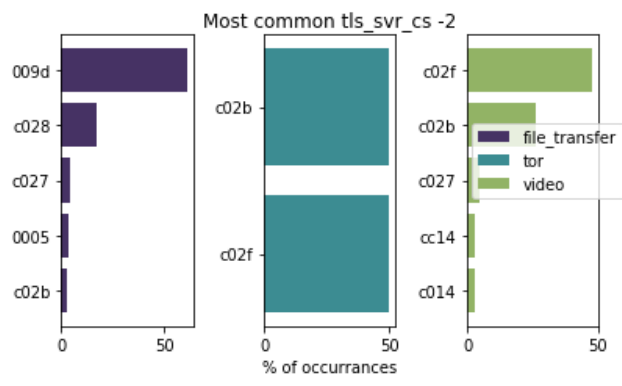}}
\subfigure{\includegraphics[height=1.2in,width=1.65in]{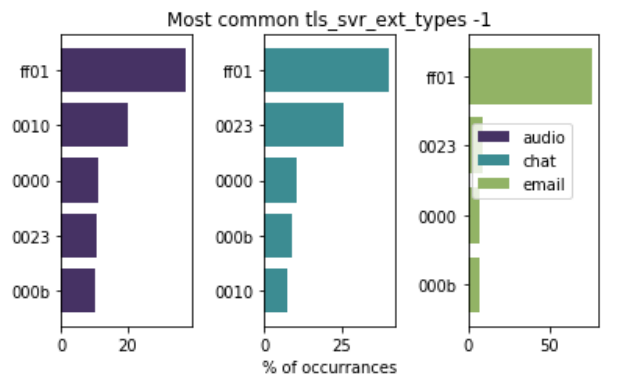}}
\subfigure{\includegraphics[height=1.2in,width=1.65in]{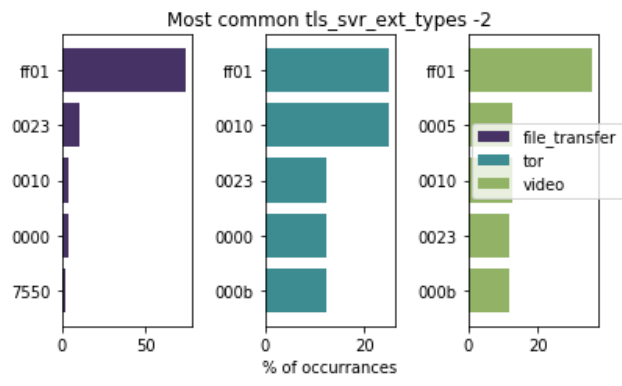}}
\caption{Most common TLS ciphersuites and extention types \\ (non-vpn2016 dataset)}
\label{common_tls_nonvpn}
\end{figure}

\begin{figure}[t]
\subfigure{\includegraphics[height=1.2in,width=1.65in]{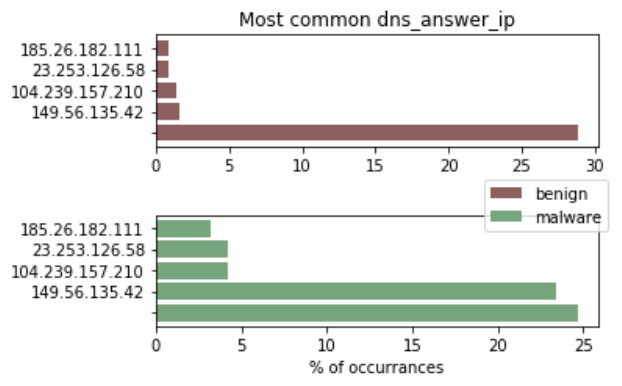}}
\subfigure{\includegraphics[height=1.2in,width=1.65in]{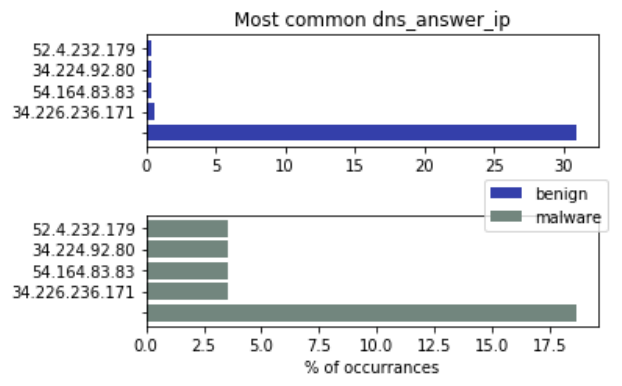}}
\subfigure{\includegraphics[height=1.2in,width=1.65in]{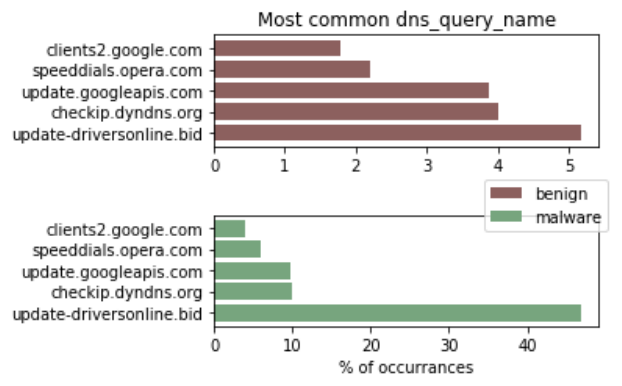}}
\subfigure{\includegraphics[height=1.2in,width=1.65in]{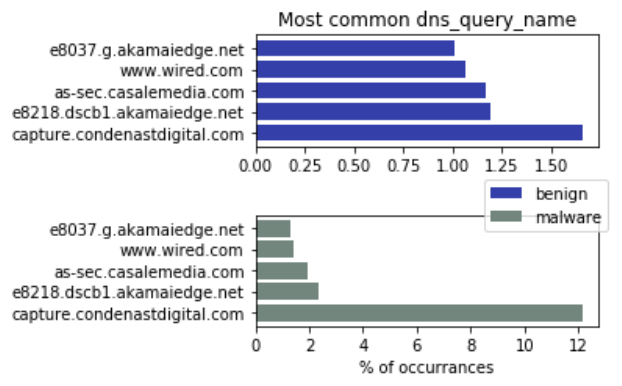}}
\caption{Most common DNS answer IP and query names \\ (Left: NetML dataset , Right: CICIDS2017 dataset)}
\label{common_dns_malware}
\vspace{5in}
\end{figure}

\begin{figure}[t]
\subfigure{\includegraphics[height=3.5in,width=2.5in]{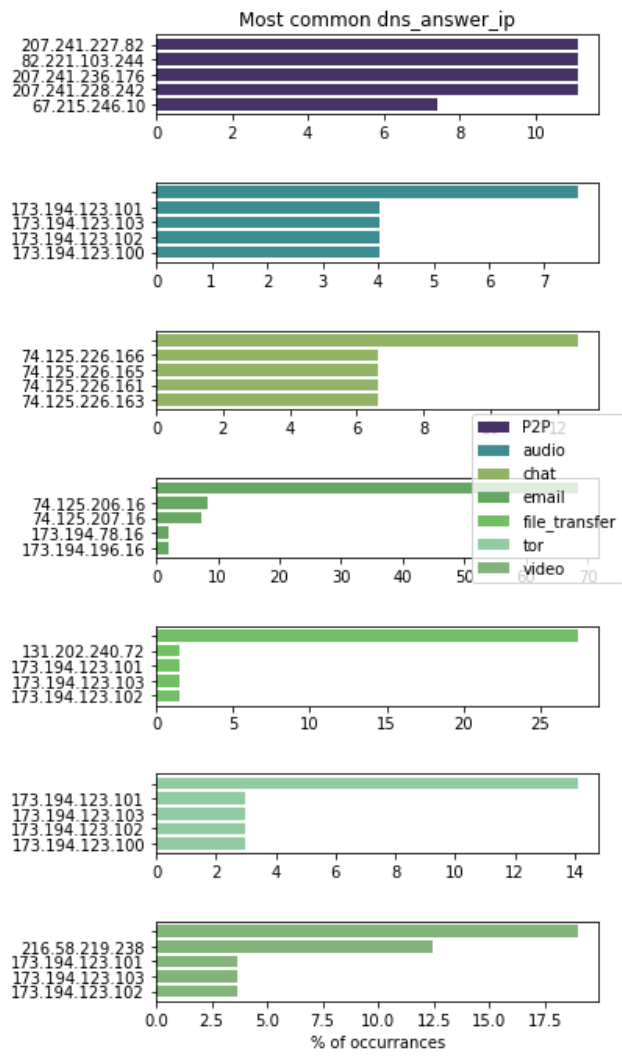}}
\subfigure{\includegraphics[height=3.5in,width=2.5in]{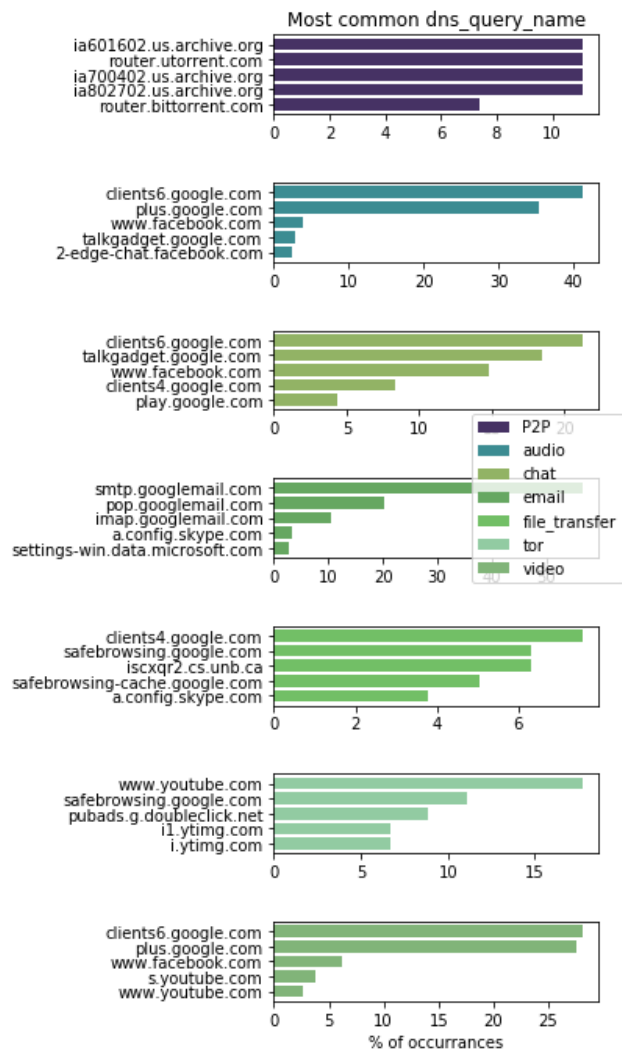}}
\caption{Most common DNS answer IP and query names \\ (non-vpn2016 dataset)}
\label{common_dns_nonvpn}
\vspace{5in}
\end{figure}

\begin{figure}[t]
\subfigure{\includegraphics[height=1.2in,width=1.65in]{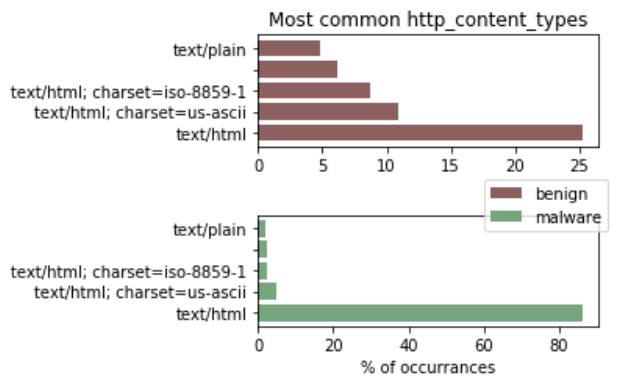}}
\subfigure{\includegraphics[height=1.2in,width=1.65in]{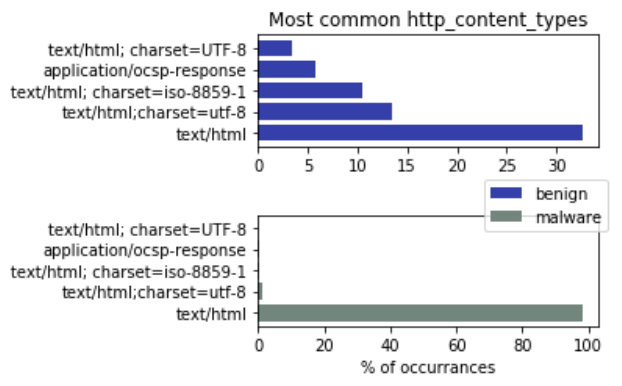}}
\subfigure{\includegraphics[height=1.2in,width=1.65in]{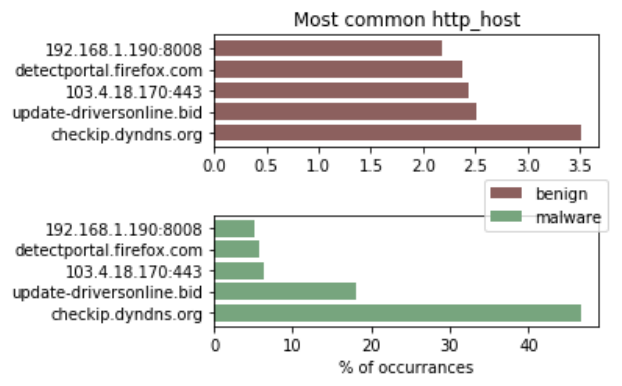}}
\subfigure{\includegraphics[height=1.2in,width=1.65in]{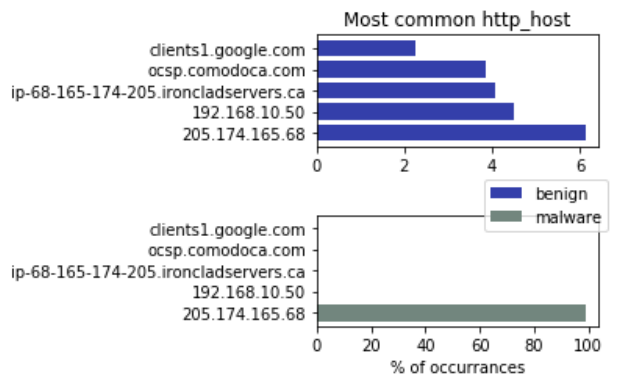}}
\subfigure{\includegraphics[height=1.2in,width=1.65in]{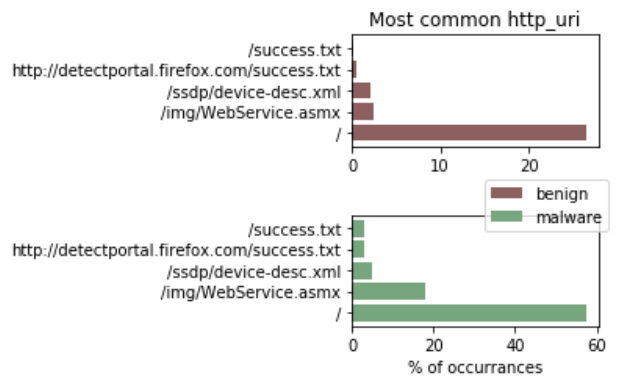}}
\subfigure{\includegraphics[height=1.2in,width=1.65in]{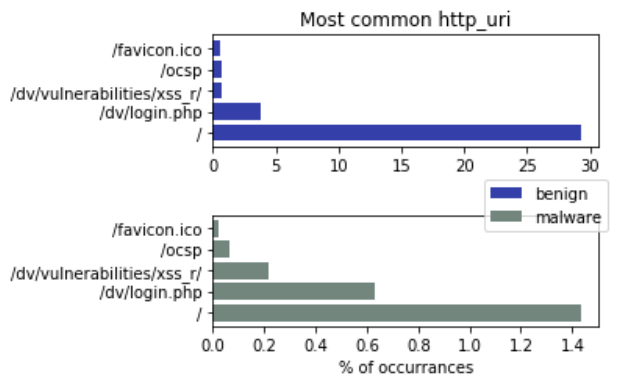}}
\caption{Most common HTTP content type, host and uri \\ (Left: NetML dataset , Right: CICIDS2017 dataset)}
\label{common_http_malware}
\end{figure}

\begin{figure}[t]
\subfigure{\includegraphics[height=1.2in,width=1.65in]{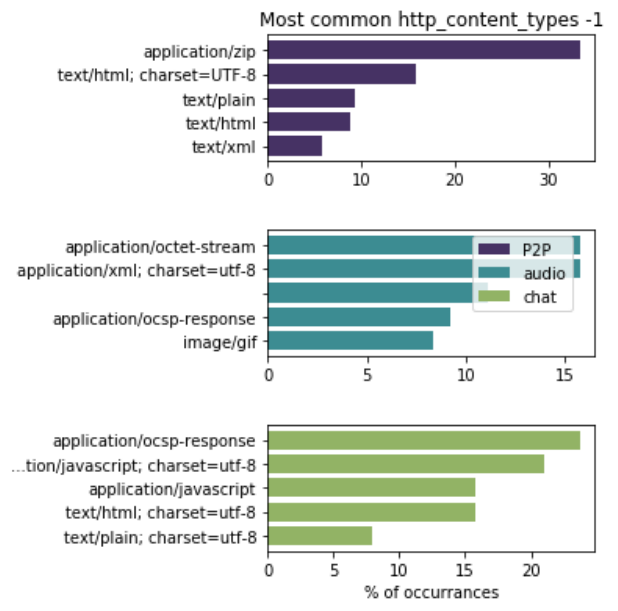}}
\subfigure{\includegraphics[height=1.2in,width=1.65in]{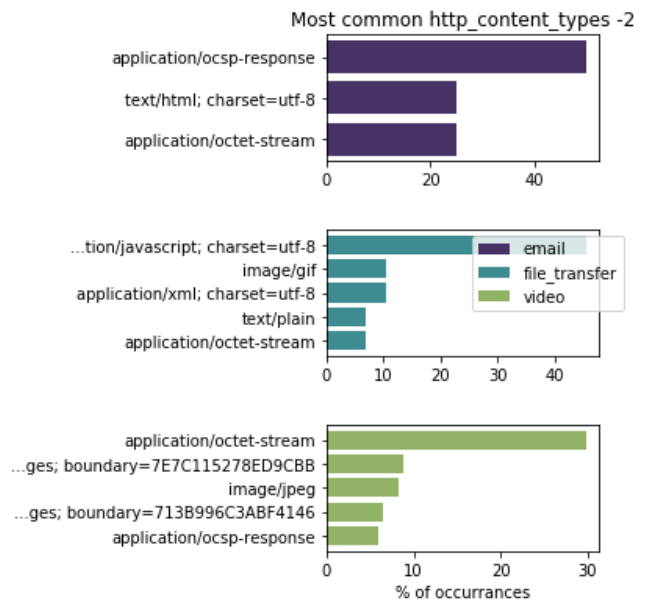}}
\subfigure{\includegraphics[height=1.2in,width=1.65in]{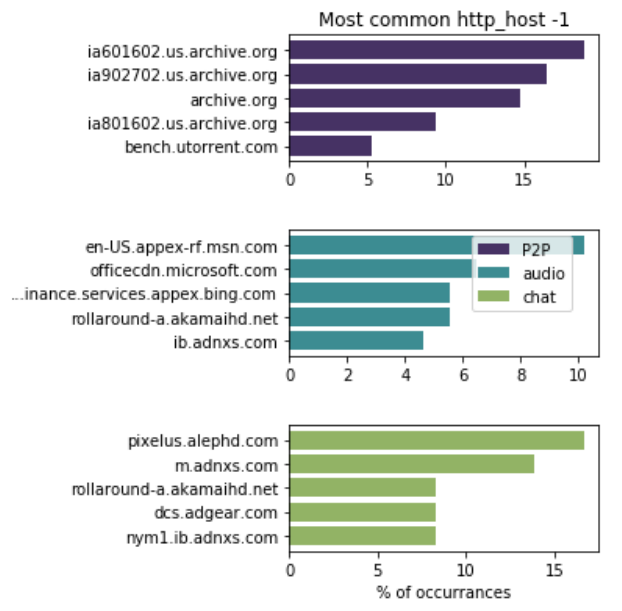}}
\subfigure{\includegraphics[height=1.2in,width=1.65in]{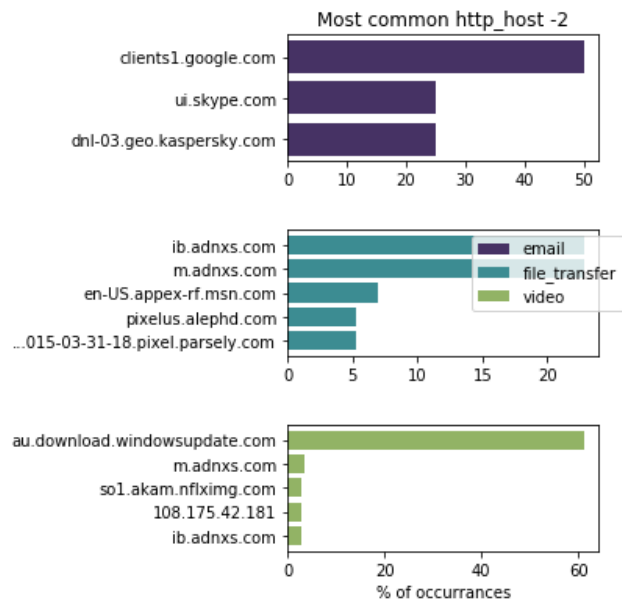}}
\subfigure{\includegraphics[height=1.2in,width=1.65in]{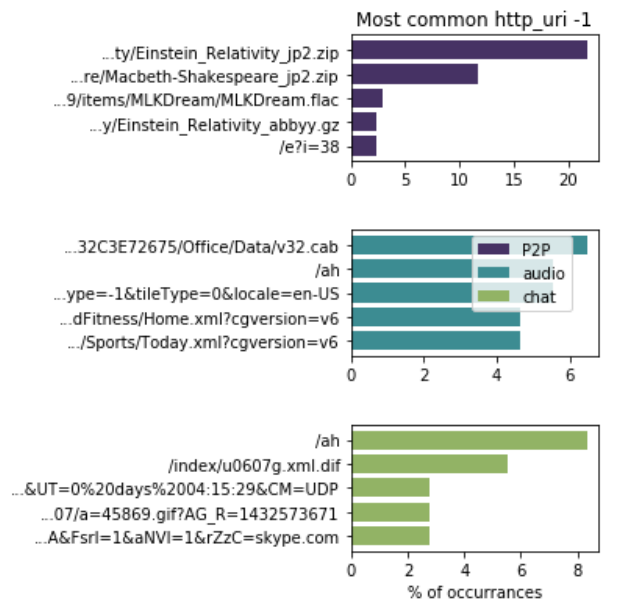}}
\subfigure{\includegraphics[height=1.2in,width=1.65in]{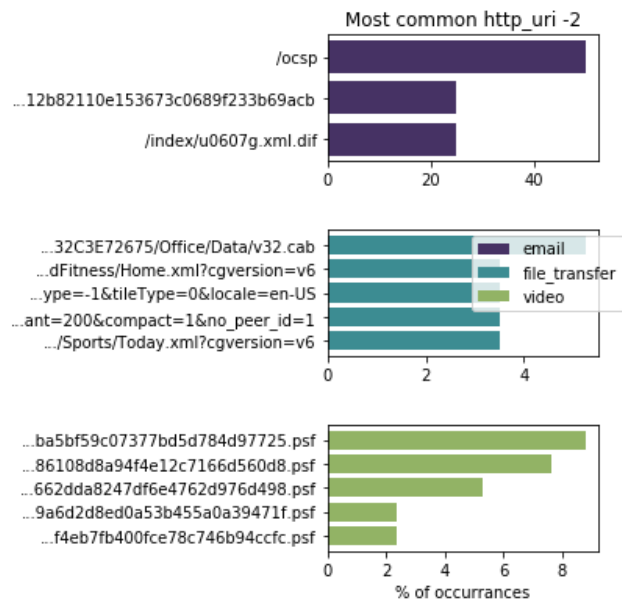}}
\caption{Most common HTTP content type, host and uri \\ (non-vpn2016 dataset)}
\label{common_http_nonvpn}
\end{figure}

\end{document}